\documentclass[3p,times]{elsarticle}

\usepackage{amssymb}
\usepackage{amsmath}
\usepackage{soul,xcolor}
\usepackage{caption}
\usepackage{subcaption}
\usepackage{graphicx}
\usepackage{tikz}
\usepackage{tkz-euclide}
\usepackage{tikz-dimline}
\usepackage{pgfplots}
\usetikzlibrary{patterns, pgfplots.fillbetween, bending, shapes}
\usepackage{color}
\usepackage{hyperref}
\usepackage[ruled,vlined]{algorithm2e}
\usepackage{comment}
\usepackage{listings}
\usepackage{seqsplit}

\usepackage{graphicx} 
\usepackage{amsmath}
\usepackage{xcolor}
\usepackage{tikz}
\usetikzlibrary{calc}
\usepackage{pgfplots}
\pgfplotsset{compat=1.17} 
\usepackage{xcolor}
\usetikzlibrary{arrows.meta, calc, positioning}
\usetikzlibrary{arrows}
\usepackage{stanli}
\usepackage{tikz-3dplot}
\usepackage{adjustbox}

\usepackage{xcolor}
\usepackage{listings}

\definecolor{CodeBackground}{RGB}{246,246,240}
\definecolor{CodeGreen}{RGB}{0,140,0}
\definecolor{CodeMagenta}{RGB}{220,0,140}
\definecolor{LineNumberGray}{RGB}{120,120,120}

\lstdefinestyle{remotecontrol}{
    language=Python,
    backgroundcolor=\color{CodeBackground},
    basicstyle=\ttfamily\small,
    keywordstyle=\color{CodeMagenta}\bfseries,
    commentstyle=\color{CodeGreen}\bfseries,
    stringstyle=\ttfamily,
    numbers=left,
    numberstyle=\color{LineNumberGray}\scriptsize,
    stepnumber=1,
    numbersep=8pt,
    frame=none,
    columns=fullflexible,
    keepspaces=true,
    showstringspaces=false,
    breaklines=true,
    tabsize=4,
    captionpos=b,
    xleftmargin=1.5em,
    framexleftmargin=1.2em
}

\definecolor{AnnotGray}{RGB}{90,90,90}

\tikzset{
  dimArrow/.style={
    draw=AnnotGray,
    line width=0.75pt,
    <->, >=stealth
  }
}

\pgfdeclareverticalshading{distance}{100pt}{
  rgb(0pt)=(1,0.8,0.8); 
  rgb(5pt)=(1,1,1); 
  rgb(50pt)=(0.2,0.2,1) 
}

\definecolor{codebg}{rgb}{0.97, 0.97, 0.97}

\lstdefinestyle{kratos}{
  language=Python,
  basicstyle=\ttfamily\scriptsize,       
  backgroundcolor=\color{codebg},        
  keywordstyle=\color{blue}\bfseries,    
  commentstyle=\color{gray},             
  stringstyle=\color{red},               
  numbers=left,                          
  numberstyle=\tiny\color{gray},         
  stepnumber=1,
  numbersep=5pt,
  showstringspaces=false,
  breaklines=true,
  frame=single,
  captionpos=b,
  tabsize=4,
  morekeywords={Parameters, Model}
}

\usetikzlibrary{shapes.misc}
\tikzset{cross/.style={cross out, draw=black, minimum size=2*(#1-\pgflinewidth), inner sep=0pt, outer sep=0pt}, cross/.default={1pt}}

\newdefinition{definition}{Definition}
\newdefinition{rem}{Remark}

\newcommand{\ignore}[1]{}

\journal{Engineering with Computers}

\begin{document}

\begin{frontmatter}



\title{Partitioned Co-Simulation for CAD-integrated Vibroacoustic Problems in Unbounded Domains}

\author[statik_tum]{Juan Ignacio Camarotti\fnref{equal}\corref{ca}}
\ead{juan.camarotti@tum.de}
\author[deme,flanders]{Philip Le\fnref{equal}}
\ead{philip.le@kuleuven.be}
\author[demel,flanders]{Yinshan Cai}
\ead{yinshan.cai@kuleuven.be}
\author[statik_tum]{Ricky Aristio}
\ead{ricky.aristio@tum.de}
\author[deSE,flanders]{Dionysios Panagiotopoulos}
\ead{dionysios.panagiotopoulos@kuleuven.be}
\author[deme,flanders]{Elke Deckers}
\ead{elke.deckers@kuleuven.be}
\author[statik_tum]{Roland Wüchner}
\ead{wuechner@tum.de}

\fntext[equal]{Juan Ignacio Camarotti and Philip Le share first authorship and contributed equally to this work.}
\address[statik_tum]{Chair of Structural Analysis, Technical University of Munich, Arcisstr. 21, 80333 München, Germany}
\address[deme]{Department of Mechanical Engineering, Campus Diepenbeek, KU Leuven, Wetenschapspark 27, Diepenbeek, B-3590, Belgium}
\address[demel]{Department of Mechanical Engineering, KU Leuven, Celestijnenlaan 300, B-3001, Heverlee, Belgium}
\address[flanders]{FlandersMake @KU Leuven, Leuven, Belgium}
\address[deSE]{Department of Mechanical Engineering, Campus De Nayer, KU Leuven, Jan Pieter de Nayerlaan 5, B-2680, St. Katelijne Waver, Belgium}
\cortext[ca]{Corresponding author}

\begin{abstract}

Vibroacoustic analysis often requires coupling structural and acoustic solvers based on different numerical formulations and discretizations, making monolithic implementations intrusive and limiting software modularity and reuse. This work presents a partitioned co-simulation framework for exterior vibroacoustic analysis that couples an Isogeometric boundary representation analysis (IBRA) structural solver with an isogeometric boundary element method (IGA-BEM) acoustic solver. The methodology operates directly on the computer-aided design (CAD) boundary representation, preserving the exact geometry throughout the analysis and supporting both weak and strong coupling between non-conforming discretizations.
A key contribution is the extension of the Aitken dynamic  relaxation and Interface Quasi-Newton with Inverse Least-Squares (IQN-ILS) convergence accelerators to complex-valued interface quantities, allowing the coupling iterations to account directly for both amplitude and phase information.

The approach is validated using one-way and two-way coupled vibroacoustic benchmark problems involving thin-shell structures and exterior acoustic domains. The results show excellent agreement with monolithic reference solutions, while the proposed complex-valued convergence accelerators improve
the robustness and convergence behavior of the strongly coupled solution
procedure without compromising solution accuracy. These results demonstrate that the proposed approach provides an accurate, robust, and modular approach for CAD-integrated frequency-domain vibroacoustic analysis. 

\end{abstract}




\end{frontmatter}


\section{Introduction}
\label{sec:introduction}
The control of structural vibration and noise radiation has become an increasingly important aspect of modern engineering design in sectors such as automotive, aerospace, transportation, and renewable energy. The widespread adoption of lightweight structures further increases the importance of accurately predicting structural--acoustic interactions, as these systems are generally more susceptible to vibroacoustic excitation. Consequently, reliable computational tools have become essential for evaluating and optimizing the dynamic and acoustic performance of engineering systems during the virtual design stage, reducing the reliance on costly experimental prototypes. In many industrial applications, frequency-domain analyses are of particular interest because they enable the characterization of steady-state structural and acoustic responses over broad frequency ranges, providing valuable information for noise, vibration, and harshness (NVH) assessment and product optimization. 


For structural dynamics and bounded acoustic domains, the Finite Element Method (FEM)~\cite{zienkiewicz2005finite,bathe2006finite,fish2007first} provides a flexible and widely adopted numerical method. However, exterior acoustic problems require special treatment of the unbounded domain, typically through domain truncation techniques such as infinite elements~\cite{burnett1994three,astley2000numerical} or perfectly matched layers (PML)~\cite{berenger1994perfectly}. In contrast, the Boundary Element Method (BEM)~\cite{brebbia1978boundary,banerjee1981boundary} inherently satisfies the Sommerfeld radiation condition while reducing the problem dimensionality to the boundary. Therefore, coupled FEM--BEM formulations have become an established approach for exterior structural-acoustic interaction problems. These formulations are, however, commonly implemented using monolithic solution strategies, where the structural and acoustic systems are assembled into a single coupled system.

In parallel with these developments, significant advances have also been achieved in CAD-integrated numerical discretizations. Isogeometric Analysis (IGA)~\cite{HUGHES20054135} and its extension to Isogeometric B-Rep Analysis (IBRA)~\cite{Breitenberger2015,Teschemacher2018,Teschemacher2022} enable simulation directly on spline-based CAD boundary representations, avoiding geometry reconstruction while preserving exact geometric representation throughout the analysis process. Compared with conventional finite and boundary element discretizations, isogeometric formulations such as IGA-FEM and IGA-BEM~\cite{SIMPSON2014265,COOX2017186} provide highly smooth approximation spaces and reduce pollution effects~\cite{diwan2019pollution}, which are particularly advantageous for wave propagation problems. The combination of CAD-based structural discretizations with IGA-BEM is especially attractive for exterior acoustic problems, since the boundary discretization required by BEM is directly available from the CAD model. Several monolithic isogeometric structural acoustic formulations have therefore been developed for frequency-domain analyses~\cite{liu2018isogeometric,wu20203d,wu2021isogeometric}.

Despite these advances, most frequency-domain vibroacoustic formulations remain based on monolithic coupling strategies~\cite{Everstine1990,JungerFeit1986}, in which the structural and acoustic governing equations are assembled into a single coupled system. Although such approaches provide accurate and mathematically consistent solutions, they require intrusive integration of the participating solvers and limit the reuse of independently developed simulation software. This restriction becomes particularly relevant in industrial workflows, where structural and acoustic analyses are often performed using dedicated commercial or proprietary solvers with different numerical formulations, meshes, and non-conforming interface discretizations. Since these solvers are often available only as executable software without access to their source code, intrusive modifications required by monolithic coupling strategies are generally infeasible. Integrating such heterogeneous software within a unified monolithic approach therefore requires substantial implementation effort and reduces software modularity, interoperability, and maintainability.

Partitioned approaches address these limitations by decomposing the coupled problem into individual physical subproblems, each governed by its own discretization and solution algorithm, while enforcing the coupling conditions through the exchange of interface quantities across the coupling interface. When these subproblems are solved using independent simulation codes whose execution, synchronization, and communication are coordinated through a dedicated coupling tool, the partitioned methodology is commonly referred to as co-simulation~\cite{SICKLINGER2015134,sicklinger2014,bucher2024}. Such a framework preserves the independence of the participating solvers while coordinating their execution and communication throughout the coupled analysis. Several partitioned structural--acoustic formulations have been investigated in recent years. Rodríguez-Tembleque et al.~\cite{RodriguezTembleque2015} developed a frequency-domain FEM-BEM partitioned formulation using mortar and localized Lagrange multiplier coupling strategies for non-matching interfaces. Bunting and Miller~\cite{Bunting2021} introduced a staggered coupling approach for \emph{time-domain} structural acoustics, while Kersschot et al.~\cite{Kersschot2020} demonstrated the applicability of partitioned co-simulation for \emph{time-domain} vibroacoustic systems by coupling a linearized Euler flow-acoustic solver with a structural solver through the preCICE communication library~\cite{preCICEv2}. Although these studies demonstrate the potential of partitioned methodologies, the integration of CAD-integrated discretizations within a co-simulation framework for \emph{frequency-domain} vibroacoustic analysis has received no attention.  Moreover, the iterative solution of strongly coupled partitioned problems poses additional challenges related to the robustness and efficiency of the coupling procedure, making effective convergence acceleration essential.

A key ingredient of strongly coupled partitioned simulations is the use of convergence acceleration techniques to stabilize and accelerate the fixed-point iterations enforcing the coupling conditions. These methods can generally be classified into two groups~\cite{Kuttler2008,DEGROOTE2010446}. The first comprises relaxation-based methods, which improve convergence by applying a global scaling factor to the interface correction, the most widely used representative being the Aitken dynamic relaxation method~\cite{Kuttler2008, irons1969}. The second comprises quasi-Newton methods, which exploit information from previous coupling iterations to approximate the interface Jacobian and construct improved interface corrections. Among these approaches, the Interface Quasi-Newton with Inverse Least-Squares (IQN-ILS) method~\cite{DEGROOTE2010446,DEGROOTE2009793} has become one of the most widely adopted techniques in partitioned multiphysics simulations.

In contrast to time-domain formulations, frequency-domain vibroacoustic analyses involve complex-valued interface quantities that simultaneously contain amplitude and phase information. However, existing implementations of Aitken, IQN-ILS, and related convergence accelerators have been developed primarily for real-valued interface vectors, requiring the real and imaginary parts to be treated separately. Such a treatment neglects the inherent complex structure of the coupled problem and may deteriorate convergence or increase the number of coupling iterations in strongly coupled vibroacoustic simulations, particularly in the vicinity of structural resonances.

In that context, this work proposes a co-simulation framework for the partitioned solution of CAD-integrated frequency-domain exterior vibroacoustic problems, combining an IBRA structural formulation with an IGA-BEM acoustic solver. The proposed framework enables the non-intrusive coupling of independently developed structural and acoustic solvers while supporting non-conforming spline-based interface discretizations through nearest-neighbor, nearest-element, and mortar-based mapping strategies. Furthermore, complex-valued extensions of the Aitken dynamic relaxation and the IQN-ILS convergence accelerators are introduced to improve the robustness and efficiency of strongly coupled frequency-domain simulations. The proposed methodology is validated through representative one-way and two-way coupled problems and compared against corresponding monolithic and analytical reference solutions.

The remainder of this paper is organized as follows. Section~\ref{sec:problem_formulation} presents the governing equations of the coupled frequency-domain vibroacoustic problem, including the structural and acoustic formulations, coupling conditions, and monolithic and partitioned solution strategies. Section~\ref{sec:partitioned_strategy} describes the proposed co-simulation framework, including the coupling algorithms, and the communication strategy between the participating solvers. Section~\ref{sec:interface} addresses the treatment of the coupling interface, presenting the considered mapping techniques for non-conforming discretizations and the convergence acceleration strategies employed for the strongly coupled partitioned solution. Section~\ref{sec:numerical_results} assesses the proposed methodology through representative one-way and two-way coupled benchmark problems and compares the obtained results against corresponding monolithic reference solutions. Finally, Section~\ref{sec:conclusions} summarizes the main findings and concludes the paper.

\section{Problem formulation}
\label{sec:problem_formulation}

This section introduces the mathematical formulation of the coupled vibroacoustic problem considered in this work. First, the FEM and indirect BEM discretized governing equations of the structural and acoustic subproblems are presented. Subsequently, the vibroacoustic coupling conditions are introduced and the corresponding monolithic and partitioned solution strategies are discussed, providing the foundation for the partitioned co-simulation framework developed in the following sections.

\subsection{Structural model}
\label{sec:structural_model}

Assuming steady-state harmonic excitation at angular frequency $\omega$, the structural response is governed by
\begin{equation}
\left(
-\omega^2 \mathbf{M}
+
\mathrm{i}\omega \mathbf{C}
+
\mathbf{K}
\right)
\hat{\mathbf{u}}(\omega)
=
\hat{\mathbf{F}}_{\mathrm{m}}(\omega)
+
\hat{\mathbf{F}}_{\mathrm{a}}(\omega).
\label{eq:structural_frequency_domain}
\end{equation}
Here, $\mathbf{M}$, $\mathbf{C}$, and $\mathbf{K} \in \mathbb{R}^{N_s \times N_s}$ denote the mass, damping, and stiffness matrices, respectively, while $\hat{\mathbf{u}}(\omega) \in \mathbb{C}^{N_s}$ contains the complex-valued structural degrees of freedom. The right-hand side consists of a mechanical load contribution $\hat{\mathbf{F}}_{\mathrm{m}}(\omega) \in \mathbb{C}^{N_s}$ and an acoustic load contribution $\hat{\mathbf{F}}_{\mathrm{a}}(\omega) \in \mathbb{C}^{N_s}$ arising from the pressure acting on the vibroacoustic interface $\Gamma_{\mathrm{sa}}$. The acoustic loading contribution is obtained from the pressure field computed by the acoustic solver and transferred to the structural interface degrees of freedom through an appropriate mapping procedure. 

\ignore{
In this paper, the structural domain is discretized using the isogeometric Kirchhoff--Love shell formulation proposed by Kiendl et al.~\cite{Kiendl2009IGA_KL}. Since transverse shear deformation is neglected, the shell is described solely by its midsurface, and the normal to the midsurface remains normal in the deformed configuration. Although the underlying formulation is geometrically nonlinear and suitable for large deformations, only its linearized form is considered in the present work. The components of the stiffness matrix $\mathbf{K}$ are obtained as follows

\begin{equation}
K_{rs}
=
\int_A
\left(
\frac{\partial \mathbf{n}}{\partial u_s}
:
\frac{\partial \varepsilon}{\partial u_r}
+
\frac{\partial \mathbf{m}}{\partial u_s}
:
\frac{\partial \kappa}{\partial u_r}
\right)
\, dA,
\end{equation}
where the first term represents the membrane contribution through membrane stresses $\mathbf{n}$ and strains $\varepsilon$, while the second term represents the bending contribution through bending stresses $\mathbf{m}$ and curvatures $\kappa$.
}

In this paper, the structural domain is discretized using spline-based isogeometric shell elements operating directly on the CAD boundary representation, thereby preserving the exact geometry throughout the analysis. Both the Kirchhoff--Love shell formulation of Kiendl et al.~\cite{Kiendl2009IGA_KL} and the Reissner--Mindlin shell formulation proposed by Benson et al.~\cite{BENSON2010276} are considered. While the former requires at least $C^1$-continuity of the displacement field both within patches and across patch interfaces, the latter requires $C^0$ across patch interfaces. Since the focus of the present work is the partitioned vibroacoustic coupling framework rather than the structural discretization itself, the interested reader is referred to~\cite{Breitenberger2015, Kiendl2009IGA_KL,BENSON2010276} for details on the employed shell formulations.



Rayleigh damping is optionally considered in this paper, with
\begin{equation}
\mathbf{C}
=
\alpha \mathbf{M}
+
\beta \mathbf{K},
\end{equation}
where $\alpha$ and $\beta$ denote the Rayleigh damping coefficients.

\subsection{Acoustic model}
\label{sec:acoustic_model}
Given the homogeneous Helmholtz equation, deploying the Green's function $G: \ \mathbb{R}^{3} \times \mathbb{R}^{3} \to \mathbb{C}$ and considering two sides of the boundary surface $\Gamma$, the indirect boundary integral equation
\begin{equation}
\hat{p}_{\mathrm{a}}(\mathbf{r})
=
\int_{\Gamma_{\mathrm{a}}}
\left[
\hat{\mu}(\mathbf{r}_{\mathrm{f}})
\frac{\partial G(\mathbf{r},\mathbf{r}_{\mathrm{f}})}{\partial \mathbf{n}}
-
G(\mathbf{r},\mathbf{r}_{\mathrm{f}})
\hat{\sigma}(\mathbf{r}_{\mathrm{f}})
\right]
\,\mathrm{d}\Gamma(\mathbf{r}_{\mathrm{f}}),
\qquad
\mathbf{r} \in \Omega \setminus \Gamma,
\end{equation}
can be derived. In this equation, $\hat{p}_{\mathrm{a}}(\mathbf{r})$ denotes the acoustic pressure at the field point $\mathbf{r}$ in domain $\Omega$ and $\mathbf{r}_{f}$ is the source point on $\Gamma$. In addition, the single $\sigma(\mathbf{r}_{f})$ and double layer potential $\mu(\mathbf{r}_{f})$ are the difference of the normal pressure gradient and pressures between two sides of $\Gamma$, respectively.

Within the coupled vibroacoustic problem, the structural and acoustic subdomains exchange information through the vibroacoustic interface $\Gamma_{\mathrm{sa}}$. The normal structural velocity $\hat{v}_n$ is communicated to the acoustic solver and imposed as a Neumann boundary condition,

\begin{equation}
\frac{\partial \hat{p}_{\mathrm{a}}}{\partial \mathbf{n}}
=
-\mathrm{i}\omega\rho_{\mathrm{a}} \hat{v}_n,
\qquad
\mathbf{r} \in \Gamma_{\mathrm{sa}},
\end{equation}
where $\rho_{\mathrm{a}}$ denotes the fluid density and $\omega \in \Psi$ is the angular frequency with $\Psi:=[\omega_{\mathrm{min}},\omega_{\mathrm{max}}]$. 

The acoustic boundary is represented using spline-based isogeometric discretizations, ensuring an exact geometric description and a geometrically consistent interface with the structural model. Practical CAD models are typically composed of multiple NURBS patches with independent parameterizations and discretizations. Because neighboring patches are generally non-conforming, continuity across shared interfaces must be enforced through an appropriate coupling strategy. For acoustic analyses, at least $C^0$-continuity is required throughout the computational domain. In this work, this requirement is fulfilled using the strong patch coupling technique for non-conforming NURBS surfaces proposed by Coox et al.~\cite{COOX2017235}.

Following a Galerkin discretization of the boundary integral equations, the acoustic problem is expressed as

\begin{equation}
\mathbf{A}(\omega)\,\mathbf{x}(\omega)
=
\mathbf{b}(\omega),  \qquad \omega \in \Psi,
\end{equation}
where $\mathbf{A}: \Psi \to \mathbb{C}^{N \times N}$ ,
$\mathbf{x}: \Psi \to \mathbb{C}^{N}$ contains the discrete unknowns associated with the
layer potentials, and $\mathbf{b}: \Psi \to \mathbb{C}^{N}$. Further details on indirect IGA-BEM formulations for acoustic analysis can be found in~\cite{COOX2017186}.

After solving the acoustic problem, the interface loads resulting from the
acoustic pressure field are transferred to the structural model through the
interface mapping procedure and applied to the shell mid-surface. Further
details on the transfer of the interface quantities and the resulting
vibroacoustic coupling are provided in
Section~\ref{sec:vibroacoustic_coupling}.

\subsection{Vibroacoustic coupling conditions and solution strategies}
\label{sec:vibroacoustic_coupling}

The vibroacoustic interaction between the structural and acoustic domains is enforced through coupling conditions prescribed on the common interface $\Gamma_{\mathrm{sa}}$, ensuring consistency of kinematic and dynamic quantities in the frequency domain. These interface conditions can be enforced either within a monolithic formulation, where both fields are solved simultaneously, or within a partitioned framework, where independent structural and acoustic solvers exchange interface quantities.

The kinematic coupling condition enforces continuity of the normal velocity at the interface. The normal particle velocity of the acoustic medium is prescribed by the structural motion according to

\begin{equation}
\hat{v}_n(\mathbf{r},\omega)
=
\mathrm{i}\omega\,\hat{u}_n(\mathbf{r},\omega),
\qquad
\mathbf{r} \in \Gamma_{\mathrm{sa}},
\end{equation}
where $\hat{u}_n$ denotes the normal component of the structural displacement field and $\hat{v}_n$ is imposed as a Neumann boundary condition on the acoustic problem.

The dynamic coupling condition enforces equilibrium of tractions at the vibroacoustic interface. For thin-walled structures, the structural loading arises from the acoustic pressure acting on the shell surface. The resulting acoustic traction is given by

\begin{equation}
\hat{\mathbf{t}}_{\mathrm{a}}(\mathbf{r})
=
\hat{\mu}(\mathbf{r})\,\mathbf{n},
\qquad
\mathbf{r} \in \Gamma_{\mathrm{sa}},
\end{equation}
where $\mathbf{n}$ denotes the outward unit normal vector of the structural domain. The corresponding discrete load contribution reads

\begin{equation}
\mathbf{F}_{\mathrm{a}}
=
\int_{\Gamma_{\mathrm{sa}}}
\hat{\mu}(\mathbf{r})
\,\mathbf{n}\,
N_i(\mathbf{r})
\,\mathrm{d}\Gamma,
\end{equation}
with $N_i$ denoting the acoustic basis functions.

Based on these coupling conditions, vibroacoustic systems can be solved using either monolithic or partitioned strategies. In a monolithic formulation, the structural and acoustic equations are assembled into a single coupled system and solved simultaneously. After spatial discretization, the frequency-domain problem can be written as

\begin{equation}
\begin{bmatrix}
\mathbf{A}_{\mathrm{s}}(\omega) & \mathbf{C}_{\mathrm{sa}}(\omega) \\
\mathbf{C}_{\mathrm{as}}(\omega) &-\frac{1}{\omega^2\rho_a} \mathbf{A}_{\mathrm{a}}(\omega)
\end{bmatrix}
\begin{bmatrix}
\hat{\mathbf{u}}(\omega) \\
\hat{\mathbf{x}}(\omega)
\end{bmatrix}
=
\begin{bmatrix}
\hat{\mathbf{F}}_{\mathrm{s}}(\omega) \\
-\frac{1}{\omega^2\rho_a}\hat{\mathbf{F}}_{\mathrm{a}}(\omega)
\end{bmatrix}.
\end{equation}
Here, $\mathbf{A}_{\mathrm{s}}(\omega)=
-\omega^2\mathbf{M}
+\mathrm{i}\omega\mathbf{C}
+\mathbf{K}$ denotes the structural dynamic stiffness operator introduced in Eq.~\eqref{eq:structural_frequency_domain}, while $\mathbf{A}_{\mathrm{a}}(\omega)$ represents the acoustic operator. The matrices $\mathbf{C}_{\mathrm{sa}}$ and $\mathbf{C}_{\mathrm{as}}$ are the discrete coupling operators enforcing the vibroacoustic interface conditions. Monolithic formulations provide a mathematically consistent reference solution in which compatibility and equilibrium are satisfied implicitly through the global system.

Alternatively, partitioned strategies retain separate specialized solvers for the structural and acoustic subproblems and enforce the coupling through an iterative exchange of interface quantities. The coupled problem can then be written as

\begin{align}
\mathbf{A}_{\mathrm{s}}(\omega)\hat{\mathbf{u}}^{(k)}(\omega)
&=
\hat{\mathbf{F}}_{\mathrm{s}}(\omega)
+
\mathbf{C}_{\mathrm{sa}}
\hat{\mathbf{x}}^{(k-1)}(\omega), \\
\mathbf{A}_{\mathrm{a}}(\omega)\hat{\mathbf{x}}^{(k)}(\omega)
&=
\hat{\mathbf{F}}_{\mathrm{a}}(\omega)
+
\mathbf{C}_{\mathrm{as}}
\hat{\mathbf{u}}^{(k)}(\omega),
\end{align}
where $(k)$ denotes the coupling iteration. A single exchange per frequency step yields a weakly coupled strategy, whereas iterative enforcement of the interface conditions results in a strongly coupled scheme.

In practical vibroacoustic applications, the structural and acoustic interface discretizations are generally non-conforming. Consequently, for partitioned approaches, the exchanged interface quantities cannot be transferred directly and require dedicated mapping operators. Within the present framework, all interface data transfers are performed through the Kratos \texttt{MappingApplication} \cite{dadvand2010,dadvand2013}. Structural displacements are transferred from the structural to the acoustic interface, while acoustic forces are mapped back conservatively to the structural side. The employed mapping techniques are described in detail in Section~\ref{sec:data_mapping}.

While monolithic approaches provide a valuable reference solution, partitioned formulations preserve solver modularity and enable the non-intrusive coupling of independent structural and acoustic solvers. This is particularly attractive for FE--BE vibroacoustic analyses, where specialized solvers can be reused without modification. The present work therefore adopts a partitioned co-simulation framework in which the structural and acoustic solvers remain fully independent and communicate exclusively through mapped interface quantities.

\section{Co-simulation strategies for partitioned vibroacoustic systems}
\label{sec:partitioned_strategy}

This section presents the partitioned co-simulation framework adopted in the present work. First, the considered coupling algorithms and the associated vibroacoustic workflow are introduced. Subsequently, the remote-controlled co-simulation strategy and the communication mechanism between the structural and acoustic solvers are described. Particular attention is devoted to the transfer of interface quantities between non-conforming discretizations through dedicated mapping operators. Finally, the fixed-point formulation of the coupled problem and the employed complex-valued convergence acceleration techniques are presented.

\subsection{Coupling algorithms and workflow}
\label{sec:coupling_methods}

In coupled multi-physics simulations, the participating subproblems interact through the exchange of interface quantities. The coupling algorithm defines how this interaction is resolved and, in particular, how interface conditions are satisfied across the coupled domains. Depending on the strength of the interaction and the desired accuracy, different coupling strategies can be employed. In the following, we distinguish between one-way (unidirectional) coupling and two-way (bidirectional) coupling. For two-way coupling, both weak (explicit) and strong (implicit) schemes are considered. In addition, the order in which interface data are exchanged and incorporated into each solver can be described by different communication patterns, most commonly Jacobi- and Gauss--Seidel-type coupling. These concepts form the basis of the partitioned vibroacoustic framework considered in this work and are discussed in detail in the following.

\ignore{
\begin{figure}[h!]
    \centering
    \definecolor{CouplingBlue}{RGB}{0,92,170}
    \definecolor{CouplingBlue}{RGB}{0,92,170}
\definecolor{DestOrange}{RGB}{220,120,0}
\definecolor{ParamGreen}{RGB}{0,140,90}

\begin{tikzpicture}[
  scale=0.86,
  transform shape,
  >={Latex[length=2.1mm]},
  box/.style={
    rectangle,
    rounded corners=2pt,
    minimum width=28mm,
    minimum height=8mm,
    align=center,
    line width=0.9pt,
    font=\small
  },
  rootbox/.style={
    box,
    draw=CouplingBlue,
    fill=CouplingBlue!10,
    text=CouplingBlue!90!black,
    font=\small\bfseries,
    minimum width=34mm
  },
  mainbox/.style={
    box,
    draw=CouplingBlue,
    fill=CouplingBlue!6,
    text=CouplingBlue!90!black,
    minimum width=27mm
  },
  weakbox/.style={
    box,
    draw=DestOrange,
    fill=DestOrange!10,
    text=DestOrange!80!black,
    minimum width=28mm
  },
  strongbox/.style={
    box,
    draw=ParamGreen,
    fill=ParamGreen!10,
    text=ParamGreen!70!black,
    minimum width=28mm
  },
  commbox/.style={
    box,
    draw=black!45,
    fill=black!4,
    text=black!80,
    minimum width=58mm,
    minimum height=9mm
  },
  line/.style={
    ->,
    line width=0.9pt,
    draw=CouplingBlue!75
  },
  softline/.style={
    line width=0.9pt,
    draw=CouplingBlue!60
  }
]

\node[rootbox] (coupling) {Coupling methods};

\node[mainbox, below left=11mm and 17mm of coupling] (oneway) {One-way};
\node[mainbox, below right=11mm and 17mm of coupling] (twoway) {Two-way};

\node[weakbox, below=13mm of twoway, xshift=-19mm] (weak)
{Weak\\[-1pt]\scriptsize explicit};

\node[strongbox, below=13mm of twoway, xshift=+19mm] (strong)
{Strong\\[-1pt]\scriptsize implicit};

\node[commbox, below=12mm of weak, xshift=19mm] (comm)
{Communication pattern: Jacobi / Gauss--Seidel};

\draw[line] (coupling.south) -- ++(0,-6mm) coordinate (h1);
\draw[line] (h1) -- (oneway.north);
\draw[line] (h1) -- (twoway.north);

\draw[line] (twoway.south) -- ++(0,-6mm) coordinate (h2);
\draw[line] (h2) -- (weak.north);
\draw[line] (h2) -- (strong.north);

\coordinate (join) at ($(comm.north)+(0,7mm)$);

\draw[softline] (weak.south) to[out=-90,in=180] (join);
\draw[softline] (strong.south) to[out=-90,in=0] (join);

\draw[line] (join) -- (comm.north);

\end{tikzpicture}
    \caption{Classification of coupling strategies. One-way coupling is unidirectional. Two-way coupling is characterized by a coupling strength (weak/explicit or strong/implicit) and a communication pattern (Jacobi or Gauss--Seidel).}
    \label{fig:coupling_tree}
\end{figure}
}

\subsubsection{Coupling strategies}

In one-way coupling, the interaction is assumed to be unidirectional, meaning that one subproblem influences the other while the feedback effect is neglected. In this case, interface data are transferred from the driving subproblem to the driven subproblem, but no information is transferred back. One-way coupling is computationally efficient and can be sufficient when the neglected feedback remains small compared to the dominant interaction. 

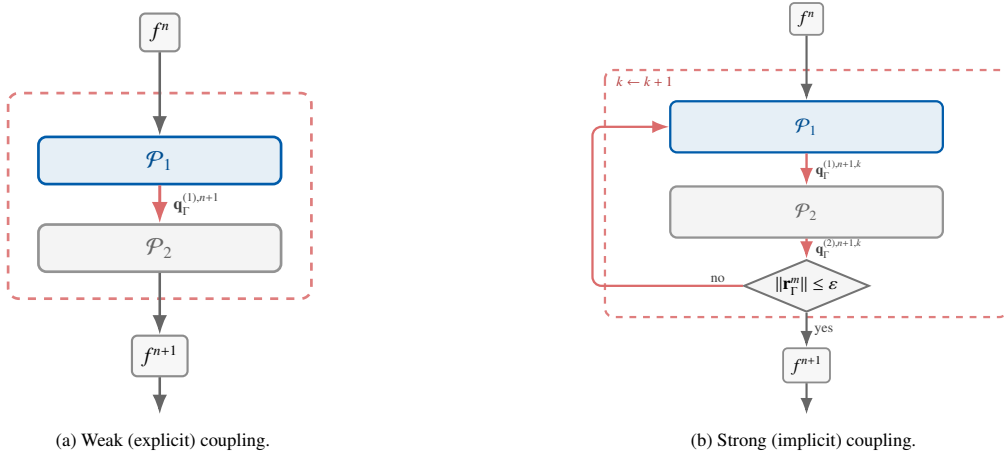
\begin{figure}[h!]
    \centering
    
    \definecolor{MyBlue}{RGB}{0,92,170}
    \definecolor{MyGray}{RGB}{120,120,120}
    \definecolor{MyRed}{RGB}{210,70,70}
    
    \begin{subfigure}{0.41\linewidth}
        \centering
        \adjustbox{valign=c}{
            \resizebox{0.6\linewidth}{!}{%
\begin{tikzpicture}[
  scale=0.88,
  transform shape,
  >={Latex[length=2.2mm]},
  timebox/.style={
    draw=black!45,
    fill=black!3,
    rounded corners=2pt,
    line width=0.8pt,
    inner sep=4pt,
    font=\small\bfseries
  },
  couplingbox/.style={
    draw=MyRed!70,
    dashed,
    rounded corners=4pt,
    line width=1.0pt
  },
  solverA/.style={
    draw=MyBlue,
    fill=MyBlue!10,
    rounded corners=3pt,
    line width=1.0pt,
    minimum width=36mm,
    minimum height=7mm,
    align=center,
    text=MyBlue!85!black,
    font=\bfseries
  },
  solverB/.style={
    draw=MyGray!80,
    fill=MyGray!8,
    rounded corners=3pt,
    line width=1.0pt,
    minimum width=36mm,
    minimum height=7mm,
    align=center,
    text=MyGray!85!black,
    font=\bfseries
  },
  timearrow/.style={
    -Latex,
    line width=0.95pt,
    draw=black!60
  },
  dataarrow/.style={
    -Latex,
    line width=1.1pt,
    draw=MyRed!80
  }
]

\node[timebox] (tk) at (0,3.0) {$f^n$};

\node[timebox] (tkp1) at (0,-1.8) {$f^{n+1}$};

\draw[couplingbox]
(-2.25,2.10) rectangle (2.25,-0.95);

\node[solverA] (P1) at (0,1.10)
{$\mathcal{P}_1$};

\node[solverB] (P2) at (0,-0.20)
{$\mathcal{P}_2$};

\draw[timearrow]
(tk.south) -- (P1.north);

\draw[timearrow]
(P2.south) -- (tkp1.north);

\draw[timearrow]
(tkp1.south) -- ++(0,-0.55);

\draw[dataarrow]
(P1.south) -- (P2.north);

\node[
  font=\scriptsize,
  text=black!75,
  anchor=west
]
at ($(P1.south)!0.50!(P2.north)+(0.08,0)$)
{$\mathbf{q}_{\Gamma}^{(1),n+1}$};

\end{tikzpicture}%
}
        }
        \caption{Weak (explicit) coupling.}
        \label{fig:weak_coupling_panel}
    \end{subfigure}
    \hfill
    \begin{subfigure}{0.56\linewidth}
        \centering
        \adjustbox{valign=c}{
            \resizebox{0.6\linewidth}{!}{%
\begin{tikzpicture}[
  scale=0.88,
  transform shape,
  >={Latex[length=2.2mm]},
  timebox/.style={
    draw=black!45,
    fill=black!3,
    rounded corners=2pt,
    line width=0.8pt,
    inner sep=4pt,
    font=\small\bfseries
  },
  couplingbox/.style={
    draw=MyRed!70,
    dashed,
    rounded corners=4pt,
    line width=1.0pt
  },
  solverA/.style={
    draw=MyBlue,
    fill=MyBlue!10,
    rounded corners=3pt,
    line width=1.0pt,
    minimum width=48mm,
    minimum height=9mm,
    align=center,
    text=MyBlue!85!black,
    font=\bfseries
  },
  solverB/.style={
    draw=MyGray!80,
    fill=MyGray!8,
    rounded corners=3pt,
    line width=1.0pt,
    minimum width=48mm,
    minimum height=9mm,
    align=center,
    text=MyGray!85!black,
    font=\bfseries
  },
  decision/.style={
    draw=black!55,
    fill=black!4,
    diamond,
    aspect=2.4,
    line width=0.9pt,
    align=center,
    font=\small\bfseries,
    inner sep=1.5pt
  },
  timearrow/.style={
    -Latex,
    line width=0.95pt,
    draw=black!60
  },
  dataarrow/.style={
    -Latex,
    line width=1.05pt,
    draw=MyRed!80
  },
  iterarrow/.style={
    -Latex,
    line width=1.0pt,
    draw=MyRed!80,
    rounded corners=4pt
  },
  label/.style={
    font=\scriptsize,
    text=black!75,
    inner sep=1pt
  }
]

\node[timebox] (tk) at (0,3.55) {$f^n$};
\node[timebox] (tkp1) at (0,-2.55) {$f^{n+1}$};

\draw[couplingbox] (-3.55,2.65) rectangle (3.55,-1.70);

\node[solverA] (P1) at (0,1.65) {$\mathcal{P}_1$};
\node[solverB] (P2) at (0,0.15) {$\mathcal{P}_2$};

\node[decision] (conv) at (0,-1.15)
{$\|\mathbf{r}_{\Gamma}^{m}\| \leq \varepsilon$};

\draw[timearrow] (tk.south) -- (P1.north);

\draw[dataarrow]
(P1.south) -- (P2.north);

\node[label, anchor=west]
at ($(P1.south)!0.50!(P2.north)+(0.12,0)$)
{$\mathbf{q}_{\Gamma}^{(1),n+1,k}$};

\draw[dataarrow]
(P2.south) -- (conv.north);

\node[label, anchor=west]
at ($(P2.south)!0.50!(conv.north)+(0.12,0)$)
{$\mathbf{q}_{\Gamma}^{(2),n+1,k}$};

\draw[timearrow]
(conv.south) -- (tkp1.north);

\node[label, anchor=west]
at ($(conv.south)!0.50!(tkp1.north)+(0.10,0)$)
{yes};

\draw[timearrow] (tkp1.south) -- ++(0,-0.55);

\draw[iterarrow]
(conv.west) -- ++(-2.65,0)
-- ++(0,2.80)
-- (P1.west);

\node[label, anchor=east]
at ($(conv.west)+(-0.28,0.12)$)
{no};

\node[
  font=\scriptsize,
  text=MyRed!85!black,
  anchor=south,
  align=center
]
at (-2.85,2.25)
{$k \leftarrow k+1$};

\end{tikzpicture}%
}
        }
        \caption{Strong (implicit) coupling.}
        \label{fig:strong_coupling_panel}
    \end{subfigure}
    
    \caption{
    Weak (explicit) and strong (implicit) coupling schemes using a
    Gauss--Seidel execution pattern.
    }
    \label{fig:weak_strong_coupling_schemes}

\end{figure}

In two-way coupled simulations, both subproblems influence each other through the bidirectional exchange of interface quantities. Weak coupling, also referred to as \emph{explicit}, \emph{staggered}, or \emph{loose} coupling, performs a single exchange of interface data and solves each subproblem once per frequency step (see Figure ~\ref{fig:weak_strong_coupling_schemes}(a)). The absence of an iterative enforcement of the interface conditions can reduce robustness and accuracy, particularly in strongly coupled configurations or when the coupled response is highly sensitive to the interface quantities.

\ignore{
\begin{figure}[h!]
    \centering
    \definecolor{MyBlue}{RGB}{0,92,170}
\definecolor{MyRed}{RGB}{210,70,70}
\definecolor{MyGray}{RGB}{120,120,120}
    
\begin{tikzpicture}[
      scale=0.85, transform shape,
      >={Latex[length=2.2mm]},
      panel/.style={draw=MyBlue, rounded corners=2pt, line width=1.0pt, inner sep=6pt},
      header/.style={draw=MyBlue, rounded corners=2pt, line width=1.0pt,
                     minimum width=0.44\linewidth, minimum height=7mm,
                     align=center, text=MyBlue, font=\bfseries},
      solverA/.style={draw=MyBlue, fill=MyBlue!15, rounded corners=3pt,
                      line width=1.0pt, minimum width=28mm, minimum height=10mm,
                      align=center, text=MyBlue, font=\small\bfseries},
      solverB/.style={draw=MyGray, fill=MyGray!10, rounded corners=3pt,
                      line width=1.0pt, minimum width=28mm, minimum height=10mm,
                      align=center, text=MyGray, font=\small\bfseries},
      timearrow/.style={-Latex, line width=1.1pt, draw=MyBlue},
      dataarrow/.style={-Latex, line width=1.0pt, draw=MyBlue},
      feedbackarrow/.style={-Latex, line width=1.0pt, draw=MyBlue},
      note/.style={draw=MyBlue, rounded corners=2pt, line width=1.0pt,
                   align=center, text=black, fill=white,
                   minimum width=0.44\linewidth, inner sep=5pt, font=\footnotesize}
    ]
    
    \node[header] (h1) {(a) One-way coupling};
    
    \node[solverA, below=10mm of h1] (A1) {Participant 1};
    \node[solverB, below=10mm of A1] (B1) {Participant 2};
    
    \draw[timearrow] ($(A1.west)+(-18mm,0)$) -- (A1.west)
      node[midway, above] {\scriptsize $t^n$};
    \draw[timearrow] (A1.east) -- ($(A1.east)+(18mm,0)$)
      node[midway, above] {\scriptsize $t^{n+1}$};

    \draw[timearrow] ($(B1.west)+(-18mm,0)$) -- (B1.west)
      node[midway, below] {\scriptsize $t^n$};
    
    \draw[timearrow] (B1.east) -- ($(B1.east)+(18mm,0)$)
      node[midway, below] {\scriptsize $t^{n+1}$};
    
    \draw[dataarrow] (B1.north) -- node[right, text=MyBlue, font=\scriptsize] {Data transfer} (A1.south);
    
    \node[note, below=5mm of B1] (note1)
    {This approach is used when the feedback effect\\
    from the driven subsystem is \textcolor{MyRed}{negligible}.};
    
    
    \node[header, right=14mm of h1] (h2) {(b) Two-way coupling};
    
    \node[solverA, below=10mm of h2] (A2) {Participant 1};
    \node[solverB, below=10mm of A2] (B2) {Participant 2};
    
    \draw[timearrow] ($(A2.west)+(-18mm,0)$) -- (A2.west)
      node[midway, above] {\scriptsize $t^n$};
    \draw[timearrow] (A2.east) -- ($(A2.east)+(18mm,0)$)
      node[midway, above] {\scriptsize $t^{n+1}$};

    \draw[timearrow] ($(B2.west)+(-18mm,0)$) -- (B2.west)
      node[midway, below] {\scriptsize $t^n$};
    
    \draw[timearrow] (B2.east) -- ($(B2.east)+(18mm,0)$)
      node[midway, below] {\scriptsize $t^{n+1}$};
    
    \def\dArrow{10mm}
    
    \draw[dataarrow]
      ([xshift=+\dArrow]A2.south) -- ([xshift=+\dArrow]B2.north);
    
    \draw[feedbackarrow]
      ([xshift=-\dArrow]B2.north) -- ([xshift=-\dArrow]A2.south);
    
    \node[text=MyBlue, font=\scriptsize] at ($(A2.south)!0.5!(B2.north)$) {Data transfer};

    \node[note, below=5mm of B2] (note2)
    {This approach is used when the feedback effect\\
    between subsystems is \textcolor{MyRed}{relevant}.};
    
    
\end{tikzpicture}
    \caption{Schematic comparison of one-way and two-way coupling strategies.}
    \label{fig:oneway_twoway_schematic}
\end{figure}
}

Strong coupling (Figure~\ref{fig:weak_strong_coupling_schemes}(b)), also referred to as \emph{implicit} or \emph{iterative} coupling, aims to enforce the interface conditions by iterating between the participating subproblems until convergence is achieved. In this setting, the coupled solution is obtained by minimizing an interface residual that measures the violation of interface equilibrium and/or compatibility. Let $\mathbf{y}_{\Gamma}^{(k)} \in \mathbb{C}^{N_\Gamma}$ denote the interface quantities exchanged between the coupled subproblems at coupling iteration $k$, where $N_\Gamma$ is the number of interface degrees of freedom. The corresponding interface residual is denoted by $\mathbf{r}_{\Gamma}^{(k)} \in \mathbb{C}^{N_\Gamma}$, and convergence is achieved when its norm falls below the prescribed tolerance $\varepsilon$. 


Strong coupling generally improves stability and accuracy compared to weak coupling, particularly in problems with pronounced interaction between the participating fields, at the expense of increased computational cost due to multiple solver evaluations per coupling step. In practice, relaxation or acceleration techniques can be employed to reduce the number of coupling iterations required for convergence.

Besides the choice of coupling strategy, the coupling procedure can also
be characterized by the employed communication pattern
\cite{MatthiesSteindorf2003,Mehl2016}. In a Jacobi-type communication
pattern, each participant uses interface data from the previous coupling
iteration to advance its solution, such that the participants can be
evaluated independently. In contrast, a Gauss--Seidel-type communication
pattern uses the latest available interface information, meaning that the
subproblems are executed sequentially and downstream solvers immediately
incorporate the most recent updates provided by upstream solvers. Jacobi
patterns can offer higher potential for parallel execution, whereas
Gauss--Seidel patterns often exhibit improved convergence properties due
to the use of up-to-date interface information
\cite{MatthiesSteindorf2003,Mehl2016}.


In the present work, one-way coupling as well as two-way weak and strong coupling strategies are investigated. For the two-way coupled simulations, a Gauss--Seidel communication pattern is employed.

\subsubsection{Vibroacoustic coupling workflow}

\begin{figure}[h!]
    \centering
    \includegraphics[width=0.50\linewidth]{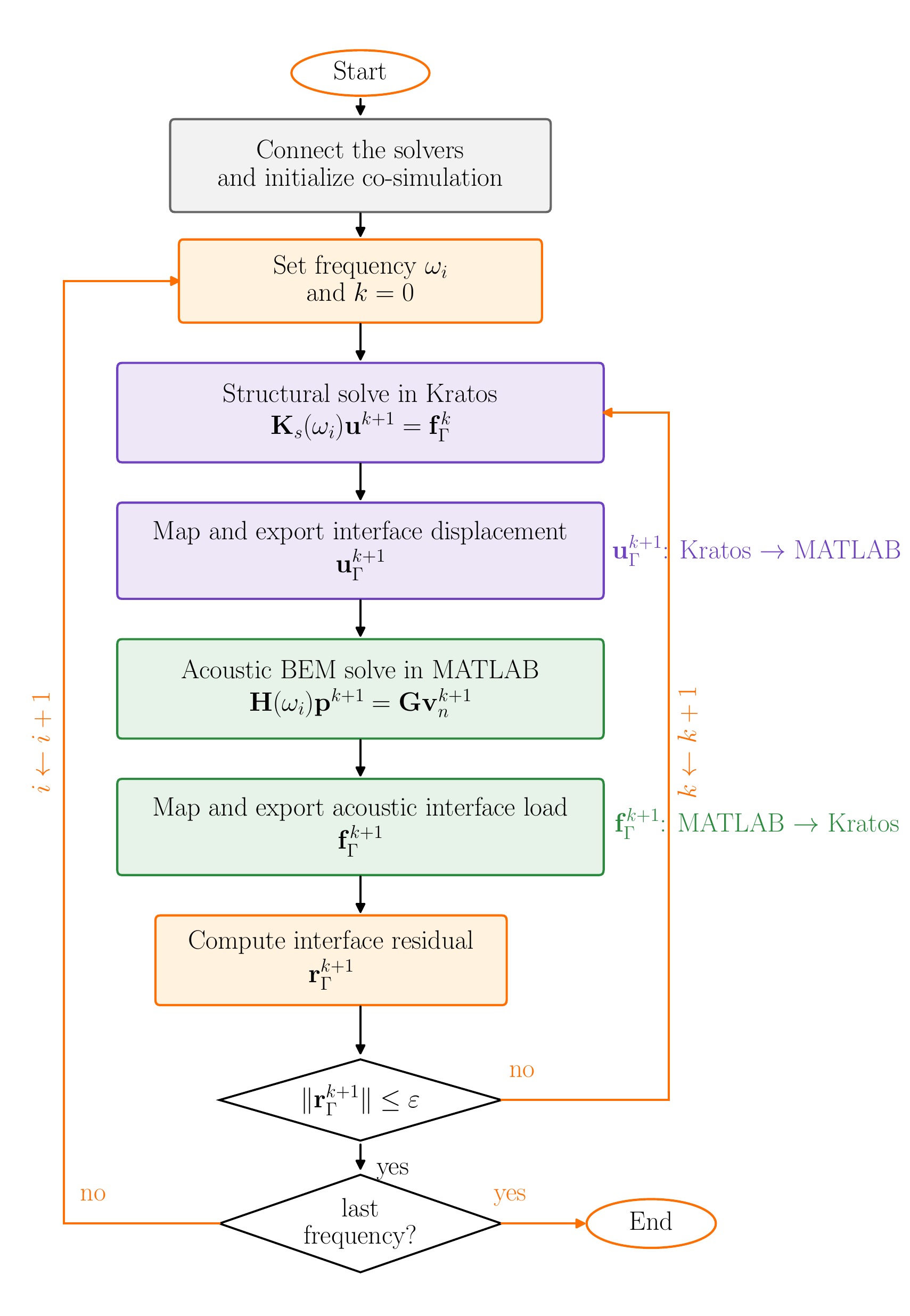}
    \caption{Workflow of the partitioned frequency-domain vibroacoustic coupling strategy.}
    \label{fig:coupling_workflow}
\end{figure}

The overall partitioned frequency-domain vibroacoustic workflow is illustrated in Figure~\ref{fig:coupling_workflow}. The previously introduced coupling strategies are applied through a staggered exchange of interface quantities between the structural and acoustic solvers. 

For a given excitation frequency, the isogeometric structural problem is solved in the open-source multiphysics framework Kratos
Multiphysics~\cite{dadvand2010,dadvand2013}, while the acoustic
Helmholtz problem is solved using a MATLAB-based isogeometric boundary element solver. The structural interface displacements are transferred
to the acoustic solver and imposed as boundary conditions. The
resulting acoustic pressures are converted into equivalent loads acting
on the structure and transferred back to the structural solver. Since
the structural and acoustic interfaces may be discretized
independently, the exchanged quantities are transferred through the
mapping operators described in Section~\ref{sec:data_mapping}. For weak coupling, the interface quantities are exchanged only once per
frequency step. In contrast, strong coupling iteratively updates the
exchanged quantities until the prescribed interface convergence
criterion is satisfied.

\subsection{Integration of the acoustic solver through a remote-controlled approach}
\label{sec:remote_controlled_co_sim}

Since the external acoustic solver is not natively accessible by the
Kratos co-simulation framework, a dedicated integration mechanism is
required to enable its participation in the coupled simulation.

To this end, the acoustic solver is integrated through a
remote-controlled solver wrapper following the concept introduced by
Bucher~\cite{bucher2024}. In this approach, the external solver does
not implement any coupling logic. Instead, it exposes a set of
callback routines associated with different stages of the solution
procedure, such as importing interface data, solving the acoustic
problem, exporting interface quantities, and performing
post-processing operations. These routines are registered during the
initialization stage and subsequently invoked by the co-simulation
controller according to the selected coupling algorithm. After the
registration stage, the external acoustic solver relinquishes control
to the co-simulation framework and remains idle until the execution
of a registered operation is requested.

The main advantage of this approach is the complete separation between
the solver implementation and the coupling algorithm. As a result, the
same acoustic solver wrapper can be used for both weakly and strongly
coupled simulations without modifying the MATLAB code. Changes in the
coupling strategy therefore only require modifications within the
co-simulation framework rather than within the external solver
implementation. Furthermore, the centralized management of the
execution sequence reduces the risk of synchronization issues and
deadlocks when compared to classical co-simulation approaches.

A representative pseudo-code implementation of the remote-controlled
solver wrapper used for the acoustic participant is provided in
\ref{app:remote_controlled_participant}.

\subsection{Data exchange between the solvers}
\label{sec:data_exchange}

\begin{figure}[h!]
    \centering
    \resizebox{1\linewidth}{!}{%
    \begin{tikzpicture}[
            font=\small,
            >=Latex,
            node distance=24mm and 28mm,
            box/.style={
                draw=black!80,
                very thick,
                rounded corners=4pt,
                align=center,
                inner sep=7pt
            },
            struct/.style={box, fill=blue!10, minimum width=4.6cm, minimum height=1.8cm},
            acoustic/.style={box, fill=orange!10, minimum width=4.6cm, minimum height=1.8cm},
            coupling/.style={box, fill=gray!10, minimum width=2.8cm, minimum height=3.2cm},
            link/.style={thick, black!70, <->}
        ]
        \scaling{0.3};
        
        \node[struct] (S) {
        \textbf{Structural solver}\\
        Harmonic analysis\\
        Kratos \texttt{StructuralMechanicsApplication}
        };
        
        \node[coupling, right=of S] (C) {
        \textbf{Coupling interface}\\[2pt]
        Kratos \texttt{CoSimulationApplication}
        };
        
        \node[acoustic, right=of C] (A) {
        \textbf{Acoustic solver}\\
        BEM Helmholtz solver\\
        \emph{MATLAB}
        };
        
        \draw[link] (S.east) -- (C.west)
        node[midway, above, font=\scriptsize] {IPC (\texttt{CoSimIO})};
        
        \draw[link] (C.east) -- (A.west)
        node[midway, above, font=\scriptsize] {IPC (\texttt{CoSimIO})};
    
    \end{tikzpicture}
}
    \caption{Data exchange architecture in the partitioned vibroacoustic co-simulation: interface quantities are exchanged between the structural harmonic solver in Kratos and the MATLAB-based acoustic BEM Helmholtz solver through \texttt{CoSimIO}.}
    \label{fig:data_exchange}
\end{figure}
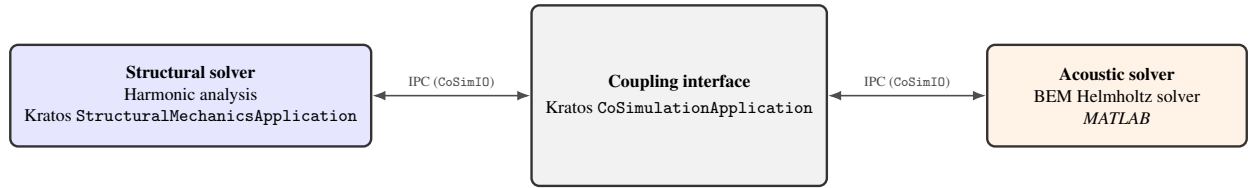

Since the structural and acoustic solvers execute as independent
processes, an interprocess communication (IPC) mechanism is required
for the exchange of interface quantities. In the present work, this
functionality is provided by the tool \texttt{CoSimIO}, which acts as a
detached communication interface between the participating solvers.
The resulting data exchange architecture is illustrated in
Figure~\ref{fig:data_exchange}.

In the present implementation, a file-based communication approach is
employed. Interface quantities are exchanged through a dedicated
communication directory, while synchronization between both
participants is ensured through auxiliary control files. This
guarantees that data are accessed only after being completely written
by the corresponding solver and thereby prevents race conditions
during the coupling procedure.

The choice of a file-based IPC approach is motivated by its robustness,
simplicity, and portability across different programming languages and
execution environments. Although this strategy introduces a higher communication
latency compared to shared-memory or socket-based approaches, the
associated overhead remains acceptable for the frequency-domain analyses
considered in this work.

\section{Interface treatment}  
\label{sec:interface}

In partitioned vibroacoustic simulations, the enforcement of the
coupling conditions across non-conforming interfaces relies on two key
ingredients. First, interface quantities must be transferred
consistently between the generally different discretizations employed by
the structural and acoustic solvers. Second, for strongly coupled
problems, the resulting fixed-point iterations require robust
convergence acceleration techniques to ensure stability and efficiency.
The present section therefore addresses both aspects of the interface
treatment, namely the transfer of interface quantities between
non-conforming discretizations and the convergence acceleration
strategies employed in the partitioned solution procedure.

\subsection{Data mapping between non-conforming discretizations}
\label{sec:data_mapping}

In partitioned vibroacoustic co-simulation, the structural and acoustic solvers generally employ independent discretizations of the vibroacoustic interface. As a consequence, the corresponding interface meshes are in general non-conforming, differing in topology, resolution, and numerical representation. To enable a consistent exchange of interface quantities between the two solvers, a data transfer operator is required to map fields between the respective interface discretizations.

Although the presented framework is not restricted to a specific discretization technology, the considered mapping procedures are particularly well suited for partitioned IGA--IGA coupling, where both structural and acoustic interfaces are represented by spline-based geometries. In such settings, independently parameterized interfaces and non-conforming discretizations naturally arise, making robust and geometrically consistent transfer operators especially important for the coupled analysis procedure. In contrast to standard FEM discretizations, where nodal coordinates are typically located on the physical interface geometry, the control points defining spline-based geometries generally do not lie on the geometry itself. As a consequence, direct control-point-to-control-point transfer strategies commonly employed in FEM--FEM coupling are, in general, not suitable for isogeometric interface coupling.

From an algebraic perspective, the data transfer between the interface
discretizations can be expressed as a linear mapping. Let
$\mathbf{u}_{\mathrm{o}} \in \mathbb{C}^{N_{\mathrm{o}}}$ denote a
vector collecting interface quantities defined on the origin
discretization and
$\mathbf{u}_{\mathrm{d}} \in \mathbb{C}^{N_{\mathrm{d}}}$ the
corresponding quantities on the destination discretization, where
$N_{\mathrm{o}}$ and $N_{\mathrm{d}}$ denote the dimensions of the
origin and destination interface vectors, respectively. The mapping
operation is written as

\begin{equation}
\mathbf{u}_{\mathrm{d}}
=
\mathbf{H}\,\mathbf{u}_{\mathrm{o}},
\end{equation}
where
$\mathbf{H} \in \mathbb{R}^{N_{\mathrm{d}}\times N_{\mathrm{o}}}$ is the
mapping operator whose entries depend on the geometric relationship
between the two interface discretizations and on the chosen mapping
strategy. For vector-valued interface quantities, such as structural
displacements and forces, the mapping is applied independently to each
component using the same mapping operator.

To ensure consistency of the coupled problem, force quantities are
transferred using a conservative mapping. Let
$\mathbf{F}_{\mathrm{d}} \in \mathbb{C}^{N_{\mathrm{d}}}$ denote a
vector of forces defined on the destination interface. The
corresponding forces on the origin discretization,
$\mathbf{F}_{\mathrm{o}} \in \mathbb{C}^{N_{\mathrm{o}}}$, are obtained
as

\begin{equation}
\mathbf{F}_{\mathrm{o}}
=
\mathbf{H}^{\mathrm{T}}\,\mathbf{F}_{\mathrm{d}}.
\end{equation}

This choice guarantees that the virtual work associated with the
interface quantities is preserved under the mapping,

\begin{equation}
\mathbf{F}_{\mathrm{o}}^{\mathrm{T}} \mathbf{u}_{\mathrm{o}}
=
\mathbf{F}_{\mathrm{d}}^{\mathrm{T}} \mathbf{u}_{\mathrm{d}},
\end{equation}
and is therefore essential for maintaining physical consistency in the
partitioned vibroacoustic coupling.

\begin{figure}[h!]
    \centering
    \includegraphics[width=0.70\linewidth]{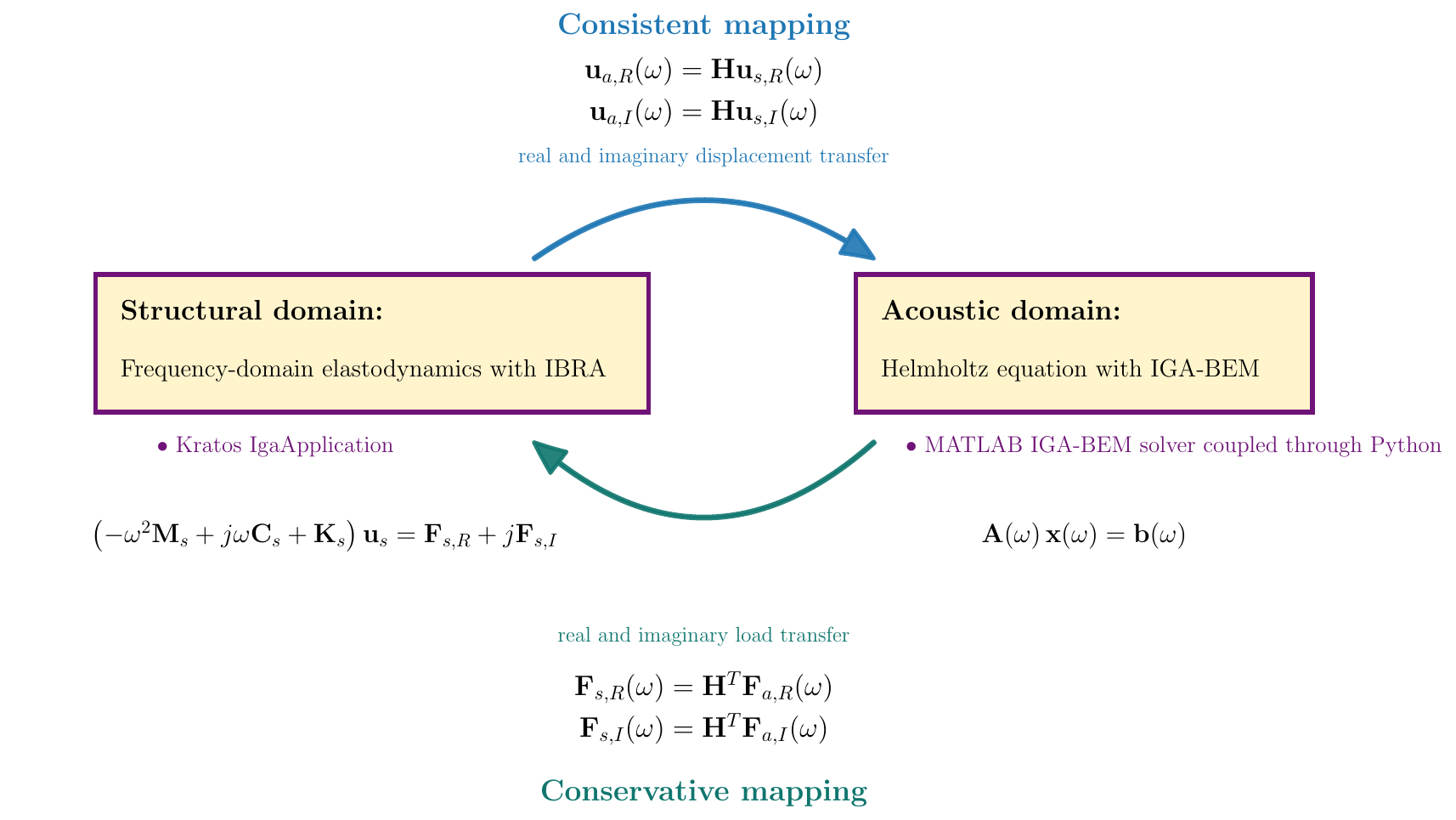}
    \caption{Overview of the proposed partitioned vibroacoustic co-simulation framework. The structural harmonic solver and the acoustic Helmholtz solver exchange real and imaginary interface quantities through consistent displacement transfer and conservative load mapping operators.}
    \label{fig:partitioned_vibroacoustic_framework}
\end{figure}

The overall partitioned vibroacoustic coupling framework together with
the transfer of interface quantities between the structural and acoustic
domains is illustrated in Figure~\ref{fig:partitioned_vibroacoustic_framework}. In the present framework, different mapping strategies are explored for the transfer of interface quantities between the acoustic and structural interfaces. The nearest-neighbor and nearest-element mappers perform pointwise transfers between interface integration points based on geometric proximity and projection operations on the underlying spline interface representations. In contrast, the mortar mapper formulates the transfer as a weak variational projection between the interface discretizations, leading to a control-point-to-control-point transfer operator assembled through numerical integration over the spline coupling interface.

\subsubsection{Nearest-neighbor mapping}
\label{sec:nearest_neighbor_mapping}

\begin{figure}[h!]
    \centering
    \includegraphics[width=0.55\linewidth]{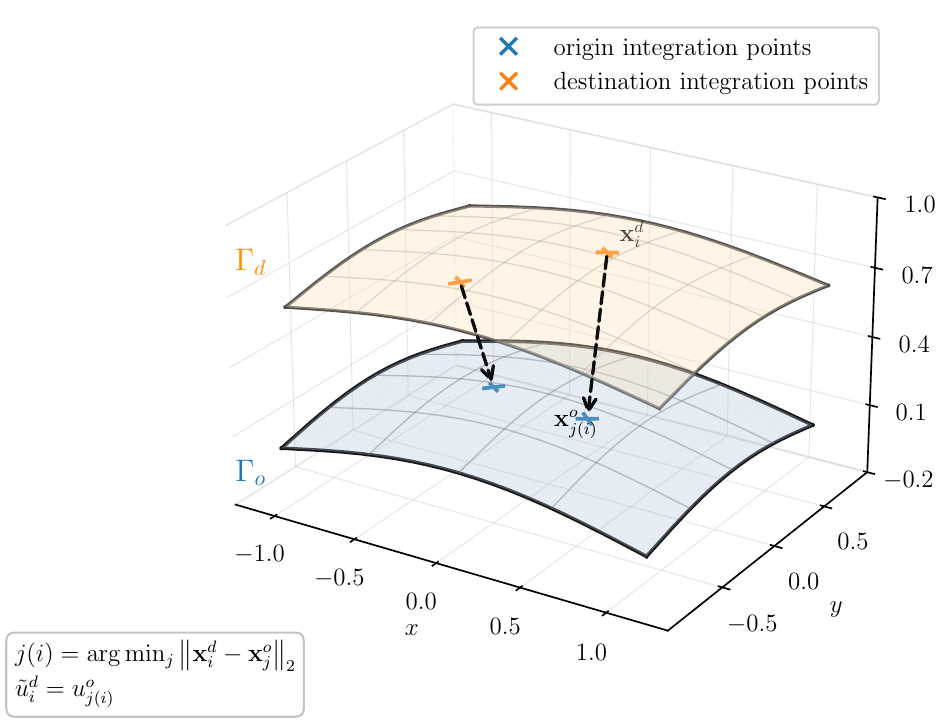}
    \caption{Nearest-neighbor mapping between non-matching origin and destination interfaces.}
    \label{fig:nearest_neighbor_mapping}
\end{figure}

In the nearest-neighbor mapping, each destination integration point is
associated with the spatially closest integration point on the origin
interface (Figure~\ref{fig:nearest_neighbor_mapping}). Let
$\mathbf{x}_{\mathrm{d}}, \mathbf{x}_{\mathrm{o}} \in \mathbb{R}^{3}$
denote points on the destination and origin interfaces, respectively,
and let $\mathcal{O}$ represent the set of origin integration points.
The mapped value $\tilde{u}(\mathbf{x}_{\mathrm{d}})$ is obtained from
the value at the closest origin integration point,

\begin{equation}
\tilde{u}(\mathbf{x}_{\mathrm{d}})
=
u(\mathbf{x}_{\mathrm{o}}),
\qquad
\mathbf{x}_{\mathrm{o}}
=
\arg\min_{\mathbf{x}_i \in \mathcal{O}}
\left\|
\mathbf{x}_{\mathrm{d}} - \mathbf{x}_i
\right\|.
\end{equation}

From an algebraic perspective, the nearest-neighbor mapper defines a
sparse binary transfer matrix
$\mathbf{H}\in\mathbb{R}^{N_{\mathrm{d}}\times N_{\mathrm{o}}}$,
where $N_{\mathrm{o}}$ and $N_{\mathrm{d}}$ denote the dimensions of
the origin and destination interface vectors, respectively. Each
destination degree of freedom is associated with exactly one origin
integration point. As a result, each row of $\mathbf{H}$ contains a
single nonzero entry equal to unity, corresponding to the nearest
origin point, while all remaining entries vanish.

The nearest-neighbor mapping is straightforward to implement, since the
transfer is based solely on the spatial proximity of interface
integration points. In practice, the closest-point search can be
performed efficiently using spatial search structures such as k-d trees \cite{bentley1975kdtree}
or octrees \cite{meagher1982octree}. However, the method does not account for the local element
geometry or interpolation within the origin discretization. As a
consequence, the mapping accuracy may deteriorate for coarse or highly
non-conforming interfaces, particularly when transferring strongly
varying interface fields.

\subsubsection{Nearest-element mapping}
\label{sec:nearest_element_mapping}

\begin{figure}[h!]
    \centering
     \includegraphics[width=0.55\linewidth]{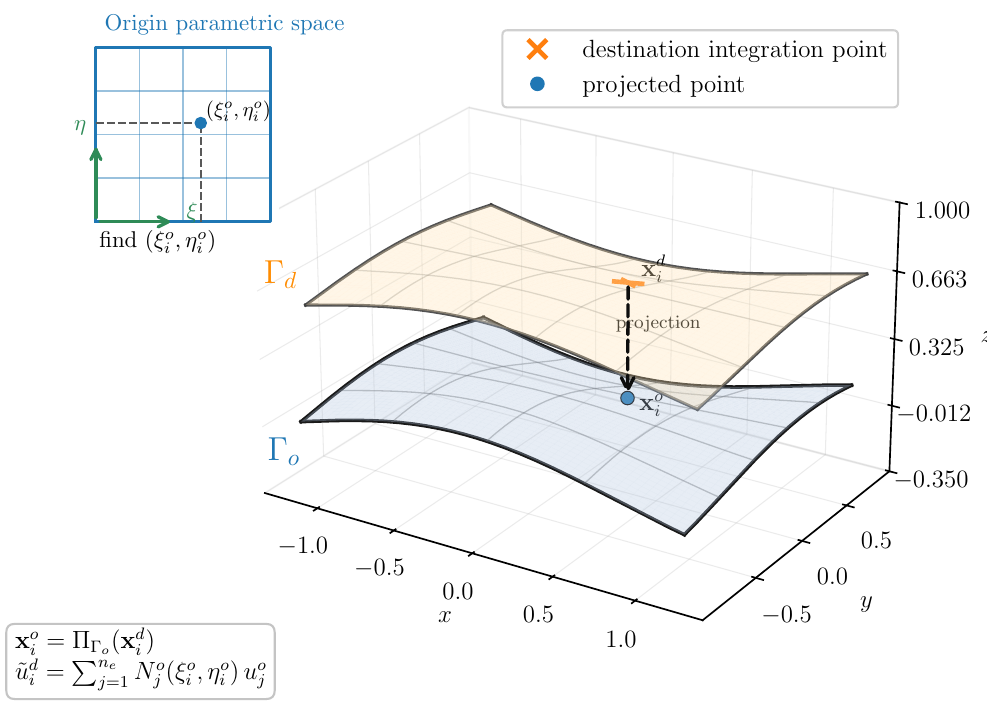}
    \caption{Nearest Element (closest geometric projection) mapper for IGA: the destination integration point is projected onto the origin IGA patch to obtain $(\xi_i^o,\eta_i^o)$, and the mapped value is evaluated as $\tilde{\mathbf{u}}_i^{d}=\mathbf{N}^o(\xi_i^o,\eta_i^o)\mathbf{u}^o$.}
    \label{fig:nearest_element_iga_schematic}
\end{figure}

In the nearest-element mapping (Figure ~\ref{fig:nearest_element_iga_schematic}), each destination integration point on
the interface $\Gamma_{\mathrm{d}}$ is mapped by geometrically projecting it onto
the origin interface $\Gamma_{\mathrm{o}}$, which is discretized using
isogeometric analysis. For a given destination integration point
$\mathbf{x}_{\mathrm{d}} \in \mathbb{R}^{3}$, the closest origin
surface patch is first identified and $\mathbf{x}_{\mathrm{d}}$ is
projected onto this patch to obtain the corresponding parametric
coordinates $\left(\xi^o_{i},\eta^o_{i}\right)\in\Omega^o$.

The mapped quantity is then evaluated by interpolating the origin field
at the projected location using the isogeometric shape functions of the
associated origin patch. The mapped value can be expressed as
\begin{equation}
\tilde{u}(\mathbf{x}_{\mathrm{d}})
=
\sum_{j=1}^{n_{\mathrm{e}}}
N^{\mathrm{o}}_{j}
\left(
\xi^o_{i}, \eta^o_{i}
\right)
u^{\mathrm{o}}_{j},
\end{equation}
where $N^{\mathrm{o}}_{j}$ denote the shape functions of the origin
discretization, $u^{\mathrm{o}}_{j}$ are the corresponding origin
degrees of freedom, and
$\left(\xi^o_{i},\eta^o_{i}\right)$ are the parametric
coordinates associated with the geometric projection of
$\mathbf{x}_{\mathrm{d}}$ onto the origin surface.

From an algebraic perspective, the nearest-element mapping defines a
sparse transfer operator
$\mathbf{H}\in\mathbb{R}^{N_{\mathrm{d}}\times N_{\mathrm{o}}}$, where
$N_{\mathrm{o}}$ and $N_{\mathrm{d}}$ denote the dimensions of the
origin and destination interface vectors, respectively. Each
destination integration point is associated with the local basis
functions of the projected origin element. In contrast to the
nearest-neighbor mapper, multiple nonzero entries may appear in each
row of $\mathbf{H}$, corresponding to the active shape functions of the
origin patch at the projected parametric location. The values of these
entries are given by the evaluated isogeometric basis functions
$N^{\mathrm{o}}_{j}(\xi^o_{i},\eta^o_{i})$.

Compared to the nearest-neighbor approach, the nearest-element mapping
accounts for the interpolation properties of the
origin discretization, resulting in a smoother and more accurate
transfer of interface quantities. This is particularly advantageous for
curved IGA interfaces and strongly non-conforming discretizations,
where pointwise nearest-neighbor transfers may introduce interpolation
errors and discontinuities.

\subsubsection{IGA-IGA Mortar Mapper}
\label{sec:iga_iga_mortar_mapping}

In contrast to the nearest-element mapping, where the transfer is
performed pointwise through geometric projection (closest-point projection), the mortar mapper formulates the interface transfer as a variational projection problem over the
coupling interface. The objective is to determine a destination field
that minimizes the mismatch with the origin field in an integral sense,
thereby providing a consistent transfer between non-matching
discretizations. The present formulation extends the mortar-based
isogeometric mapping concepts proposed in~\cite{Apostolatos2021}
towards partitioned IGA--IGA vibroacoustic coupling involving
non-conforming NURBS interfaces.

Let $\Gamma_{\mathrm{o}}$ and $\Gamma_{\mathrm{d}}$ denote the origin
and destination interfaces, respectively. Furthermore, let
$u_{\mathrm{o}}:\Gamma_{\mathrm{o}}\subset\mathbb{R}^{3}\rightarrow\mathbb{C}$
and
$u_{\mathrm{d}}:\Gamma_{\mathrm{d}}\subset\mathbb{R}^{3}\rightarrow\mathbb{C}$
denote the complex-valued scalar interface fields on the origin and destination
interfaces. The mapped destination field $u_{\mathrm{d}}$ is obtained
by minimizing the $L_2$-norm of the error between the origin and
destination fields over the interface,
\begin{equation}
\Psi(u_{\mathrm{d}})
=
\frac{1}{2}
\int_{\Gamma}
\left(
u_{\mathrm{d}}(\mathbf{x})
-
u_{\mathrm{o}}(\mathbf{x})
\right)^2
\,\mathrm{d}\Gamma .
\end{equation}

Enforcing stationarity of the functional leads to the weak form
\begin{equation}
\int_{\Gamma}
N_{\mathrm{d}}(\mathbf{x})
\left(
u_{\mathrm{d}}(\mathbf{x})
-
u_{\mathrm{o}}(\mathbf{x})
\right)
\,\mathrm{d}\Gamma
=
0 , \label{eq:weak}
\end{equation}
where $N_{\mathrm{d}}$ denotes the shape functions of the destination
interface.

The origin and destination fields are approximated using their
corresponding isogeometric basis functions,
\begin{align}
u_{\mathrm{d}}(\mathbf{x})
&\approx
u_{\mathrm{d}}^{h}(\mathbf{x})
=
\mathbf{N}_{\mathrm{d}}^{T}(\mathbf{x})
\mathbf{u}_{\mathrm{d}},
\\
u_{\mathrm{o}}(\mathbf{x})
&\approx
u_{\mathrm{o}}^{h}(\mathbf{x})
=
\mathbf{N}_{\mathrm{o}}^{T}(\mathbf{x})
\mathbf{u}_{\mathrm{o}},
\end{align}
where
$\mathbf{N}_{\mathrm{o}}(\mathbf{x}) \in \mathbb{R}^{N_{\mathrm{o}}}$
and
$\mathbf{N}_{\mathrm{d}}(\mathbf{x}) \in \mathbb{R}^{N_{\mathrm{d}}}$
collect the active isogeometric basis functions of the origin and
destination interfaces, respectively, while
$\mathbf{u}_{\mathrm{o}} \in \mathbb{C}^{N_{\mathrm{o}}}$ and
$\mathbf{u}_{\mathrm{d}} \in \mathbb{C}^{N_{\mathrm{d}}}$ are the
corresponding vectors of interface degrees of freedom.

Substituting the discrete approximations into the weak form (Eq.~\ref{eq:weak}) yields the
mortar system
\begin{equation}
\int_{\Gamma_{\mathrm{d}}}
\mathbf{N}_{\mathrm{d}}
\mathbf{N}_{\mathrm{d}}^{T}
\,\mathrm{d}\Gamma
\,
\mathbf{u}_{\mathrm{d}}
=
\int_{\Gamma_{\text{do}}}
\mathbf{N}_{\mathrm{d}}
\mathbf{N}_{\mathrm{o}}^{T}
\,\mathrm{d}\Gamma
\,
\mathbf{u}_{\mathrm{o}},
\end{equation}
which can be written in matrix form as

\begin{equation}
\mathbf{M}_{\mathrm{DD}}
\mathbf{u}_{\mathrm{d}}
=
\mathbf{M}_{\mathrm{DO}}
\mathbf{u}_{\mathrm{o}},
\end{equation}
where
$\mathbf{M}_{\mathrm{DD}}\in
\mathbb{R}^{N_{\mathrm{d}}\times N_{\mathrm{d}}}$ and
$\mathbf{M}_{\mathrm{DO}}\in
\mathbb{R}^{N_{\mathrm{d}}\times N_{\mathrm{o}}}$ are the mortar
matrices resulting from the integration of the destination basis
functions and the destination-origin basis function products,
respectively.

Assuming that the matrix $\mathbf{M}_{\mathrm{DD}}$ is invertible, the
mapping operator is obtained as
\begin{equation}
\mathbf{u}_{\mathrm{d}}
=
\mathbf{H}_{\mathrm{DO}}
\mathbf{u}_{\mathrm{o}},
\qquad
\mathbf{H}_{\mathrm{DO}}
=
\mathbf{M}_{\mathrm{DD}}^{-1}
\mathbf{M}_{\mathrm{DO}}.
\end{equation}

For the IGA--IGA mortar mapper, the evaluation of the coupling
integrals requires the construction of a common integration domain
between both interfaces ($\Gamma_{\text{do}}$). Since the origin and destination IGA patches
generally possess different parametric discretizations, the knot spans
of the destination interface are first mapped to the physical space and
subsequently projected onto the parametric space of the origin
interface. This procedure defines overlapping regions where the mortar
integrals are evaluated consistently.

\begin{figure}[h!]
    \centering
    \includegraphics[width=0.70\linewidth]{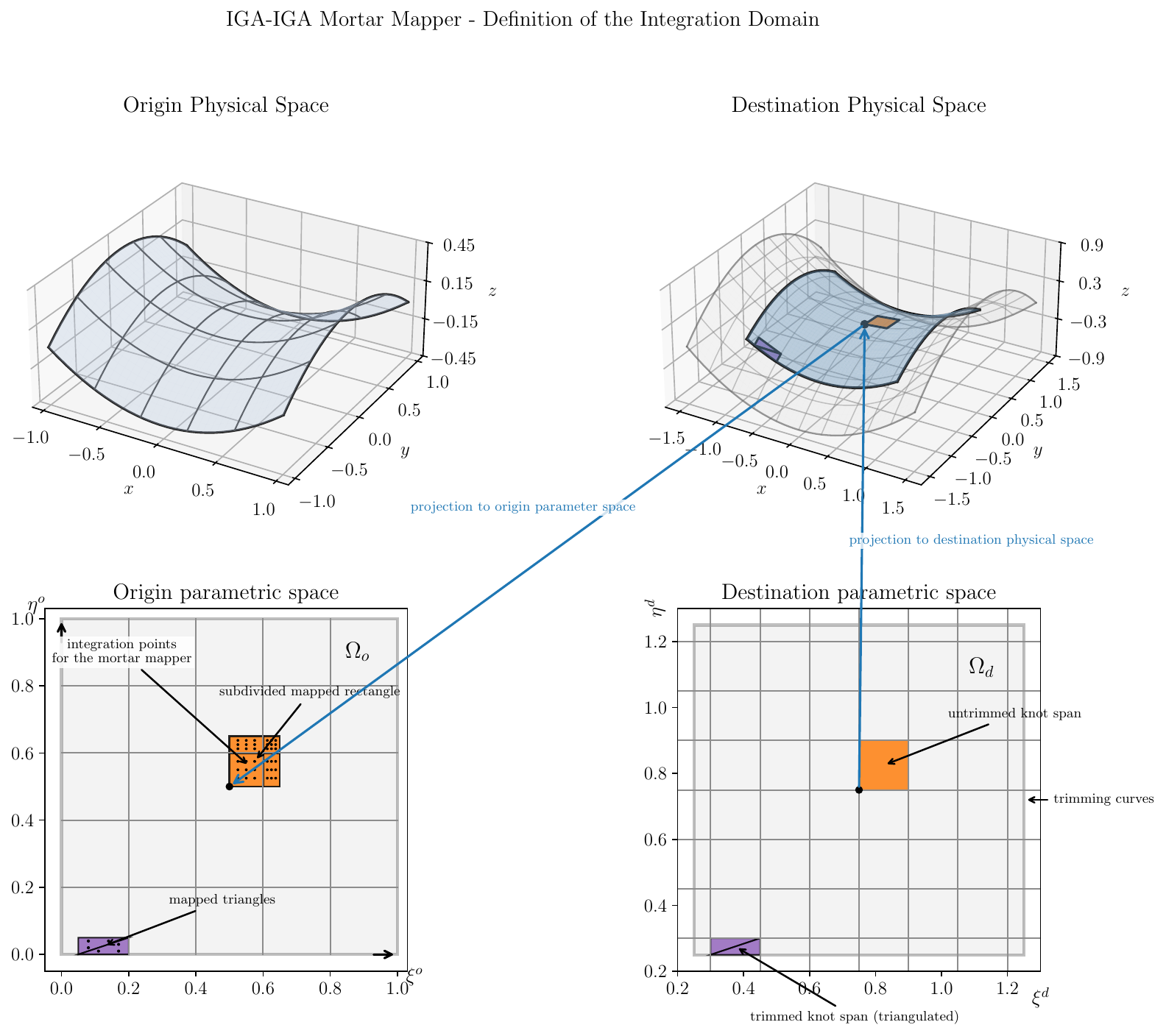}
    \caption{Construction of the integration domain for the IGA-IGA mortar mapper. Destination knot spans are projected onto the origin parameter space and subdivided along the origin knot lines, yielding conforming integration subdomains used for the evaluation of the mortar coupling integrals.}
    \label{fig:iga_iga_mortar_mapper}
\end{figure}

Figure~\ref{fig:iga_iga_mortar_mapper} illustrates this procedure. The
destination knot spans are mapped from the destination parametric space
to the destination physical surface and then projected onto the origin
parametric space. The resulting projected regions are subdivided along
the knot lines of the origin discretization, generating integration
subdomains that are conforming with both parameterizations. Numerical
quadrature points are then constructed within these subdomains in the
origin parametric space. To evaluate the destination basis functions
$\mathbf{N}_{\mathrm{d}}$, each quadrature point is mapped to the
physical interface and subsequently projected back onto the destination
parametric space. The corresponding origin and destination basis
functions, $\mathbf{N}_{\mathrm{o}}$ and $\mathbf{N}_{\mathrm{d}}$,
are then evaluated at their respective parametric coordinates and used
to assemble the mortar matrices $\mathbf{M}_{\mathrm{DD}}$ and
$\mathbf{M}_{\mathrm{DO}}$.

\ignore{Compared to the nearest-element mapping, the mortar formulation does
not rely on pointwise interpolation alone, but instead enforces the
transfer in a variational sense over the interface. This results in a
more consistent and conservative transfer, particularly for strongly
non-matching discretizations and curved IGA geometries.}

\subsection{Fixed-point problem and convergence accelerators}
\label{sec:convergence_accelerators}

In strongly coupled partitioned vibroacoustic simulations, the
interface equilibrium conditions are enforced iteratively through the
exchange of interface quantities between the participating solvers. As
a consequence, the coupled problem can be interpreted as a fixed-point
problem defined on the vibroacoustic interface.

Let $\Gamma$ denote the coupling interface between the structural and
acoustic domains. Furthermore, let
$u_\Gamma:\Gamma\subset\mathbb{R}^{3}\rightarrow\mathbb{C}^{3}$ and
$F_\Gamma:\Gamma\subset\mathbb{R}^{3}\rightarrow\mathbb{C}^{3}$ denote
the complex-valued interface displacement and load fields,
respectively. For a given interface displacement field
$u_\Gamma$, the acoustic solver computes the corresponding acoustic
pressure field and the resulting interface load,

\begin{equation}
F_\Gamma = \mathcal{A}(u_\Gamma),
\end{equation}
where $\mathcal{A}(\cdot)$ denotes the acoustic solution operator.

Similarly, for a given interface load $F_\Gamma$, the structural
solver computes the corresponding structural response and the updated
interface displacement,

\begin{equation}
\tilde{u}_\Gamma = \mathcal{S}(F_\Gamma),
\end{equation}
where $\mathcal{S}(\cdot)$ represents the structural solution
operator.

The coupled vibroacoustic problem can therefore be written as

\begin{equation}
u_\Gamma
=
\mathcal{S}
\left(
\mathcal{A}(u_\Gamma)
\right),
\end{equation}
which corresponds to a fixed-point problem on the interface
unknowns. Introducing the residual operator

\begin{equation}
r(u_\Gamma)
=
\mathcal{S}
\left(
\mathcal{A}(u_\Gamma)
\right)
-
u_\Gamma,
\end{equation}
the coupled problem is equivalently written as

\begin{equation}
r(u_\Gamma)=0.
\end{equation}

At the discrete level, the interface displacement, load, and residual
fields are represented by the complex-valued vectors
$\mathbf{u}_\Gamma\in\mathbb{C}^{N_x}$,
$\mathbf{F}_\Gamma\in\mathbb{C}^{N_x}$, and
$\mathbf{r}_\Gamma\in\mathbb{C}^{N_x}$, respectively, where
$N_x=N_{\mathrm{o}}$ or $N_x=N_{\mathrm{d}}$ depending on the selected
interface representation used by the coupling algorithm.

A standard Gauss--Seidel partitioned iteration corresponds to the
fixed-point update
\begin{equation}
\mathbf{u}_\Gamma^{(k+1)}
=
\mathcal{S}
\left(
\mathcal{A}(\mathbf{u}_\Gamma^{(k)})
\right).
\end{equation}
where $k$ denotes the coupling iteration index.

Although this procedure preserves the modularity of the individual
solvers, the fixed-point iterations may converge slowly or even
diverge for strongly coupled vibroacoustic configurations,
particularly in the vicinity of resonance frequencies where the
interaction between the structural and acoustic fields becomes more
pronounced. In such cases, convergence acceleration techniques become
essential to stabilize and accelerate the iterative coupling
procedure.

Convergence accelerators generally operate on the residual of the fixed-point
iteration. Let the interface residual at coupling iteration $k$ be
defined as

\begin{equation}
\mathbf{r}^{(k)}
=
\tilde{\mathbf{x}}^{(k+1)}
-
\mathbf{x}^{(k)},
\qquad
\mathbf{r}^{(k)},\mathbf{x}^{(k)}
\in
\mathbb{C}^{N_x}.
\end{equation}

Here, $\mathbf{x}^{(k)}$ denotes the current interface solution and
$\tilde{\mathbf{x}}^{(k+1)}$ the interface values obtained after one
complete coupling step. The vector $\mathbf{x}$ is used as a generic
notation for the accelerated interface quantity and may represent
either interface displacements or interface loads, depending on the
selected coupling formulation.

The standard Aitken relaxation updates the interface solution
according to~\cite{irons1969}
\begin{equation}
\mathbf{x}^{(k+1)}
=
\mathbf{x}^{(k)}
+
\alpha^{(k)} \mathbf{r}^{(k)},
\end{equation}
where $\alpha^{(k)}$ is a
dynamically updated relaxation factor. In the classical real-valued
formulation, the relaxation parameter is computed as
\begin{equation}
\alpha^{(k)}
=
-\alpha^{(k-1)}
\frac{
\left(\mathbf{r}^{(k-1)}\right)^T
\left(
\mathbf{r}^{(k)}
-
\mathbf{r}^{(k-1)}
\right)
}{
\left(
\mathbf{r}^{(k)}
-
\mathbf{r}^{(k-1)}
\right)^T
\left(
\mathbf{r}^{(k)}
-
\mathbf{r}^{(k-1)}
\right)
}.
\end{equation}

The IQN-ILS method follows a different strategy~\cite{DEGROOTE2010446,DEGROOTE2009793}.
Instead of applying a scalar relaxation to the interface residual,
IQN-ILS constructs an approximation of the inverse interface
Jacobian directly from previous coupling iterations.

The method assumes that changes in the interface solution and
the corresponding residual variations approximately satisfy
\begin{equation}
\Delta \mathbf{x}
\approx
\mathbf{J}^{-1}
\Delta \mathbf{r},
\qquad
\mathbf{J}
=
\frac{\partial \mathbf{r}}{\partial \mathbf{x}},
\end{equation}
where
$\mathbf{J}\in\mathbb{C}^{N_x\times N_x}$ denotes the Jacobian of the
discrete residual with respect to the interface solution vector.

The differences between successive coupling iterations are defined as
\begin{equation}
\Delta \mathbf{r}_i
=
\mathbf{r}^{(i+1)}
-
\mathbf{r}^{(i)},
\qquad
\Delta \tilde{\mathbf{x}}_i
=
\tilde{\mathbf{x}}^{(i+1)}
-
\tilde{\mathbf{x}}^{(i)}.
\end{equation}

These vectors are assembled into the matrices
\begin{equation}
\mathbf{V}
=
\begin{bmatrix}
\Delta \mathbf{r}_1 &
\Delta \mathbf{r}_2 &
\cdots &
\Delta \mathbf{r}_m
\end{bmatrix},
\qquad
\mathbf{W}
=
\begin{bmatrix}
\Delta \tilde{\mathbf{x}}_1 &
\Delta \tilde{\mathbf{x}}_2 &
\cdots &
\Delta \tilde{\mathbf{x}}_m
\end{bmatrix}.
\end{equation}
where
\(\mathbf{V},\mathbf{W}\in\mathbb{C}^{N_x\times m}\),
\(\Delta\mathbf{r}_i\in\mathbb{C}^{N_x}\), and
\(\Delta\tilde{\mathbf{x}}_i\in\mathbb{C}^{N_x}\).
The parameter \(m\) denotes the number of residual and solution
difference vectors retained from previous coupling iterations, i.e.,
the number of columns of the history matrices \(\mathbf{V}\) and
\(\mathbf{W}\).

The current residual is then approximated as a linear combination
of previously observed residual differences by solving the
least-squares problem
\begin{equation}
\mathbf{V}\mathbf{c}
\approx
-\mathbf{r}^{(k)},
\end{equation}

Finally, the interface correction is computed as
\begin{equation}
\mathbf{x}^{(k+1)}
=
\mathbf{x}^{(k)}
+
\mathbf{W}\mathbf{c}
+
\mathbf{r}^{(k)}.
\end{equation}

Since most co-simulation infrastructures exchange interface quantities
only as real-valued vectors, complex-valued interface fields arising
in frequency-domain simulations are commonly transferred as separated
real and imaginary components. A straightforward extension therefore
consists of applying conventional real-valued convergence accelerators
independently to both fields. However, this treatment neglects the
intrinsic coupling between amplitude and phase information and may
lead to deteriorated convergence behavior. To address this issue, the
present work reconstructs the interface residual directly in the
complex domain and performs the convergence acceleration using
complex-valued interface quantities.

For the Aitken method, two different complex-valued formulations are
investigated. The first formulation computes the relaxation parameter
directly from the complex-valued residual using a Hermitian inner
product, while the second formulation evaluates the relaxation
parameter from the component-wise modulus of the complex residual. In
the following, these approaches are referred to as the
\emph{complex Aitken} and \emph{complex modulus Aitken} methods,
respectively. In contrast, the IQN-ILS method is extended by
constructing the quasi-Newton approximation directly in the complex
domain.

For both Aitken and IQN-ILS methods, the interface residual is reconstructed as a
single complex-valued quantity,
\begin{equation}
\mathbf{r}^{(k)}_{\mathrm{c}}
=
\mathbf{r}^{(k)}_{\mathrm{real}}
+
\mathrm{i}\,\mathbf{r}^{(k)}_{\mathrm{imag}},
\qquad
\mathbf{r}^{(k)}_{\mathrm{c}}\in
\mathbb{C}^{N_x},
\mathbf{r}^{(k)}_{\mathrm{real}},
\mathbf{r}^{(k)}_{\mathrm{imag}}
\in
\mathbb{R}^{N_x}.
\end{equation}

For the \emph{complex Aitken}, the relaxation parameter is then
computed using the Hermitian inner product
$\langle \mathbf{a},\mathbf{b} \rangle = \mathbf{a}^{H}\mathbf{b}$,
where $(\cdot)^H$ denotes the conjugate transpose, which yields the
complex-valued Aitken update
\begin{equation}
\alpha^{(k)}
=
-\alpha^{(k-1)}
\frac{
\left(\mathbf{r}^{(k-1)}\right)^H
\left(
\mathbf{r}^{(k)}
-
\mathbf{r}^{(k-1)}
\right)
}{
\left(
\mathbf{r}^{(k)}
-
\mathbf{r}^{(k-1)}
\right)^H
\left(
\mathbf{r}^{(k)}
-
\mathbf{r}^{(k-1)}
\right)
}.
\end{equation}

Since the relaxation parameter is complex-valued, the interface
correction
\begin{equation}
\Delta \mathbf{x}^{(k)}_{\mathrm{c}}
=
\alpha^{(k)}
\mathbf{r}^{(k)}_{\mathrm{c}}.
\end{equation}
can be interpreted geometrically as a simultaneous scaling and
rotation of the complex residual vector in the complex plane. The
magnitude of $\alpha^{(k)}$ controls the amplitude of the interface
correction, while its phase determines the rotation applied to the
correction direction.

In contrast to applying two independent real-valued convergence
accelerators separately to the real and imaginary interface fields,
the present formulation treats the interface quantities as a unified
complex-valued residual field. Consequently, a single relaxation
parameter is applied simultaneously to both components, thereby
preserving the intrinsic coupling between amplitude and phase
information.

Although the relaxation parameter is computed from complex-valued
residuals, the numerical experiments indicate that unrestricted phase
rotations of the scalar relaxation coefficient may deteriorate the
robustness of the fixed-point iterations, particularly near resonance
frequencies. Improved convergence behavior was obtained when the
relaxation parameter remained predominantly aligned with the real
axis. Therefore, the phase of the complex relaxation parameter is
regularized by constraining it within a prescribed interval around the
real axis. More specifically, the phase of $\alpha^{(k)}$ is bounded
around either $0$ or $\pi$, depending on the sign of its real part. These observations also motivated the investigation of an alternative
formulation in which the relaxation parameter remains purely
real-valued while still being applied to the reconstructed
complex-valued residual field.

Figure~\ref{fig:complex_alpha_phase_restriction} illustrates the
admissible region of the complex relaxation parameter in the complex
plane. The admissible phase interval is restricted to sectors of
opening angle $\Delta\phi_{\max}$ around the positive and negative
real axes, thereby preventing excessively large rotations of the
interface correction direction during the fixed-point iterations.

\begin{figure}[h!]
    \centering
    \begin{tikzpicture}[scale=2.25]

    \def\r{1.0}
    \def\theta{20}

    \fill[blue!12]
        (0,0) -- (\theta:\r)
        arc(\theta:-\theta:\r) -- cycle;

    \fill[blue!12]
        (0,0) -- ({180-\theta}:\r)
        arc({180-\theta}:{180+\theta}:\r) -- cycle;

    \draw[->, thick]
        (-1.35,0) -- (1.35,0)
        node[right] {$\operatorname{Re}(\alpha)$};

    \draw[->, thick]
        (0,-1.05) -- (0,1.05)
        node[above] {$\operatorname{Im}(\alpha)$};

    \draw[line width=1.0pt, blue!70]
        (-1.08,0) -- (1.08,0);

    \draw[dashed, thin]
        (0,0) -- (\theta:1.08);

    \draw[dashed, thin]
        (0,0) -- (-\theta:1.08);

    \draw[dashed, thin]
        (0,0) -- ({180-\theta}:1.08);

    \draw[dashed, thin]
        (0,0) -- ({180+\theta}:1.08);

    \draw[->, very thick, red!75]
    (0,0) -- (0.82,0.22);

    \node[red!75] at (0.97,0.24)
        {$\alpha^{(k)}$};

    \node at (0.58,0.45)
        {$\theta^\circ=\Delta\phi_{\max}$};

    \node at (-0.58,0.45)
        {$\theta^\circ=\Delta\phi_{\max}$};

    \node[align=center] at (0,-1.28)
    {
        $
        \mathcal{A}_{\theta}
        =
        \left\{
        \alpha\in\mathbb{C}
        \;:\;
        |\arg(\alpha)|\leq\theta
        \;\lor\;
        |\arg(-\alpha)|\leq\theta
        \right\}
        $
    };

\end{tikzpicture}
    \caption{
    Schematic representation of the phase regularization applied to the
    complex-valued Aitken relaxation parameter. The admissible phase
    variation is restricted to sectors of opening angle
    $\Delta\phi_{\max}$ around the positive and negative real axes in
    order to avoid excessively large rotations of the interface
    correction direction.
    }
    \label{fig:complex_alpha_phase_restriction}
\end{figure}

This strategy preserves the coupled complex-valued treatment of the
interface residual while preventing excessively large phase rotations
of the scalar relaxation parameter. The resulting formulation can
therefore be interpreted as a phase-regularized extension of the
classical real-valued Aitken relaxation method.

An alternative relaxation strategy is also investigated in which the
Aitken relaxation parameter is computed from the component-wise
magnitude of the complex residual vector. In this formulation, the
complex residual is transformed into the real-valued residual
\begin{equation}
\mathbf{r}^{(k)}_{\mathrm{abs}}
=
\left|
\mathbf{r}^{(k)}_{\mathrm{c}}
\right|.
\end{equation}
where the absolute value is applied component-wise.

The relaxation parameter is then computed using the classical
real-valued Aitken expression,
\begin{equation}
\alpha^{(k)}
=
-\alpha^{(k-1)}
\frac{
\left(\mathbf{r}^{(k-1)}_{\mathrm{abs}}\right)^T
\left(
\mathbf{r}^{(k)}_{\mathrm{abs}}
-
\mathbf{r}^{(k-1)}_{\mathrm{abs}}
\right)
}{
\left(
\mathbf{r}^{(k)}_{\mathrm{abs}}
-
\mathbf{r}^{(k-1)}_{\mathrm{abs}}
\right)^T
\left(
\mathbf{r}^{(k)}_{\mathrm{abs}}
-
\mathbf{r}^{(k-1)}_{\mathrm{abs}}
\right)
}.
\end{equation}

The resulting scalar relaxation factor remains real-valued, but is
applied to the full complex residual field,
\begin{equation}
\Delta \mathbf{x}^{(k)}_{\mathrm{c}}
=
\alpha^{(k)}
\mathbf{r}^{(k)}_{\mathrm{c}}.
\end{equation}

The interface correction is subsequently decomposed into its real and
imaginary contributions,
\begin{equation}
\Delta \mathbf{x}^{(k)}_{\mathrm{real}}
=
\Re\!\left(
\Delta \mathbf{x}^{(k)}_{\mathrm{c}}
\right),
\qquad
\Delta \mathbf{x}^{(k)}_{\mathrm{imag}}
=
\Im\!\left(
\Delta \mathbf{x}^{(k)}_{\mathrm{c}}
\right).
\end{equation}
which are transferred back to the individual solvers as independent
real-valued interface updates through the co-simulation communication
routines.

A similar extension is introduced for the IQN-ILS method. In this
case, the matrices of residual and solution differences are assembled
directly using complex-valued interface quantities, and the
least-squares problem is solved in the complex domain,
\begin{equation}
\mathbf{V}\mathbf{c}
\approx
-\mathbf{r}^{(k)},
\qquad
\mathbf{V}\in\mathbb{C}^{N_x\times m},
\quad
\mathbf{r}^{(k)}\in\mathbb{C}^{N_x},
\quad
\mathbf{c}\in\mathbb{C}^{m}.
\end{equation}

In contrast to scalar relaxation methods, the complex-valued IQN-ILS
formulation does not compute a single complex relaxation coefficient.
Instead, the method constructs a multi-dimensional approximation of
the inverse interface Jacobian directly in the complex domain. As a
consequence, the interface correction is not restricted to a single
scaling and rotation of the residual vector, but may involve a more
general transformation of the complex-valued interface residual based
on information gathered from previous coupling iterations.

As a consequence, the coefficients defining the quasi-Newton update
become complex-valued and naturally account for the phase relation
between the structural and acoustic fields. Although the quasi-Newton
operations are performed using complex-valued quantities, the final
interface correction is again decomposed into real and imaginary
components before being communicated to the individual solvers.

\section{Numerical results}
\label{sec:numerical_results}

This section presents three benchmark problems to assess the accuracy,
robustness, and convergence behavior of the proposed partitioned
vibroacoustic co-simulation framework. The examples are organized in a
progressive manner, increasing in complexity from one-way coupled
configurations to fully bidirectional vibroacoustic interaction problems.
Unless otherwise stated, the acoustic response is evaluated over frequency
sweeps with a frequency step of $\Delta f = 1\;\mathrm{Hz}$ and reported
in terms of the magnitude of the complex acoustic pressure, $|p|$.
All partitioned simulations employ a Gauss--Seidel communication pattern
in which the structural and acoustic solvers are executed sequentially and
the most recently available interface quantities are exchanged between
both participants during the coupling procedure.

Since the monolithic and partitioned formulations are implemented in different software environments, direct wall-clock time comparisons are not considered. The partitioned approach is instead evaluated in terms of solution accuracy, convergence behavior, and the number of coupling iterations.

\subsection{One-way coupled rectangular plate}
\label{subsec:example_1}

\begin{figure}[h!]
    \centering
    \begin{tikzpicture}
          \scaling{0.6};
        
          \definecolor{PlateBlue}{RGB}{0,92,170}
          \definecolor{PlateFill}{RGB}{0,92,170}
          \definecolor{AnnotGray}{RGB}{90,90,90}
          \definecolor{LoadRed}{RGB}{180,60,60}
          \definecolor{LoadYellow}{RGB}{255,165,0}
        
          \point{A}{0}{0}
          \point{B}{10}{0}
          \point{D}{3}{3.5}
          \point{C}{13}{3.5}
        
          \begin{scope}[fill=PlateFill, fill opacity=0.08, draw=none]
            \fill (A) -- (B) -- (C) -- (D) -- cycle;
          \end{scope}
        
          \begin{scope}[draw=PlateBlue, line width=0.9pt]
            \beam{4}{A}{B}
            \beam{4}{B}{C}
            \beam{4}{C}{D}
            \beam{4}{D}{A}
          \end{scope}
        
          \point{ABm}{5}{0}
          \point{DCm}{8}{3.5}
          \point{ADm}{1.5}{1.75}
          \point{BCm}{11.5}{1.75}
        
          \newcommand{\simplesupportdown}[1]{%
          \begin{scope}[shift={(#1)}, draw=AnnotGray, line width=0.4pt]
            \draw (0,0) -- (-0.18,-0.28) -- (0.18,-0.28) -- cycle;  
            \draw (-0.28,-0.32) -- (0.28,-0.32);                    
          \end{scope}
        }
        
          \begin{scope}[draw=AnnotGray, scale=0.42, transform shape]
          \support{2}{A}[0]   \support{2}{ABm}[0] \support{2}{B}[0]
          \support{2}{D}[0]   \support{2}{DCm}[0] \support{2}{C}[0]
          \support{2}{ADm}[0]
          \support{2}{BCm}[0]
        \end{scope}

          \begin{scope}[draw=LoadYellow, line width=0.85pt]
            \foreach \s in {0.18,0.34,0.50,0.66,0.82}{
              \foreach \t in {0.22,0.45,0.68}{
                \coordinate (P) at ($ (A)!\s!(B) + (A)!\t!(D) $);
                \draw[-{Latex[length=2.6mm,width=1.8mm]}]
                  ($(P)+(0,0.85)$) -- (P);
              }
            }
            \node[text=LoadYellow] at ($(A)!0.4!(C) + (0,1.7)$) {$p=1\;Pa$};
          \end{scope}
        
          \begin{scope}[draw=AnnotGray, line width=0.45pt]
            \draw[<->] ($(A)+(0,-1.55)$) -- ($(B)+(0,-1.55)$)
              node[midway, fill=white, inner sep=1.2pt] {$ L= 1\;m$};
            \draw ($(A)+(0,-1.15)$) -- (A);
            \draw ($(B)+(0,-1.15)$) -- (B);
        
            \def\offx{-2.5}
            \def\offy{0.15}
            \draw[<->] ($(A)+(\offx,\offy)$) -- ($(D)+(\offx,\offy)$)
              node[midway, fill=white, inner sep=1.2pt] {$w = 0.5\;m$};
            \draw ($(A)+(\offx+0.25,\offy-0.05)$) -- (A);
            \draw ($(D)+(\offx+0.25,\offy-0.05)$) -- (D);
          \end{scope}
        
        \tikzset{
          axisArrow/.style={
            draw=LoadRed,
            -{Latex[length=2.2mm,width=1.6mm]},
            line width=0.9pt
          }
        }
        
        \def\ax{1}
        \def\ay{1}
        \def\az{1}
        
        \coordinate (O) at ($(A)$);
        \fill[LoadRed] (O) circle (0.6pt);
        
        \pgfmathsetmacro{\ADx}{3}
        \pgfmathsetmacro{\ADy}{3.5}
        \pgfmathsetmacro{\ADn}{sqrt(\ADx*\ADx + \ADy*\ADy)}
        \pgfmathsetmacro{\uyx}{\ADx/\ADn}
        \pgfmathsetmacro{\uyy}{\ADy/\ADn}
        
        \draw[axisArrow] (O) -- ($(O)+(\ax,0)$)
          node[below=1.0mm, text=LoadRed] {$x$};
        
        \draw[axisArrow] (O) -- ($(O)+(\ay*\uyx,\ay*\uyy)$)
          node[left=1.0mm, text=LoadRed] {$y$};
        
        \draw[axisArrow] (O) -- ($(O)+(0,\az)$)
          node[left=1.0mm, text=LoadRed] {$z$};
        
        \definecolor{MeasBlue}{RGB}{40,120,200}
        
        \coordinate (SPLxy) at ($(A)!0.5!(C)$);
        
        \coordinate (SPL) at ($(SPLxy)+(0,2.1)$);
        
        \fill[MeasBlue] (SPL) circle (1.3pt);
        
        \draw[dashed, MeasBlue, line width=0.6pt]
          (SPLxy) -- (SPL);
        
        \node[
          anchor=west,
          text=MeasBlue,
          font=\scriptsize,
          fill=white,
          inner sep=1pt
        ] at ($(SPL)+(0.15,0)$)
        {Pressure Probe @ $[0.5\;m,\;0.25\;m,\;1.0\;m]$};
    
\end{tikzpicture}
    \caption{Simply supported rectangular plate subjected to a uniform transverse surface load}
    \label{fig:rectangular_plate_one_way}
\end{figure}
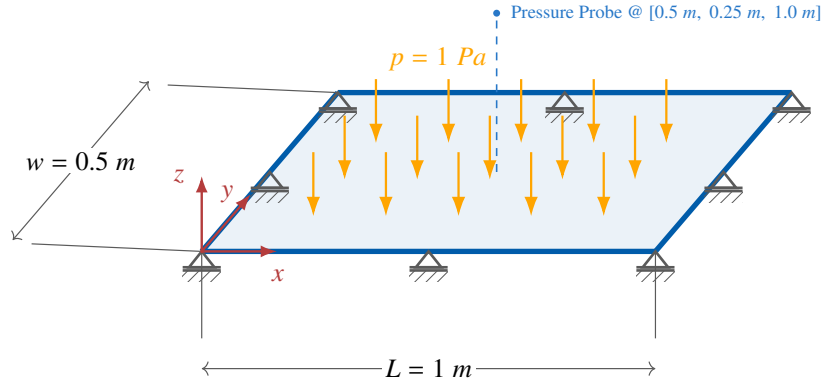

\ignore{The first numerical example considers a one-way coupled vibroacoustic problem and serves primarily to investigate the influence of the interface mapping strategy on the predicted acoustic response. In this configuration, the structural displacements at the vibroacoustic interface are transferred to the acoustic solver and imposed as boundary conditions for the Helmholtz problem, while acoustic feedback to the structural domain is neglected. Consequently, the observed differences between the monolithic and partitioned solutions can be primarily attributed to the transfer of interface quantities rather than to coupling-iteration errors. Since the coupling is unidirectional, no coupling iterations are required and, consequently, no convergence accelerator is employed in this example.

The considered problem involves a simply supported rectangular plate
subjected to a uniform harmonic surface load. The plate has
$L = 1\,\mathrm{m}$ and $w = 0.5\,\mathrm{m}$, and is simply
supported along all four edges, as illustrated in
Figure~\ref{fig:rectangular_plate_one_way}. A constant harmonic surface
load of unit amplitude $p = 1\,\mathrm{Pa}$ is applied normal to the
plate surface.}

The first numerical example considers a one-way coupled vibroacoustic benchmark to assess the influence of the interface mapping strategy on the predicted acoustic response. The structural displacements are transferred to the acoustic solver and imposed as boundary conditions for the Helmholtz problem, while acoustic feedback is neglected. Consequently, the observed differences between the monolithic and partitioned solutions can be primarily attributed to the transfer of interface quantities rather than to coupling-iteration errors. Since the coupling is unidirectional, no coupling iterations are required and, consequently, no convergence accelerator is employed in this example.

The benchmark consists of a simply supported rectangular plate with dimensions $L = 1\,\mathrm{m}$ and $w = 0.5\,\mathrm{m}$ and, subjected to a unit harmonic surface load acting normal to the plate surface, as illustrated in
Figure~\ref{fig:rectangular_plate_one_way}.

The plate is modeled as a thin, homogeneous isotropic shell with thickness $t = 5 \times 10^{-3}\,\mathrm{m}$, density $\rho = 7800\,\mathrm{kg/m^3}$, Young's modulus $E = 200 \times 10^{9}\,\mathrm{Pa}$, and Poisson's ratio $\nu = 0.3$. The surrounding acoustic medium is assumed to be air with sound speed $c = 340\,\mathrm{m/s}$ and density $\rho_f = 1.225\,\mathrm{kg/m^3}$. The vibroacoustic response is evaluated over the frequency range $0$--$1000\;\mathrm{Hz}$.

Both undamped and damped structural configurations are considered. While undamped systems are not representative of practical engineering applications, they provide a particularly demanding benchmark for partitioned coupling. For the damped case,  Rayleigh damping is introduced using
$\alpha = 6.8$ and $\beta = 4.8 \times 10^{-6}$, corresponding to
modal damping ratios of approximately $0.7$--$0.9$\% over the
frequency range investigated. 

For the initial validation of the partitioned framework, interface
quantities are transferred using the nearest-element mapping approach
described in Section~\ref{sec:nearest_element_mapping}. The influence
of alternative mapping strategies on the vibroacoustic response is
investigated subsequently in Section~\ref{sec:interface_transfer_influence}. The structural interface is discretized
using quadratic NURBS basis functions ($p=q=2$) with $30\times30$
knot spans, while the acoustic interface employs a quadratic NURBS
discretization with $18\times9$ knot spans. The acoustic discretization is selected according to~\cite{marburg2002six}. The radiated acoustic field is evaluated in terms of the acoustic pressure magnitude at the observation point $\mathbf{x} = [0.5,\; 0.25,\; 1.0]$, shown in Figure~\ref{fig:rectangular_plate_one_way}.

\begin{figure}[h!]
    \centering

    \begin{subfigure}[b]{0.49\linewidth}
        \centering
        \includegraphics[width=\linewidth]{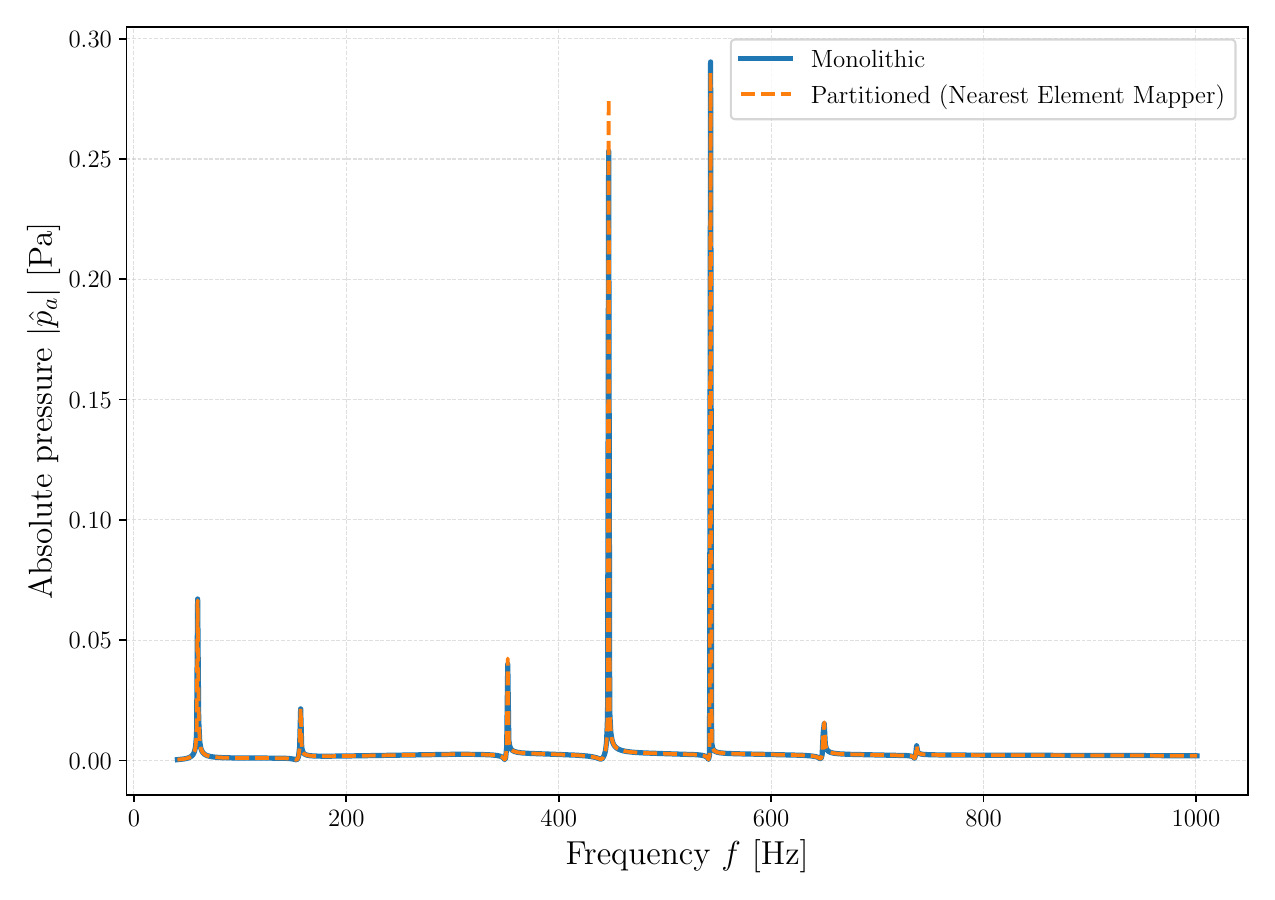}
        \caption{Without damping}
        \label{fig:spl_without_damping}
    \end{subfigure}
    \hfill
    \begin{subfigure}[b]{0.49\linewidth}
        \centering
        \includegraphics[width=\linewidth]{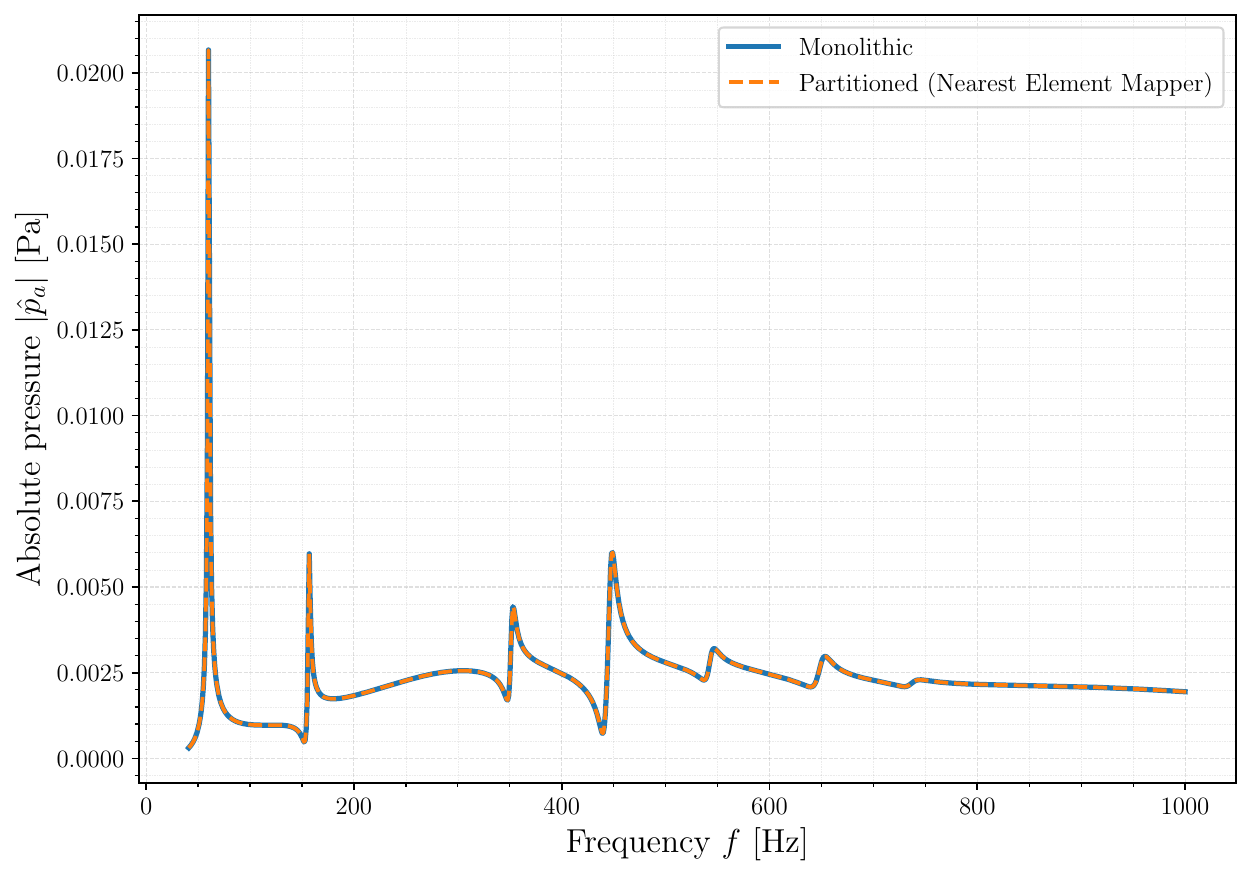}
        \caption{With damping}
        \label{fig:spl_with_damping}
    \end{subfigure}

    \caption{
    Comparison of the acoustic pressure magnitude obtained with the monolithic
    and partitioned one-way vibroacoustic coupling approaches:
    (a) undamped structural model and (b) damped structural model.
    }
    
    \label{fig:spl_plate_one_way}
\end{figure}

Figure~\ref{fig:spl_plate_one_way} compares the acoustic pressure
magnitude obtained using the monolithic reference solution and the
proposed partitioned co-simulation framework for both the undamped and
damped structural configurations. In both cases, very good agreement is
observed across the investigated frequency range, with the partitioned
solution accurately reproducing both the resonance frequencies and the
pressure amplitudes of the monolithic reference. Slightly larger
discrepancies may occur in the vicinity of resonance peaks, where small
errors introduced by the transfer of the interface quantities can be
amplified by the increased sensitivity of the acoustic response. The
damped configuration exhibits smoother resonance peaks and lower pressure
amplitudes compared to the undamped case, while maintaining the same
overall frequency response trend.

The undamped configuration is included primarily for completeness, since
some level of damping is generally present in practical systems. Moreover,
the dynamic stiffness matrix becomes singular at the exact eigenfrequencies
and ill-conditioned in their vicinity, making the response highly sensitive
to small discretization and mapping errors. Results close to the resonances
should therefore be interpreted with particular care.

\subsubsection{Influence of the interface mapper on the vibroacoustic response}
\label{sec:interface_transfer_influence}

The previous results demonstrated that the partitioned framework is able
to accurately reproduce the monolithic vibroacoustic response when
using the nearest-element mapper. In the following, the sensitivity of
the acoustic response to the selected interface transfer strategy is
investigated.

To emphasize the influence of the mapping procedure, the acoustic
interface is intentionally discretized much more coarsely than the
structural interface. The comparison is performed using a refined quadratic NURBS structural
discretization with $30\times30$ knot spans and a coarse quadratic
NURBS acoustic discretization with approximately $10\times6$ knot spans.

To quantify the influence of the interface transfer procedure, the
relative pressure amplitude error with respect to the monolithic
reference solution is computed as
\begin{equation}
    e_{\mathrm{rel}}(f)
    =
    \frac{
        \left|
        \left|\hat{p}_{\mathrm{a,part}}(f)\right|
        -
        \left|\hat{p}_{\mathrm{a,mono}}(f)\right|
        \right|
    }{
        \left|\hat{p}_{\mathrm{a,mono}}(f)\right|
    }
    \times 100\%.
    \label{eq:pressure_error}
\end{equation}

\begin{figure}[t!]
\centering

\begin{subfigure}[t]{0.48\linewidth}
    \centering
    \includegraphics[width=\linewidth]
    {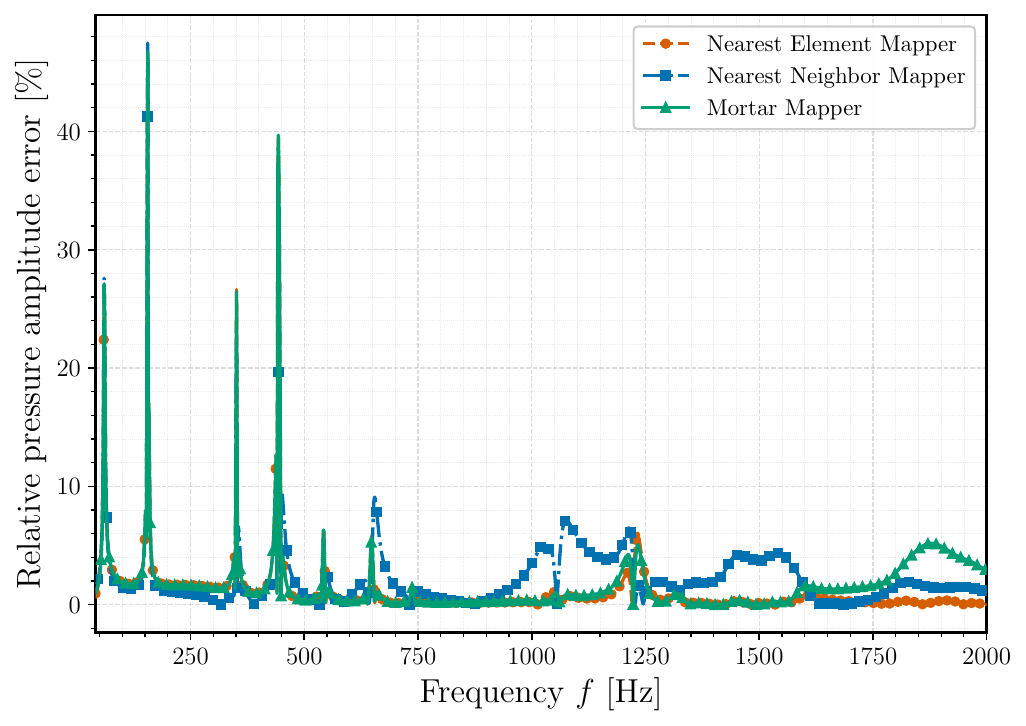}
    \caption{Coarse acoustic discretization, $10\times6$}
\end{subfigure}
\hfill
\begin{subfigure}[t]{0.48\linewidth}
    \centering
    \includegraphics[width=\linewidth]
    {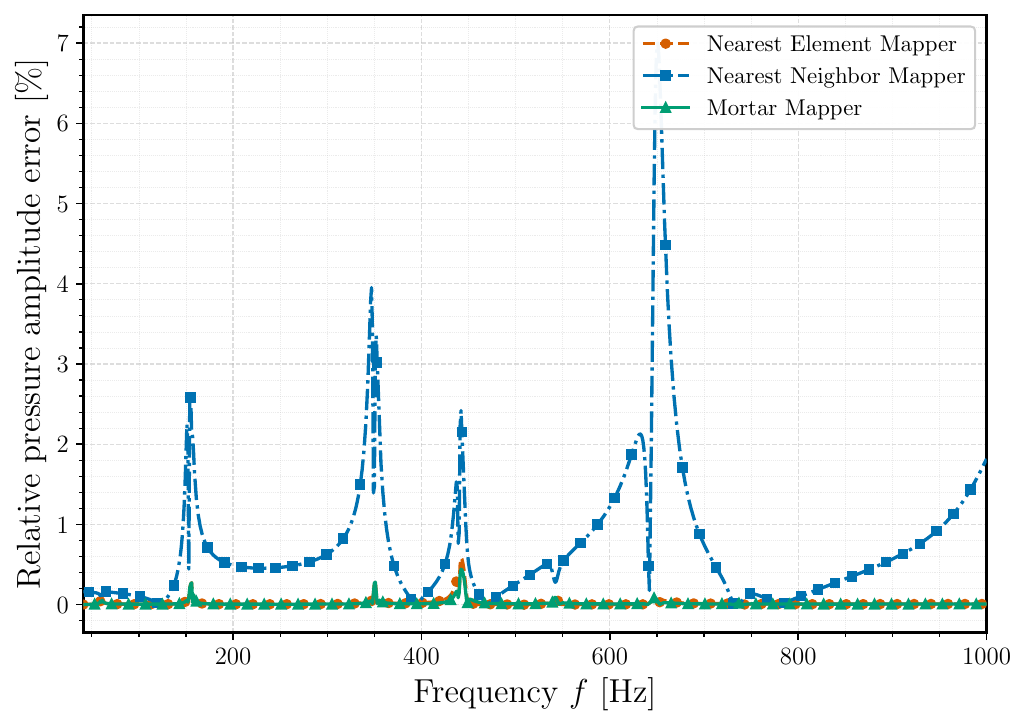}
    \caption{Refined acoustic discretization, $18\times9$}
\end{subfigure}

\caption{
Relative acoustic pressure error between the monolithic and partitioned
vibroacoustic solutions for the nearest-element, nearest-neighbor, and
mortar mapping strategies. For the strongly non-conforming $10\times6$
acoustic discretization in (a), the influence of the interface transfer
strategy is clearly visible. Refining the acoustic interface to
$18\times9$ knot spans in (b) significantly reduces the pressure errors
and the differences between the considered mapping approaches.
}
\label{fig:pressure_relative_error_mappers}

\end{figure}

Figure~\ref{fig:pressure_relative_error_mappers} compares the relative acoustic pressure error at the probe location with respect to the monolithic reference solution, computed using Eq.~\eqref{eq:pressure_error}, for the different interface transfer strategies and acoustic interface discretizations in the damped configuration.

For the strongly non-conforming $10\times6$ acoustic discretization
shown in Figure~\ref{fig:pressure_relative_error_mappers}(a), the influence
of the interface transfer strategy is clearly visible. All mapping
approaches reproduce the overall acoustic response with good accuracy over
most of the investigated frequency range, while larger deviations occur
in the vicinity of structural resonances. In these regions, the
vibroacoustic response becomes particularly sensitive to small
perturbations in the transferred interface quantities. Among the considered
approaches, the nearest-element mapper generally yields the smallest
pressure errors, whereas the nearest-neighbor mapper exhibits larger error
peaks. The mortar mapper shows a similar qualitative behavior, although
larger deviations are observed for the intentionally coarse acoustic
destination discretization.

The influence of the interface transfer strategy is substantially reduced
when the acoustic interface is refined to $18\times9$ knot spans, as shown
in Figure~\ref{fig:pressure_relative_error_mappers}(b). The relative
pressure errors decrease significantly for all considered mapping
approaches, with the nearest-element and mortar mappers remaining below
$1\%$ throughout the investigated frequency range. This reduction
demonstrates that the larger discrepancies observed for the $10\times6$
case are strongly influenced by the severe mismatch between the interface
discretizations.

It should be emphasized that the coarse configuration is deliberately
designed to amplify the influence of the interface transfer procedure and
therefore represents a particularly challenging non-conforming coupling
case. The reported errors should consequently not be interpreted as the
accuracy expected for practical discretizations. Rather, the comparison
demonstrates that the choice of interface mapper can noticeably influence
the predicted acoustic response for strongly non-conforming interfaces,
whereas these differences become considerably smaller as the acoustic
interface discretization is refined.

\subsection{Two-way coupled rectangular plate submerged in water}
\label{subsec:example_2}

The second example extends the previous configuration to a fully
two-way coupled vibroacoustic problem. The same simply supported
rectangular plate is subjected to a harmonic structural surface load
and to the acoustic excitation generated by a monopole source located
within the surrounding water domain (Figure~\ref{fig:example_2}). The
structural and acoustic fields interact through the bidirectional
exchange of interface quantities, which are transferred using the
nearest-element mapping approach described in
Section~\ref{sec:nearest_element_mapping}.

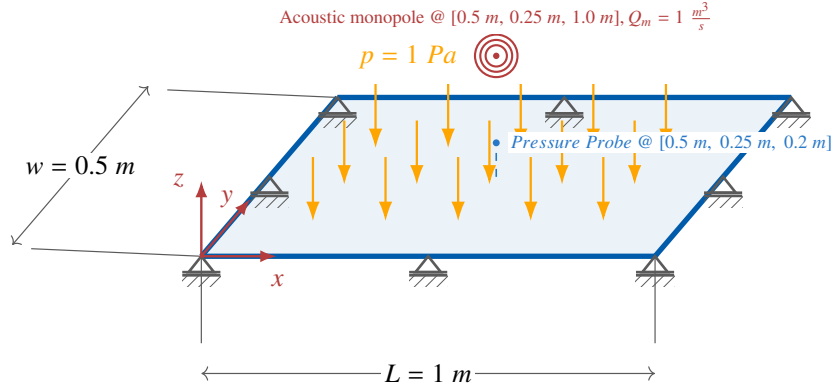
\begin{figure}[h!]
    \centering
    \begin{tikzpicture}
          \scaling{0.6};
        
          \definecolor{PlateBlue}{RGB}{0,92,170}
          \definecolor{PlateFill}{RGB}{0,92,170}
          \definecolor{AnnotGray}{RGB}{90,90,90}
          \definecolor{LoadRed}{RGB}{180,60,60}
          \definecolor{LoadYellow}{RGB}{255,165,0}
        
          \point{A}{0}{0}
          \point{B}{10}{0}
          \point{D}{3}{3.5}
          \point{C}{13}{3.5}
        
          \begin{scope}[fill=PlateFill, fill opacity=0.08, draw=none]
            \fill (A) -- (B) -- (C) -- (D) -- cycle;
          \end{scope}
        
          \begin{scope}[draw=PlateBlue, line width=0.9pt]
            \beam{4}{A}{B}
            \beam{4}{B}{C}
            \beam{4}{C}{D}
            \beam{4}{D}{A}
          \end{scope}
        
          \point{ABm}{5}{0}
          \point{DCm}{8}{3.5}
          \point{ADm}{1.5}{1.75}
          \point{BCm}{11.5}{1.75}
        
          \newcommand{\simplesupportdown}[1]{%
          \begin{scope}[shift={(#1)}, draw=AnnotGray, line width=0.4pt]
            \draw (0,0) -- (-0.18,-0.28) -- (0.18,-0.28) -- cycle;  
            \draw (-0.28,-0.32) -- (0.28,-0.32);                    
          \end{scope}
        }
        
          \begin{scope}[draw=AnnotGray, scale=0.42, transform shape]
          \support{2}{A}[0]   \support{2}{ABm}[0] \support{2}{B}[0]
          \support{2}{D}[0]   \support{2}{DCm}[0] \support{2}{C}[0]
          \support{2}{ADm}[0]
          \support{2}{BCm}[0]
        \end{scope}

          \begin{scope}[draw=LoadYellow, line width=0.85pt]
            \foreach \s in {0.18,0.34,0.50,0.66,0.82}{
              \foreach \t in {0.22,0.45,0.68}{
                \coordinate (P) at ($ (A)!\s!(B) + (A)!\t!(D) $);
                \draw[-{Latex[length=2.6mm,width=1.8mm]}]
                  ($(P)+(0,0.85)$) -- (P);
              }
            }
            \node[text=LoadYellow] at ($(A)!0.35!(C) + (0,1.9)$) {$p=1\;Pa$};
          \end{scope}
        
          \begin{scope}[draw=AnnotGray, line width=0.45pt]
            \draw[<->] ($(A)+(0,-1.55)$) -- ($(B)+(0,-1.55)$)
              node[midway, fill=white, inner sep=1.2pt] {$ L= 1\;m$};
            \draw ($(A)+(0,-1.15)$) -- (A);
            \draw ($(B)+(0,-1.15)$) -- (B);
        
            \def\offx{-2.5}
            \def\offy{0.15}
            \draw[<->] ($(A)+(\offx,\offy)$) -- ($(D)+(\offx,\offy)$)
              node[midway, fill=white, inner sep=1.2pt] {$w = 0.5\;m$};
            \draw ($(A)+(\offx+0.25,\offy-0.05)$) -- (A);
            \draw ($(D)+(\offx+0.25,\offy-0.05)$) -- (D);
          \end{scope}
        
        \tikzset{
          axisArrow/.style={
            draw=LoadRed,
            -{Latex[length=2.2mm,width=1.6mm]},
            line width=0.9pt
          }
        }
        
        \def\ax{1}
        \def\ay{1}
        \def\az{1}
        
        \coordinate (O) at ($(A)$);
        \fill[LoadRed] (O) circle (0.6pt);
        
        \pgfmathsetmacro{\ADx}{3}
        \pgfmathsetmacro{\ADy}{3.5}
        \pgfmathsetmacro{\ADn}{sqrt(\ADx*\ADx + \ADy*\ADy)}
        \pgfmathsetmacro{\uyx}{\ADx/\ADn}
        \pgfmathsetmacro{\uyy}{\ADy/\ADn}
        
        \draw[axisArrow] (O) -- ($(O)+(\ax,0)$)
          node[below=1.0mm, text=LoadRed] {$x$};
        
        \draw[axisArrow] (O) -- ($(O)+(\ay*\uyx,\ay*\uyy)$)
          node[left=1.0mm, text=LoadRed] {$y$};
        
        \draw[axisArrow] (O) -- ($(O)+(0,\az)$)
          node[left=1.0mm, text=LoadRed] {$z$};
        
        \definecolor{MeasBlue}{RGB}{40,120,200}
        
        \coordinate (SPLxy) at ($(A)!0.5!(C)$);
        
        \coordinate (SPL) at ($(SPLxy)+(0,0.45)$);
        
        \fill[MeasBlue] (SPL) circle (1.3pt);
        
        \draw[dashed, MeasBlue, line width=0.6pt]
          (SPLxy) -- (SPL);
        
        \node[
          anchor=west,
          text=MeasBlue,
          font=\scriptsize,
          fill=white,
          inner sep=1pt
        ] at ($(SPL)+(0.15,0)$)
        {$Pressure\;Probe \;@ \;[0.5 \;m,\;0.25\;m,\;0.2\;m]$};

        
        \coordinate (MonoBase) at ($ (A)!0.5!(B) + (A)!0.5!(D) $);
        
        \coordinate (Monopole) at ($(MonoBase)+(0,1.6)$);
        
        \begin{scope}[draw=LoadRed, line width=0.8pt]
          \fill[LoadRed] (Monopole) circle (1.2pt);
        
          \draw (Monopole) circle (4pt);
          \draw (Monopole) circle (6pt);
          \draw (Monopole) circle (8pt);
        \end{scope}
        
        \node[
          above=8pt,
          fill=white,
          inner sep=1.5pt,
          text=LoadRed,
          font=\scriptsize
        ] at (Monopole)
        {Acoustic monopole @ $[0.5\;m,\,0.25\;m,\,1.0\;m], Q_{m}=1\;\frac{m^3}{s}$};

    \end{tikzpicture}
    \caption{Two-way coupled vibroacoustic benchmark problem consisting of a simply supported rectangular plate subjected to a harmonic surface load and coupled with an acoustic monopole source located in the surrounding water domain. The structural and acoustic fields interact through the bidirectional exchange of interface quantities, while the sound pressure level (SPL) is evaluated at a probe position located above the plate surface.}
    \label{fig:example_2}
\end{figure}

In contrast to the one-way coupled configuration, the present problem
introduces a strong vibroacoustic interaction and therefore constitutes
a significantly more demanding partitioned solution procedure. Owing to
the substantially higher density of water compared to air, the acoustic
feedback acting on the structure is considerably stronger, making this
benchmark particularly suitable for assessing the robustness of the
proposed partitioned framework and the effectiveness of the complex
convergence acceleration strategies introduced in
Section~\ref{sec:convergence_accelerators}.

The structural model employs the same material properties as in
Example~1. However, to obtain a representative submerged configuration,
the plate thickness is increased to $t=0.01\,\mathrm{m}$. Only the
damped structural configuration is considered in this example.
Structural damping is modeled using the same Rayleigh damping
coefficients as in Example~1, namely $\alpha=6.8$ and
$\beta=4.8\times10^{-6}$.

For all strongly coupled simulations presented in this example,
convergence is assessed using the absolute norm of the interface
residual according to

\begin{equation}
\|r^{(k)}\| \leq 10^{-6},
\end{equation}
where $r^{(k)}$ denotes the interface residual at coupling iteration
$k$. The selected tolerance was found to provide partitioned solutions
in very close agreement with the corresponding monolithic reference
while maintaining a reasonable computational cost. 

For the present strongly coupled configuration, the partitioned
fixed-point iteration without convergence acceleration fails to converge.
The use of a convergence accelerator is therefore required to obtain a
converged partitioned solution.

\begin{figure}[h!]
    \centering

    \begin{subfigure}[t]{0.47\textwidth}
        \centering
        \includegraphics[width=\textwidth]
        {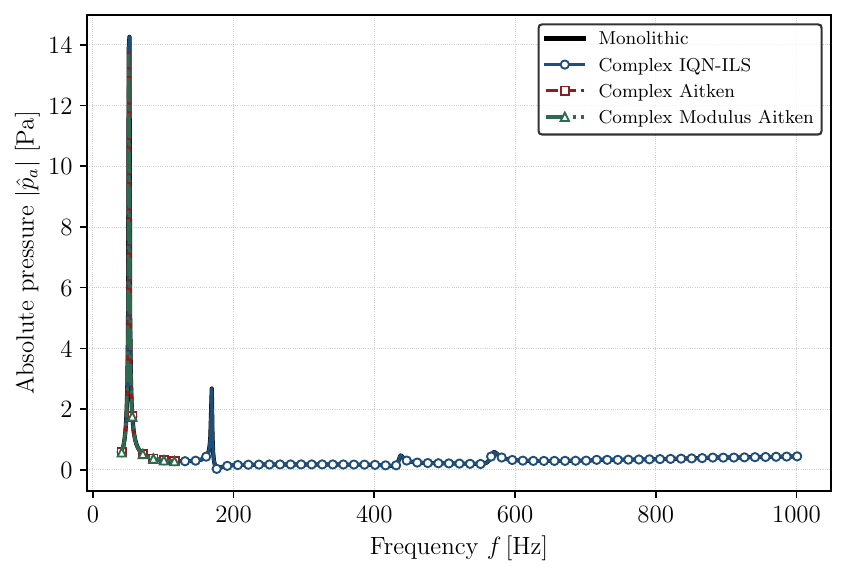}
        \caption{Acoustic pressure magnitude.}
        \label{fig:example2_complex_accelerators_pressure}
    \end{subfigure}
    \hfill
    \begin{subfigure}[t]{0.47\textwidth}
        \centering
        \includegraphics[width=\textwidth]
        {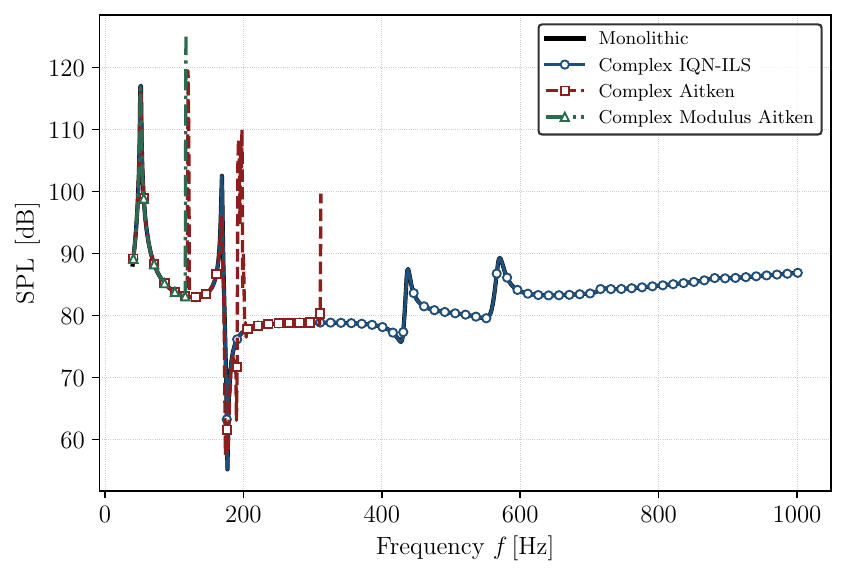}
        \caption{Sound pressure level.}
        \label{fig:example2_complex_accelerators_spl}
    \end{subfigure}

    \vspace{0.4cm}

    \begin{subfigure}[t]{0.47\textwidth}
        \centering
        \includegraphics[width=\textwidth]
        {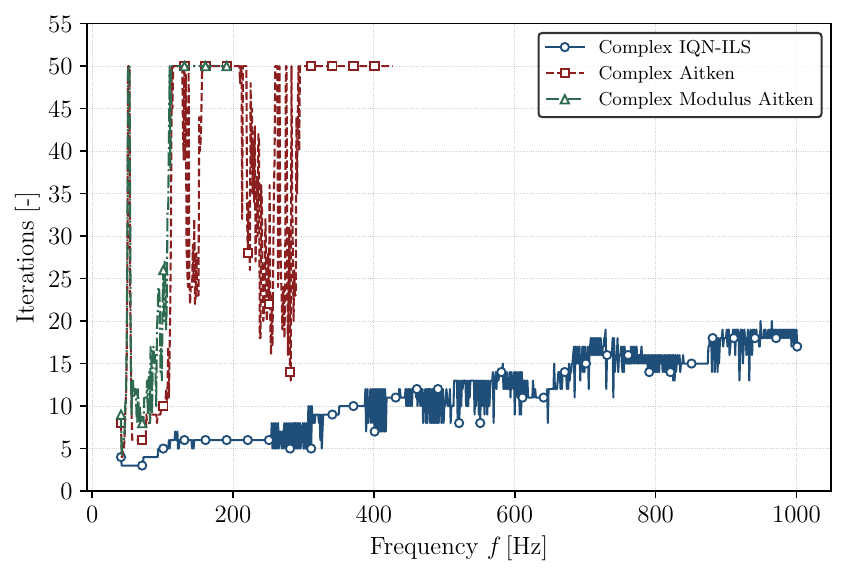}
        \caption{Number of coupling iterations.}
        \label{fig:example2_complex_accelerators_iterations}
    \end{subfigure}

    \caption{
    Comparison of the proposed complex-valued convergence accelerators
    for the submerged two-way coupled vibroacoustic benchmark:
    (a) acoustic pressure magnitude, (b) sound pressure level, and
    (c) number of coupling iterations required for convergence.
    The acoustic responses are compared with the monolithic reference
    solution, and the maximum number of coupling iterations is set to 50.
    }
    \label{fig:example2_complex_acceleration}
\end{figure}

Figure~\ref{fig:example2_complex_acceleration} compares the acoustic
response obtained with the monolithic reference solution and the strongly
coupled partitioned formulation using the proposed complex-valued IQN-ILS,
complex Aitken, and complex modulus Aitken convergence accelerators. The
response is reported in terms of both the acoustic pressure magnitude and
the sound pressure level (SPL). Although the pressure magnitude is used as
the primary quantity throughout this work, the SPL representation is
additionally included here because its logarithmic scale facilitates the
visual comparison of the different solutions over the large dynamic range
associated with the resonance peaks. The corresponding number of coupling
iterations required for convergence at each excitation frequency is also
reported.

As shown in Figures~\ref{fig:example2_complex_accelerators_pressure}
and~\ref{fig:example2_complex_accelerators_spl}, the complex IQN-ILS
accelerator remains in very close agreement with the monolithic reference
solution throughout the investigated frequency range, demonstrating the
capability of the proposed partitioned framework to accurately reproduce
the strongly coupled vibroacoustic response. The SPL representation
provides a complementary visualization of this agreement, particularly
in frequency regions where the acoustic pressure varies over several
orders of magnitude.

\begin{figure}[t!]
    \centering
    \includegraphics[width=0.55\linewidth]
    {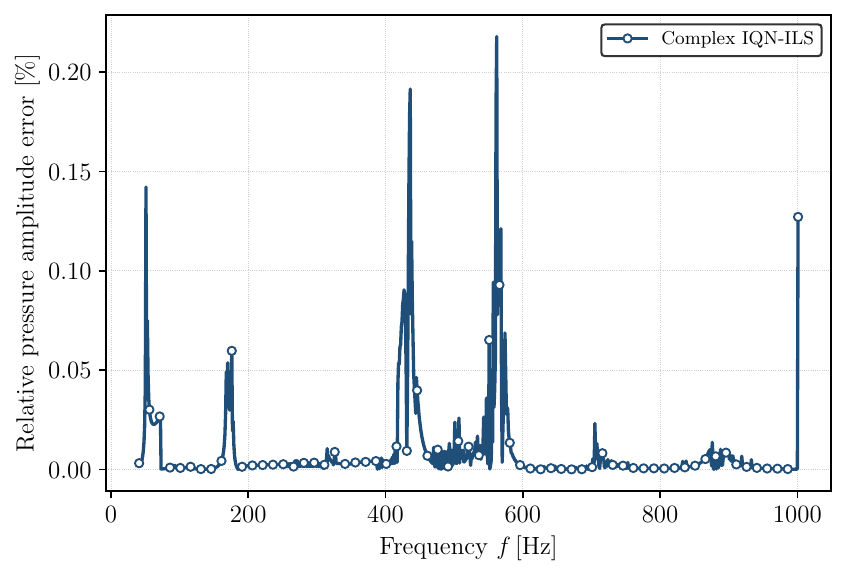}
    \caption{
    Relative acoustic pressure error between the monolithic and partitioned
    solutions obtained using the complex IQN-ILS convergence accelerator.
    }
    \label{fig:example2_complex_iqnils_pressure_error}
\end{figure}

This agreement is further quantified by the relative acoustic pressure
error shown in Figure~\ref{fig:example2_complex_iqnils_pressure_error},
computed with respect to the monolithic reference solution using
Eq.~\eqref{eq:pressure_error}. The error remains below 0.25\% over most
of the investigated frequency range, confirming the accuracy of the
complex IQN-ILS coupling strategy.

As shown in Figure~\ref{fig:example2_complex_accelerators_iterations}, the
number of coupling iterations required by the complex IQN-ILS method
increases gradually with frequency, reflecting the stronger
acoustic--structure interaction at higher excitation frequencies.
Nevertheless, fewer than ten iterations are required below approximately
\(400\,\mathrm{Hz}\), and convergence is achieved well below the prescribed
maximum of 50 iterations over the complete frequency sweep.

In contrast, as shown also in Figures~\ref{fig:example2_complex_accelerators_pressure} and~\ref{fig:example2_complex_accelerators_spl}, both Aitken-based accelerators exhibit significantly poorer convergence behavior. The complex Aitken produces solutions only up to approximately \(427\,\mathrm{Hz}\), while the complex modulus Aitken method does so only up to approximately \(191\,\mathrm{Hz}\). Beyond these frequencies, the fixed-point iterations become numerically unstable and eventually produce non-finite values, preventing the computation of a partitioned solution. Even within the frequency ranges where the methods remain operational, both Aitken variants generally require more coupling iterations than the complex IQN-ILS approach and frequently fail to satisfy the prescribed convergence criterion. In Figure~\ref{fig:example2_complex_accelerators_pressure}, the pressure range is limited to preserve the visibility of the monolithic and complex IQN-ILS solutions, since the very large pressure amplitudes obtained with the poorly converged Aitken solutions would otherwise dominate the linear scale. These larger deviations remain visible in the SPL representation in Figure~\ref{fig:example2_complex_accelerators_spl} due to its logarithmic scale.

The convergence behavior of the Aitken-based accelerators observed in the present study is consistent with previous findings in the partitioned FSI literature. For strongly coupled problems, particularly those exhibiting pronounced added-mass effects, scalar relaxation methods such as Aitken are known to exhibit reduced robustness and slower convergence than interface quasi-Newton methods \cite{DEGROOTE2010446,Delaisse2023QuasiNewton}. In contrast, IQN-ILS approximates the inverse interface Jacobian from previous coupling iterations, thereby providing a more robust and efficient coupling strategy for strongly coupled partitioned simulations. The behavior observed for the present submerged vibroacoustic problem is therefore in agreement with these findings from the partitioned FSI literature.

These results indicate that the proposed complex IQN-ILS formulation provides a more robust and efficient coupling strategy than the investigated Aitken-based alternatives for the present benchmark problem. In particular, it remains in very close agreement with the monolithic solution while converging over the entire investigated frequency range.

\begin{figure}[h!]
    \centering

    \begin{subfigure}[t]{0.47\textwidth}
        \centering
        \includegraphics[width=\textwidth]{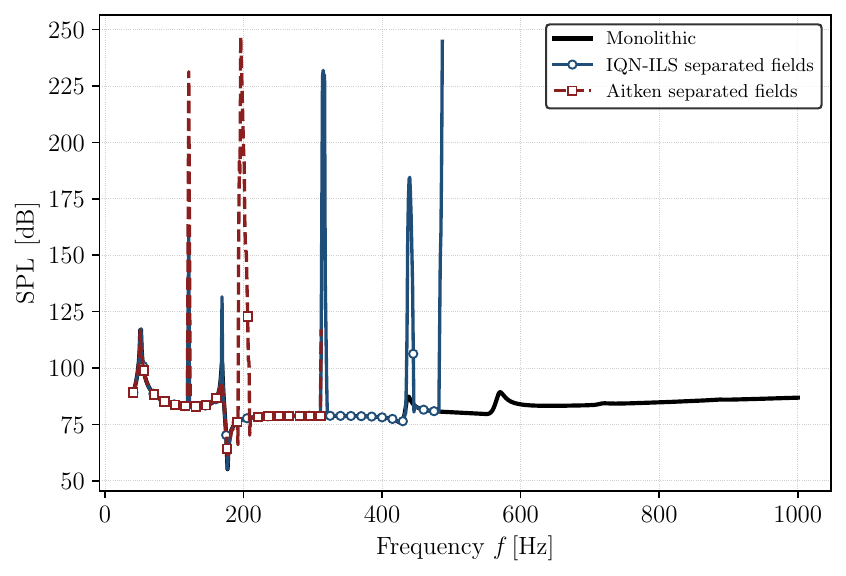}
        \caption{Comparison of the monolithic solution with the strongly coupled partitioned solutions obtained when accelerating the real and imaginary interface quantities independently.}
        \label{fig:example2_separated_accelerators_spl}
    \end{subfigure}
    \hfill
    \begin{subfigure}[t]{0.47\textwidth}
        \centering
        \includegraphics[width=\textwidth]{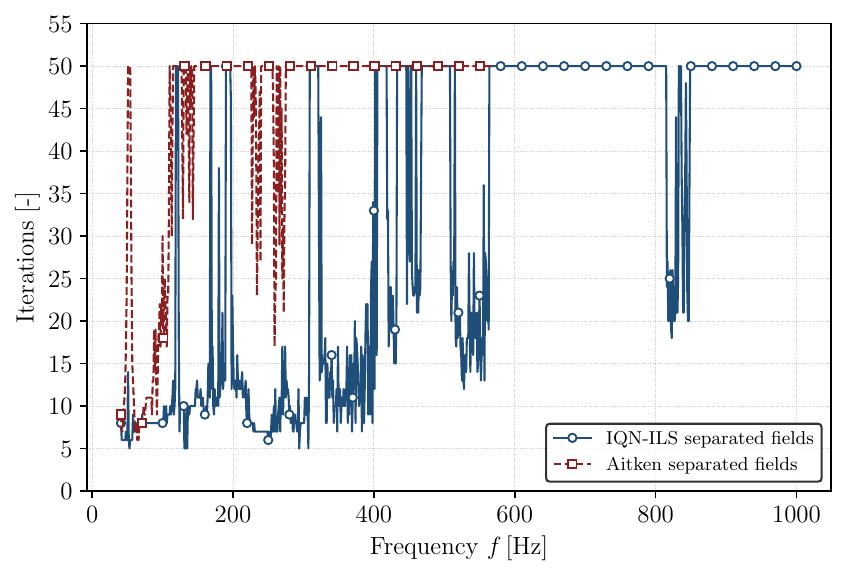}
        \caption{Number of coupling iterations required by the separated-field acceleration strategies.}
        \label{fig:example2_separated_accelerators_iterations}
    \end{subfigure}

    \caption{Performance of the separated-field acceleration strategies for the submerged two-way coupled vibroacoustic benchmark. The left figure compares the SPL response with the monolithic reference solution, while the right figure reports the corresponding number of coupling iterations required for convergence. The maximum number of coupling iterations is set to 50.}
    \label{fig:example2_separated_acceleration}
\end{figure}

To further investigate the importance of the proposed complex-valued
formulation, the results are compared against an alternative strategy in
which the real and imaginary interface quantities are accelerated
independently. Figure~\ref{fig:example2_separated_acceleration} compares
the corresponding SPL with the monolithic
reference solution and reports the associated number of coupling
iterations.

As shown in Figure~\ref{fig:example2_separated_accelerators_spl}, the
separated-field IQN-ILS formulation provides finite partitioned solutions
over the entire investigated frequency range. However, the agreement with
the monolithic reference deteriorates with increasing frequency, with
substantial deviations occurring at higher frequencies. The separated-field
Aitken formulation exhibits an even poorer behavior, with deviations from
the monolithic reference developing at lower frequencies. For clarity,
the SPL curves of both formulations are truncated once these deviations
become sufficiently large to compromise the readability of the comparison.

The convergence histories in
Figure~\ref{fig:example2_separated_accelerators_iterations} provide further
insight into this behavior. For the separated-field IQN-ILS formulation,
the number of coupling iterations increases substantially with frequency
and frequently reaches the prescribed maximum of 50 iterations. Thus,
although finite solutions are obtained over the entire frequency range,
the prescribed convergence criterion is not satisfied at many frequencies.
For the separated-field Aitken formulation, the convergence behavior is
even less robust. In addition to frequently reaching the maximum number
of coupling iterations, finite interface values are obtained only up to
approximately \(570\,\mathrm{Hz}\), beyond which the fixed-point iterations
become numerically unstable and generate non-finite values.

Compared with the genuinely complex-valued formulation, independently
accelerating the real and imaginary interface quantities leads to reduced
convergence robustness and a noticeable deterioration of the predicted
acoustic response. These results indicate that preserving the coupling
between the amplitude and phase of the interface quantities during the
convergence acceleration is important for accurately solving strongly
coupled frequency-domain vibroacoustic problems.

\ignore{
\subsection{Example 2: Two-way coupled rectangular plate}
\label{subsec:example_2}

The second example extends the previous configuration to a fully
two-way coupled vibroacoustic problem. The same simply supported
rectangular plate is subjected to a harmonic structural surface load
and to the acoustic excitation generated by a monopole source located
within the acoustic domain (Figure~\ref{fig:example_2}). The structural
and acoustic fields interact through the bidirectional exchange of
interface quantities, which are transferred using the nearest-element
mapping approach described in
Section~\ref{sec:nearest_element_mapping}.

\begin{figure}[h!]
    \centering
    \begin{tikzpicture}
          \scaling{0.6};
        
          \definecolor{PlateBlue}{RGB}{0,92,170}
          \definecolor{PlateFill}{RGB}{0,92,170}
          \definecolor{AnnotGray}{RGB}{90,90,90}
          \definecolor{LoadRed}{RGB}{180,60,60}
          \definecolor{LoadYellow}{RGB}{255,165,0}
        
          \point{A}{0}{0}
          \point{B}{10}{0}
          \point{D}{3}{3.5}
          \point{C}{13}{3.5}
        
          \begin{scope}[fill=PlateFill, fill opacity=0.08, draw=none]
            \fill (A) -- (B) -- (C) -- (D) -- cycle;
          \end{scope}
        
          \begin{scope}[draw=PlateBlue, line width=0.9pt]
            \beam{4}{A}{B}
            \beam{4}{B}{C}
            \beam{4}{C}{D}
            \beam{4}{D}{A}
          \end{scope}
        
          \point{ABm}{5}{0}
          \point{DCm}{8}{3.5}
          \point{ADm}{1.5}{1.75}
          \point{BCm}{11.5}{1.75}
        
          \newcommand{\simplesupportdown}[1]{%
          \begin{scope}[shift={(#1)}, draw=AnnotGray, line width=0.4pt]
            \draw (0,0) -- (-0.18,-0.28) -- (0.18,-0.28) -- cycle;  
            \draw (-0.28,-0.32) -- (0.28,-0.32);                    
          \end{scope}
        }
        
          \begin{scope}[draw=AnnotGray, scale=0.42, transform shape]
          \support{2}{A}[0]   \support{2}{ABm}[0] \support{2}{B}[0]
          \support{2}{D}[0]   \support{2}{DCm}[0] \support{2}{C}[0]
          \support{2}{ADm}[0]
          \support{2}{BCm}[0]
        \end{scope}

          \begin{scope}[draw=LoadYellow, line width=0.85pt]
            \foreach \s in {0.18,0.34,0.50,0.66,0.82}{
              \foreach \t in {0.22,0.45,0.68}{
                \coordinate (P) at ($ (A)!\s!(B) + (A)!\t!(D) $);
                \draw[-{Latex[length=2.6mm,width=1.8mm]}]
                  ($(P)+(0,0.85)$) -- (P);
              }
            }
            \node[text=LoadYellow] at ($(A)!0.35!(C) + (0,1.9)$) {$p=1\;Pa$};
          \end{scope}
        
          \begin{scope}[draw=AnnotGray, line width=0.45pt]
            \draw[<->] ($(A)+(0,-1.55)$) -- ($(B)+(0,-1.55)$)
              node[midway, fill=white, inner sep=1.2pt] {$ L= 1\;m$};
            \draw ($(A)+(0,-1.15)$) -- (A);
            \draw ($(B)+(0,-1.15)$) -- (B);
        
            \def\offx{-2.5}
            \def\offy{0.15}
            \draw[<->] ($(A)+(\offx,\offy)$) -- ($(D)+(\offx,\offy)$)
              node[midway, fill=white, inner sep=1.2pt] {$w = 0.5\;m$};
            \draw ($(A)+(\offx+0.25,\offy-0.05)$) -- (A);
            \draw ($(D)+(\offx+0.25,\offy-0.05)$) -- (D);
          \end{scope}
        
        \tikzset{
          axisArrow/.style={
            draw=LoadRed,
            -{Latex[length=2.2mm,width=1.6mm]},
            line width=0.9pt
          }
        }
        
        \def\ax{1}
        \def\ay{1}
        \def\az{1}
        
        \coordinate (O) at ($(A)$);
        \fill[LoadRed] (O) circle (0.6pt);
        
        \pgfmathsetmacro{\ADx}{3}
        \pgfmathsetmacro{\ADy}{3.5}
        \pgfmathsetmacro{\ADn}{sqrt(\ADx*\ADx + \ADy*\ADy)}
        \pgfmathsetmacro{\uyx}{\ADx/\ADn}
        \pgfmathsetmacro{\uyy}{\ADy/\ADn}
        
        \draw[axisArrow] (O) -- ($(O)+(\ax,0)$)
          node[below=1.0mm, text=LoadRed] {$x$};
        
        \draw[axisArrow] (O) -- ($(O)+(\ay*\uyx,\ay*\uyy)$)
          node[left=1.0mm, text=LoadRed] {$y$};
        
        \draw[axisArrow] (O) -- ($(O)+(0,\az)$)
          node[left=1.0mm, text=LoadRed] {$z$};
        
        \definecolor{MeasBlue}{RGB}{40,120,200}
        
        \coordinate (SPLxy) at ($(A)!0.5!(C)$);
        
        \coordinate (SPL) at ($(SPLxy)+(0,0.45)$);
        
        \fill[MeasBlue] (SPL) circle (1.3pt);
        
        \draw[dashed, MeasBlue, line width=0.6pt]
          (SPLxy) -- (SPL);
        
        \node[
          anchor=west,
          text=MeasBlue,
          font=\scriptsize,
          fill=white,
          inner sep=1pt
        ] at ($(SPL)+(0.15,0)$)
        {$Pressure\;Probe \;@ \;[0.5 \;m,\;0.25\;m,\;0.2\;m]$};

        
        \coordinate (MonoBase) at ($ (A)!0.5!(B) + (A)!0.5!(D) $);
        
        \coordinate (Monopole) at ($(MonoBase)+(0,1.6)$);
        
        \begin{scope}[draw=LoadRed, line width=0.8pt]
          \fill[LoadRed] (Monopole) circle (1.2pt);
        
          \draw (Monopole) circle (4pt);
          \draw (Monopole) circle (6pt);
          \draw (Monopole) circle (8pt);
        \end{scope}
        
        \node[
          above=8pt,
          fill=white,
          inner sep=1.5pt,
          text=LoadRed,
          font=\scriptsize
        ] at (Monopole)
        {Acoustic monopole @ $[0.5\;m,\,0.25\;m,\,1.0\;m], Q_{m}=1\;\frac{m^3}{s}$};

    \end{tikzpicture}
    \caption{Two-way coupled vibroacoustic benchmark problem consisting of a simply supported rectangular plate subjected to a harmonic surface load and coupled with an acoustic monopole source located in the acoustic domain. The structural and acoustic fields interact through the bidirectional exchange of interface quantities, while the acoustic pressure is evaluated at a probe position located above the plate surface.}
    \label{fig:example_2}
\end{figure}

In contrast to the one-way coupled configuration, the present problem
introduces a strong vibroacoustic interaction and therefore constitutes
a significantly more demanding partitioned solution procedure. The
exchange of complex-valued interface quantities becomes particularly
sensitive near structural resonances, where conventional fixed-point
iterations may converge slowly or even diverge. This example is
therefore used to assess the robustness of the proposed framework and
the effectiveness of the different complex convergence acceleration
strategies proposed in Section~\ref{sec:convergence_accelerators}.

For all strongly coupled simulations presented in the following,
convergence is assessed using the relative norm of the interface
residual between two successive coupling iterations according to

\begin{equation}
\frac{|r^{(k)}|}{|r^{(k-1)}|} \leq 10^{-4},
\end{equation}

where $r^{(k)}$ denotes the interface residual at coupling iteration
$k$. The selected tolerance was found to provide partitioned solutions
in very close agreement with the corresponding monolithic reference
while maintaining a reasonable computational cost.

\subsubsection{Undamped structural model}

The undamped two-way coupled configuration is first used to investigate the influence of the convergence accelerator on the robustness and efficiency of the partitioned solution procedure. Figure~\ref{fig:example2_complex_acceleration_undamped} compares the SPL response obtained with the monolithic reference solution and with the strongly coupled partitioned formulation using the proposed complex-valued Aitken approaches and IQN-ILS acceleration strategies. For comparison, the weakly coupled partitioned solution is also included. The corresponding number of coupling iterations required for convergence at each excitation frequency is additionally reported.

\begin{figure}[h!]
    \centering

    \begin{subfigure}[t]{0.49\textwidth}
        \centering
        \includegraphics[width=\textwidth]{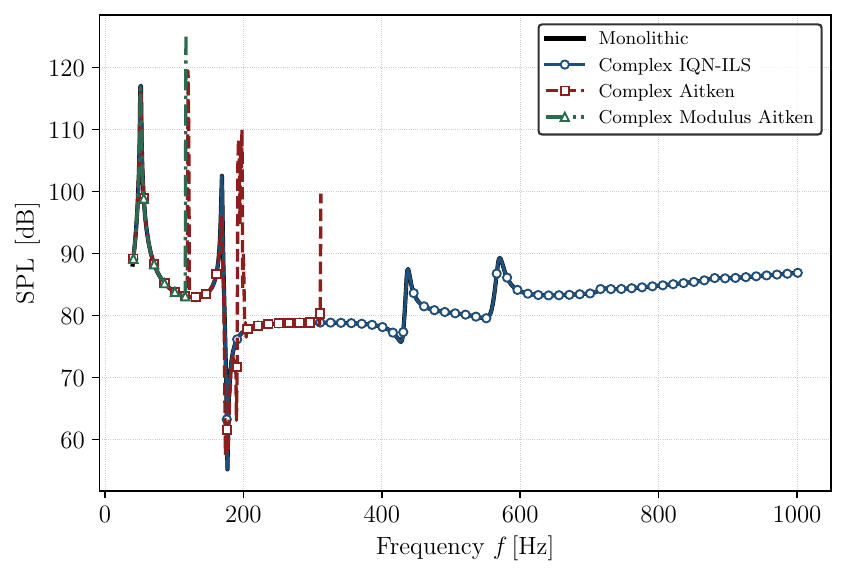}
        \caption{Comparison of the monolithic solution with the strongly coupled partitioned SPL responses obtained using the complex-valued convergence accelerators.}
        \label{fig:example2_complex_accelerators_spl}
    \end{subfigure}
    \hfill
    \begin{subfigure}[t]{0.49\textwidth}
        \centering
        \includegraphics[width=\textwidth]{example_2_complex_strong_vs_weak_spl.pdf}
        \caption{Comparison of the monolithic, weakly coupled, and strongly coupled partitioned SPL responses using the complex IQN-ILS accelerator.}
        \label{fig:example2_complex_strong_vs_weak_spl}
    \end{subfigure}

    \vspace{0.25cm}

    \begin{subfigure}[t]{0.50\textwidth}
        \centering
        \includegraphics[width=\textwidth]{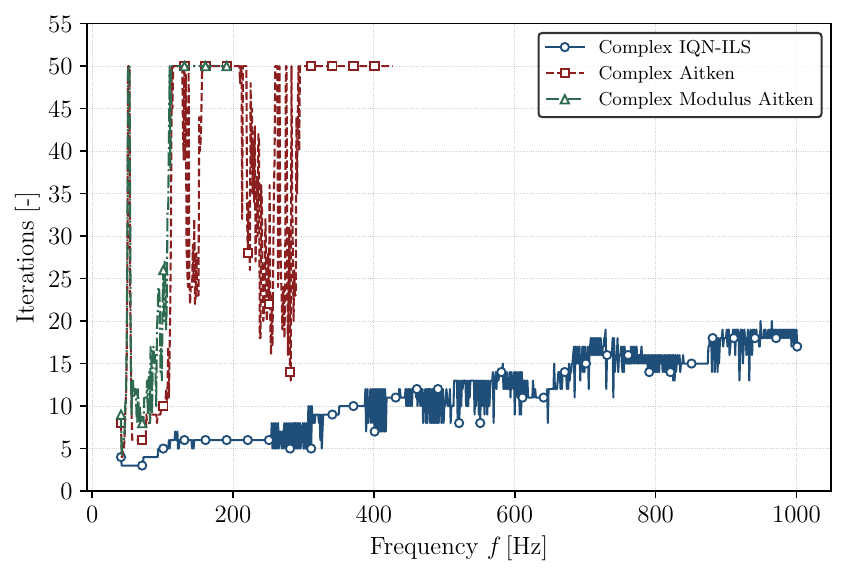}
        \caption{Corresponding number of coupling iterations required by the complex-valued convergence accelerators.}
        \label{fig:example2_complex_accelerators_iterations}
    \end{subfigure}

    \caption{Comparison of the monolithic, weakly coupled, and strongly coupled partitioned vibroacoustic solutions for the undamped configuration using complex-valued convergence accelerators. The maximum number of coupling iterations is set to 50.}
    \label{fig:example2_complex_acceleration_undamped}
\end{figure}

The weakly coupled solution already provides a good approximation of the monolithic response over a large portion of the investigated frequency range. However, noticeable deviations appear in the vicinity of the structural resonance frequencies, where the acoustic-structural interaction becomes significantly stronger and the explicit treatment of the interface conditions is no longer sufficient to accurately reproduce the coupled response. In these regions, the strongly coupled formulations provide a substantially improved agreement with the monolithic reference solution.

All the complex-valued strong coupling strategies reproduce the monolithic SPL response with very good agreement over most of the investigated frequency range. Nevertheless, clear differences appear in the convergence behavior of the coupling iterations. While both the complex Aitken methods requires a noticeable increase in the number of iterations near resonant frequencies, the complex IQN-ILS formulation maintains a significantly faster and more stable convergence behavior throughout the frequency sweep.

To further assess the importance of the complex-valued formulation, the results are compared against an alternative approach in which the real and imaginary interface fields are accelerated independently. Figure~\ref{fig:example2_separated_acceleration_undamped} shows that this separated-field treatment considerably deteriorates the robustness of the partitioned iterations, particularly in the vicinity of resonance frequencies.

\begin{figure}[h!]
    \centering

    \begin{subfigure}[t]{0.49\textwidth}
        \centering
        \includegraphics[width=\textwidth]{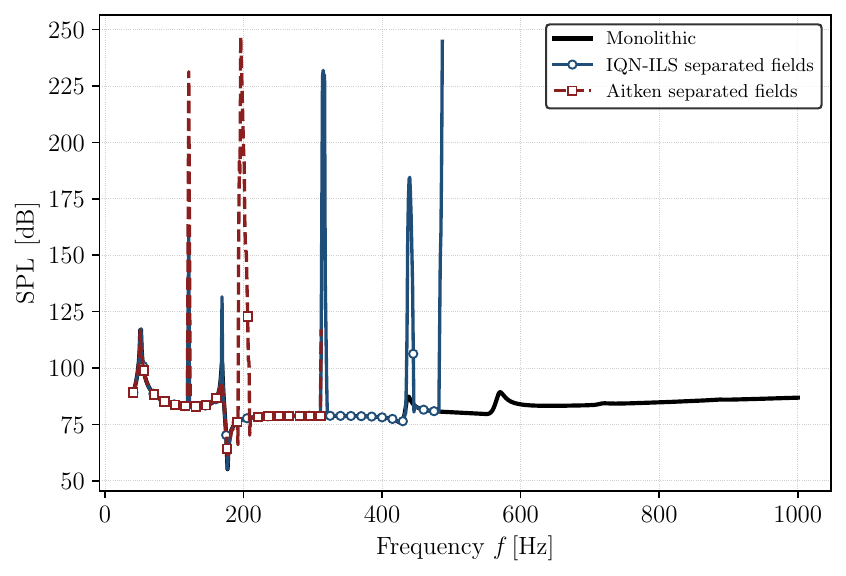}
        \caption{Comparison of the monolithic solution with the strongly coupled partitioned SPL responses obtained when accelerating the real and imaginary interface quantities independently.}
        \label{fig:example2_separated_accelerators_spl}
    \end{subfigure}
    \hfill
    \begin{subfigure}[t]{0.49\textwidth}
        \centering
        \includegraphics[width=\textwidth]{example_2_separated_strong_vs_weak_spl.pdf}
        \caption{Comparison of the monolithic, weakly coupled, and strongly coupled partitioned SPL responses for the separated-field acceleration strategy.}
        \label{fig:example2_separated_strong_vs_weak_spl}
    \end{subfigure}

    \vspace{0.25cm}

    \begin{subfigure}[t]{0.50\textwidth}
        \centering
        \includegraphics[width=\textwidth]{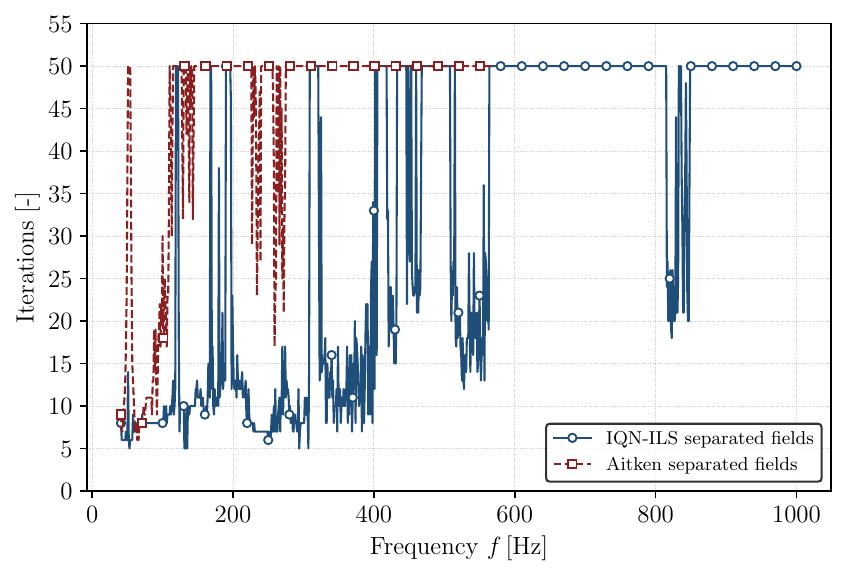}
        \caption{Corresponding number of coupling iterations required by the separated-field acceleration strategies.}
        \label{fig:example2_separated_accelerators_iterations}
    \end{subfigure}

    \caption{Comparison of the monolithic, weakly coupled, and strongly coupled partitioned vibroacoustic solutions for the undamped configuration when accelerating the real and imaginary interface quantities independently. The maximum number of coupling iterations is set to 50.}
    \label{fig:example2_separated_acceleration_undamped}
\end{figure}

The limitations of the separated-field approach become particularly evident around the structural resonance at approximately $f \approx 543\,\mathrm{Hz}$. In this region, the coupling iterations become highly sensitive to the phase relation between the structural and acoustic fields. Whereas the complex IQN-ILS method converges rapidly and remains in close agreement with the monolithic reference solution, the complex Aitken formulation exhibits a strong increase in the number of iterations and eventually fails to converge for some frequencies. The separated-field formulations perform even worse, producing significantly larger iteration counts and noticeable deviations in the predicted SPL response.

Overall, the results indicate that directly accelerating the complex-valued interface quantities is essential for obtaining robust partitioned solutions in strongly coupled frequency-domain vibroacoustic problems. By accounting simultaneously for amplitude and phase information, the complex-valued accelerators provide a substantially more stable coupling procedure than approaches based on independent treatment of the real and imaginary fields. Among the investigated strategies, the complex IQN-ILS method provides the best overall performance, combining very good agreement with the monolithic reference solution and consistently low iteration counts across the investigated frequency range.

\subsubsection{Damped structural model}

The influence of structural damping on the partitioned vibroacoustic coupling procedure is next investigated using the same Rayleigh damping model introduced previously in Example~1, with
\begin{equation}
    \alpha = 6.8,
    \qquad
    \beta = 4.8 \times 10^{-6}.
\end{equation}
These parameters correspond to a lightly damped structural response with modal damping ratios below approximately $1\%$ over the investigated frequency range.

Figure~\ref{fig:example2_damped_results} summarizes the results obtained for the damped configuration. The SPL response obtained with the monolithic reference solution is compared against both weakly and strongly coupled partitioned formulations, while the corresponding convergence histories for the investigated acceleration strategies are also reported.

\begin{figure}[h!]
    \centering

    \begin{subfigure}[t]{0.50\textwidth}
        \centering
        \includegraphics[width=\textwidth]{example_2_SPL_vs_frequency_with_damping.pdf}
        \caption{Comparison of the monolithic, weakly coupled, and strongly coupled partitioned SPL responses for the damped vibroacoustic configuration using the nearest-element mapper.}
        \label{fig:example2_damped_spl}
    \end{subfigure}

    \vspace{0.4cm}

    \begin{subfigure}[t]{0.48\textwidth}
        \centering
        \includegraphics[width=\textwidth]{example_2_convergence_histories_complex_with_damping.pdf}
        \caption{Convergence histories obtained using the complex-valued Aitken and IQN-ILS accelerators.}
        \label{fig:example2_damped_complex_iterations}
    \end{subfigure}
    \hfill
    \begin{subfigure}[t]{0.48\textwidth}
        \centering
        \includegraphics[width=\textwidth]{example_2_convergence_histories_separated_fields_with_damping.pdf}
        \caption{Convergence histories obtained when accelerating the real and imaginary interface fields independently.}
        \label{fig:example2_damped_separated_iterations}
    \end{subfigure}

    \caption{Damped two-way coupled vibroacoustic configuration. The top figure reports the SPL response over the investigated frequency range, while the bottom figures show the corresponding number of coupling iterations required for convergence using different acceleration strategies.}

    \label{fig:example2_damped_results}
\end{figure}

Compared to the undamped configuration, the introduction of structural damping significantly stabilizes the coupled problem. Both the weakly and strongly coupled partitioned formulations now remain in very close agreement with the monolithic reference solution throughout the investigated frequency range. The damping smooths the structural resonance peaks and reduces the strength of the acoustic-structural feedback at the interface, thereby alleviating the sensitivity of the coupled system to interface inconsistencies and phase errors.

The improved stability of the damped configuration is also reflected in the convergence behavior of the strong coupling iterations. All complex-valued acceleration strategies converge robustly over the entire frequency sweep and require only a relatively small number of coupling iterations compared to the undamped case. Nevertheless, the complex IQN-ILS formulation still consistently outperforms both Aitken approaches, exhibiting lower iteration counts and a more uniform convergence behavior across the investigated frequencies.

Although the damping substantially improves the robustness of the partitioned formulation, the manner in which the interface quantities are accelerated continues to play an important role. The convergence histories clearly show that accelerating the real and imaginary interface fields independently again leads to noticeably poorer convergence properties than the genuinely complex-valued formulation. Even in the presence of damping, the separated-field approaches require significantly larger iteration counts, indicating that the coupled amplitude-phase relation between the structural and acoustic responses remains essential for obtaining an efficient fixed-point iteration.

Overall, the damped configuration confirms two important observations. First, structural damping considerably alleviates the stability difficulties typically encountered in strongly coupled frequency-domain vibroacoustic simulations. Second, the proposed complex-valued acceleration techniques consistently provides superior convergence behavior compared to approaches based on independent acceleration of the real and imaginary interface fields, demonstrating its robustness for both undamped and damped vibroacoustic coupling problems.
}

\ignore{
\subsection{Example 3: Two-way coupled beam-like shell structure}
\label{subsec:example_3}

The third example considers a two-way coupled vibroacoustic problem involving a slender shell strip subjected to a harmonic surface load and coupled with an acoustic monopole source located in the acoustic domain (Figure~\ref{fig:example_3_beam}). In contrast to the previous plate configurations, the structure is supported only along the left and right edges, resulting in a predominantly beam-like bending response.

\begin{figure}[h!]
    \centering
    \begin{tikzpicture}
  \scaling{0.6};

  \definecolor{PlateBlue}{RGB}{0,92,170}
  \definecolor{PlateFill}{RGB}{0,92,170}
  \definecolor{AnnotGray}{RGB}{90,90,90}
  \definecolor{LoadRed}{RGB}{180,60,60}
  \definecolor{LoadYellow}{RGB}{255,165,0}
  \definecolor{MeasBlue}{RGB}{40,120,200}

  \point{A}{0}{0}
  \point{B}{10}{0}
  \point{D}{2.5}{4.33}
  \point{C}{12.5}{4.33}

  \begin{scope}[fill=PlateFill, fill opacity=0.08, draw=none]
    \fill (A) -- (B) -- (C) -- (D) -- cycle;
  \end{scope}

  \begin{scope}[draw=PlateBlue, line width=0.9pt]
    \beam{4}{A}{B}
    \beam{4}{B}{C}
    \beam{4}{C}{D}
    \beam{4}{D}{A}
  \end{scope}

  \coordinate (ADm) at ($(A)!0.5!(D)$);
  \coordinate (BCm) at ($(B)!0.5!(C)$);

  \begin{scope}[draw=AnnotGray, scale=0.42, transform shape]
    \support{2}{A}[0]
    \support{2}{ADm}[0]
    \support{2}{D}[0]

    \support{2}{B}[0]
    \support{2}{BCm}[0]
    \support{2}{C}[0]
  \end{scope}

  \begin{scope}[draw=LoadYellow, line width=0.85pt]

    \foreach \s in {0.12,0.24,0.36,0.48,0.60,0.72,0.84}{
      \foreach \t in {0.25,0.50,0.75}{

        \coordinate (P) at ($ (A)!\s!(B) + (A)!\t!(D) $);

        \draw[-{Latex[length=2.6mm,width=1.8mm]}]
          ($(P)+(0,0.8)$) -- (P);
      }
    }

    \node[text=LoadYellow]
      at ($(A)!0.3!(C) + (0,2.1)$)
      {$p=1\;Pa$};

  \end{scope}

  \begin{scope}[draw=AnnotGray, line width=0.45pt]

    \draw[<->]
      ($(A)+(0,-1.4)$) --
      ($(B)+(0,-1.4)$)
      node[midway, fill=white, inner sep=1.2pt]
      {$L = 1\;m$};

    \draw ($(A)+(0,-1.0)$) -- (A);
    \draw ($(B)+(0,-1.0)$) -- (B);

    \def\offx{-2.0}
    \def\offy{0.1}

    \draw[<->]
      ($(A)+(\offx,\offy)$) --
      ($(D)+(\offx,\offy)$)
      node[midway, fill=white, inner sep=1.2pt]
      {$w = 0.5\;m$};

    \draw ($(A)+(\offx+0.2,\offy-0.05)$) -- (A);
    \draw ($(D)+(\offx+0.2,\offy-0.05)$) -- (D);

  \end{scope}

  \tikzset{
    axisArrow/.style={
      draw=LoadRed,
      -{Latex[length=2.2mm,width=1.6mm]},
      line width=0.9pt
    }
  }

  \coordinate (O) at ($(A)$);
  \fill[LoadRed] (O) circle (0.6pt);

  \draw[axisArrow]
    (O) -- ($(O)+(1,0)$)
    node[below=1mm, text=LoadRed]
    {$x$};

  \draw[axisArrow]
    (O) -- ($(O)!1cm!(D)$)
    node[left=1mm, text=LoadRed]
    {$y$};

  \draw[axisArrow]
    (O) -- ($(O)+(0,1)$)
    node[left=1mm, text=LoadRed]
    {$z$};

  \coordinate (SPLxy) at ($ (A)!0.5!(B) + (A)!0.5!(D) $);
  \coordinate (SPL) at ($(SPLxy)+(0,0.45)$);

  \fill[MeasBlue] (SPL) circle (1.3pt);

  \draw[dashed, MeasBlue, line width=0.6pt]
    (SPLxy) -- (SPL);

  \node[
    anchor=west,
    text=MeasBlue,
    font=\scriptsize,
    fill=white,
    inner sep=1pt
  ]
  at ($(SPL)+(0.2,0)$)
  {$SPL\;Probe\;@\;[0.5\;m,\;0.25\;m,\;0.2\;m]$};

  \coordinate (MonoBase)
    at ($ (A)!0.5!(B) + (A)!0.5!(D) $);

  \coordinate (Monopole)
    at ($(MonoBase)+(0,1.8)$);

  \begin{scope}[draw=LoadRed, line width=0.8pt]
    \fill[LoadRed] (Monopole) circle (1.2pt);

    \draw (Monopole) circle (4pt);
    \draw (Monopole) circle (6pt);
    \draw (Monopole) circle (8pt);
  \end{scope}

  \node[
    above=8pt,
    fill=white,
    inner sep=1.5pt,
    text=LoadRed,
    font=\scriptsize
  ]
  at (Monopole)
  {Acoustic monopole @ $[0.5,\,0.25,\,1.0],\;Q_m=1\;m^3/s$};

\end{tikzpicture}
    \caption{Two-way coupled vibroacoustic benchmark consisting of a slender shell strip with beam-like response subjected to a harmonic surface load and coupled with an acoustic monopole source located in the acoustic domain. The shell is supported only along the left and right edges, leading to predominantly beam-like bending behavior. The sound pressure level (SPL) is evaluated at a probe position located near the free end of the structure.}
    \label{fig:example_3_beam}
\end{figure}

An aluminum shell with thickness $t = 2.5 \times 10^{-3}\,\mathrm{m}$, Young's modulus $E = 70 \times 10^{9}\,\mathrm{Pa}$, Poisson ratio $\nu = 0.33$, and density $\rho = 2700\,\mathrm{kg/m^3}$ is considered. The acoustic medium corresponds to air with sound speed $c = 340\,\mathrm{m/s}$ and density $\rho_f = 1.225\,\mathrm{kg/m^3}$.

The structural domain is discretized using quadratic isogeometric Kirchhoff--Love shell elements with 30 knot spans in each parametric direction, while the acoustic domain is discretized using quadratic IGA-BEM elements with 18 and 9 knot spans in the longitudinal and transverse directions, respectively. The vibroacoustic response is evaluated over the frequency range
$0 \leq f \leq 1000\,\mathrm{Hz}$.

Compared to the previous plate examples, the reduced bending stiffness and beam-type support condition lead to larger structural displacements and a stronger acoustic-structural interaction. To further increase the difficulty of the partitioned problem, no structural damping is considered in this example. The resulting sharp resonance peaks and strong phase variations make this configuration more demanding for strongly coupled partitioned simulations and therefore well suited for assessing the robustness of the proposed convergence accelerators.

Figure~\ref{fig:example_3_spl_iterations} compares the SPL response and the corresponding number of coupling iterations obtained with the different convergence acceleration strategies. For all simulations, interface quantities are transferred between the structural and acoustic discretizations using the nearest-element mapper. The top row presents the results obtained with the proposed complex-valued acceleration techniques, while the bottom row shows the corresponding results for acceleration strategies applied independently to the real and imaginary components of the interface fields. In each case, both the SPL response and the number of coupling iterations required for convergence are reported.

\begin{figure}[h!]
    \centering

    \begin{subfigure}[t]{0.49\textwidth}
        \centering
        \includegraphics[width=\textwidth]{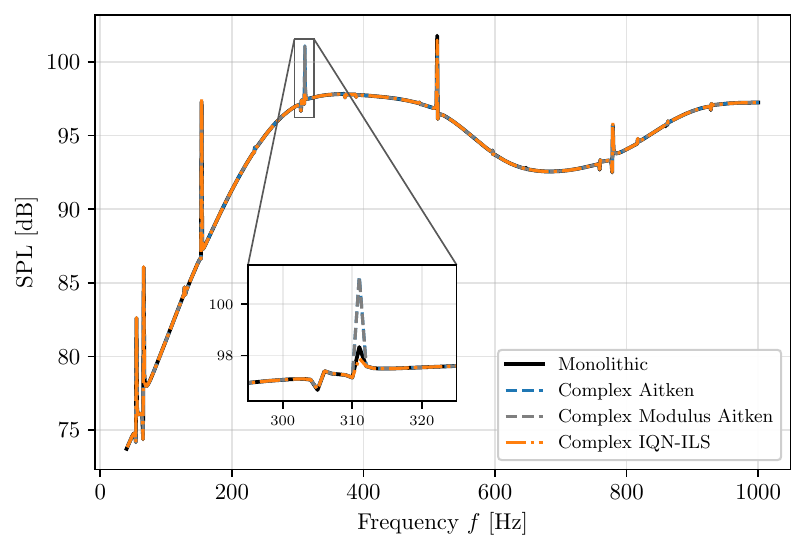}
        \caption{SPL response obtained using the complex-valued convergence accelerators.}
        \label{fig:example_3_complex_spl}
    \end{subfigure}
    \hfill
    \begin{subfigure}[t]{0.49\textwidth}
        \centering
        \includegraphics[width=\textwidth]{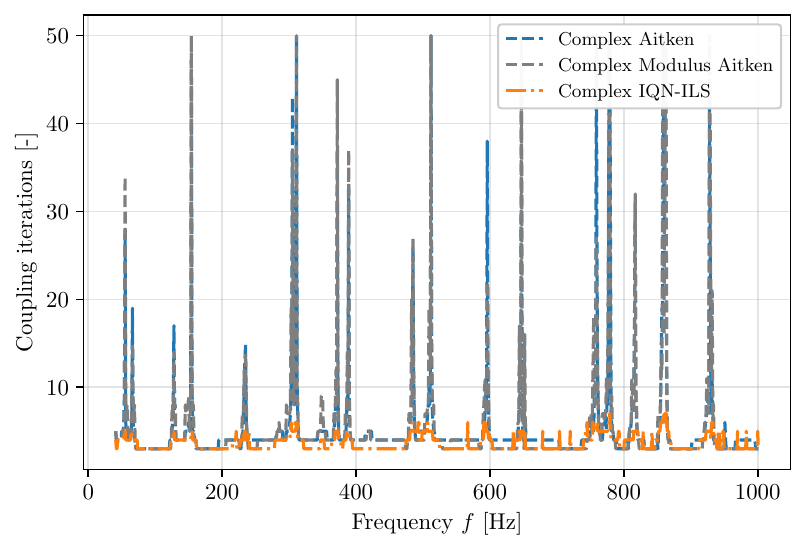}
        \caption{Corresponding number of coupling iterations for the complex-valued convergence accelerators.}
        \label{fig:example_3_complex_iterations}
    \end{subfigure}

    \vspace{0.4cm}

    \begin{subfigure}[t]{0.49\textwidth}
        \centering
        \includegraphics[width=\textwidth]{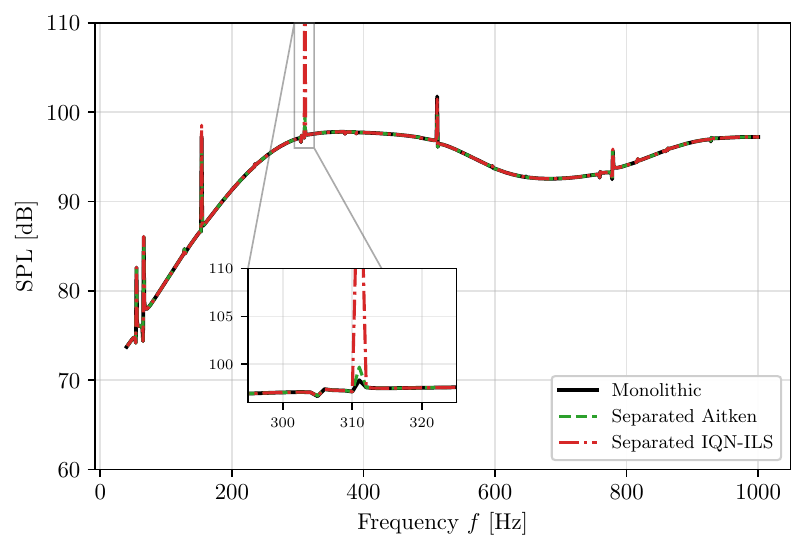}
        \caption{SPL response obtained when accelerating the real and imaginary interface quantities independently.}
        \label{fig:example_3_separated_spl}
    \end{subfigure}
    \hfill
    \begin{subfigure}[t]{0.49\textwidth}
        \centering
        \includegraphics[width=\textwidth]{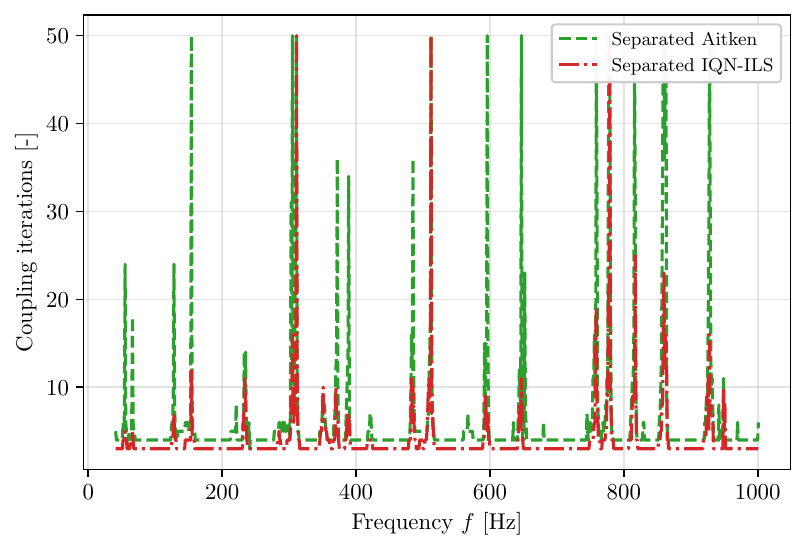}
        \caption{Corresponding number of coupling iterations for the separated-field acceleration strategies.}
        \label{fig:example_3_separated_iterations}
    \end{subfigure}

    \caption{Comparison of the monolithic and strongly coupled partitioned vibroacoustic solutions for the undamped beam-like shell configuration. The top row presents the results obtained using the complex-valued convergence accelerators, while the bottom row shows the corresponding separated-field formulations in which the real and imaginary interface quantities are accelerated independently. The left column reports the SPL response over the investigated frequency range, whereas the right column shows the number of coupling iterations required for convergence at each excitation frequency. The maximum number of coupling iterations is set to 50.}
    
    \label{fig:example_3_spl_iterations}
\end{figure}

The obtained results clearly show that the complex-valued IQN-ILS
accelerator provides the most robust convergence behavior among all
considered approaches. As observed in
Figure~\ref{fig:example_3_complex_iterations}, the number of coupling
iterations remains nearly uniform over the complete frequency sweep,
without significant peaks near the structural resonances, while
maintaining excellent agreement with the monolithic reference
solution.

In contrast, as shown in Figure~\ref{fig:example_3_separated_iterations}, the separated-field IQN-ILS formulation exhibits
pronounced iteration peaks around several resonance frequencies and
fails to accurately reproduce the monolithic SPL response at one
resonance frequency. This behavior indicates that independently
accelerating the real and imaginary interface quantities reduces the
effectiveness of the quasi-Newton approximation in strongly coupled
resonance regions.

Both Aitken-based approaches exhibit significantly poorer convergence
behavior, frequently reaching the prescribed maximum number of
coupling iterations. In particular, the complex Aitken
formulation was found to be highly sensitive to the admissible
magnitude and phase variation of the scalar relaxation parameter. The
numerical results indicate that large phase rotations of the
relaxation coefficient deteriorate the robustness of the fixed-point
iterations, whereas improved convergence behavior is obtained when the
relaxation parameter remains close to the real axis.

Figure~\ref{fig:complex_aitken_sensitivity} presents a sensitivity study of the proposed complex-valued Aitken relaxation strategy with respect to the admissible magnitude and phase variation of the complex relaxation coefficient for a fixed excitation frequency $f = 311,\mathrm{Hz}$. Here, the relaxation parameter is treated as a fully complex quantity, allowing both its magnitude and phase to evolve during the coupling iterations. This should be distinguished from the modulus-based Aitken variant already introduced, which employs the component-wise norm of the complex residual  to compute a real relaxation coefficient. The selected frequency corresponds to a resonance-dominated region of the coupled vibroacoustic response and therefore represents a particularly demanding case for the strongly coupled partitioned iterations. It should be emphasized that the observed convergence behavior is frequency-dependent and may vary for different excitation frequencies. The left column investigates the influence of the maximum admissible relaxation magnitude $|\alpha|_{\max}$ while keeping the phase variation fixed to $\Delta\phi_{\max}=0^\circ$, whereas the right column analyzes the influence of the maximum admissible phase variation $\Delta\phi_{\max}$ while fixing $|\alpha|_{\max}=2$. For all considered configurations, the initial relaxation parameter is set to $\alpha_0 = 0.01 + 0i$.

\begin{figure}[h!]
    \centering

    \begin{subfigure}[t]{0.49\textwidth}
        \centering
        \includegraphics[width=\textwidth]{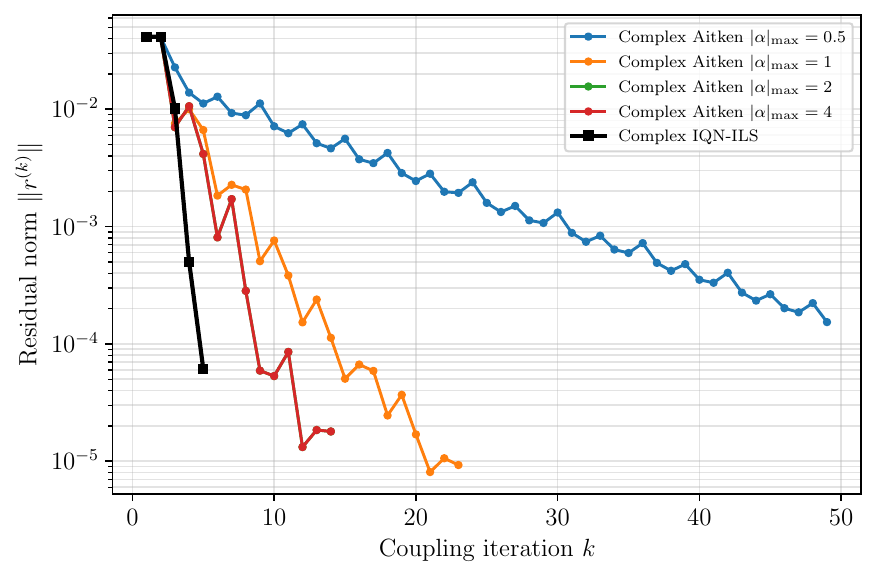}
        \caption{Influence of the maximum admissible relaxation magnitude $|\alpha|_{\max}$ on the residual convergence history.}
    \end{subfigure}
    \hfill
    \begin{subfigure}[t]{0.49\textwidth}
        \centering
        \includegraphics[width=\textwidth]{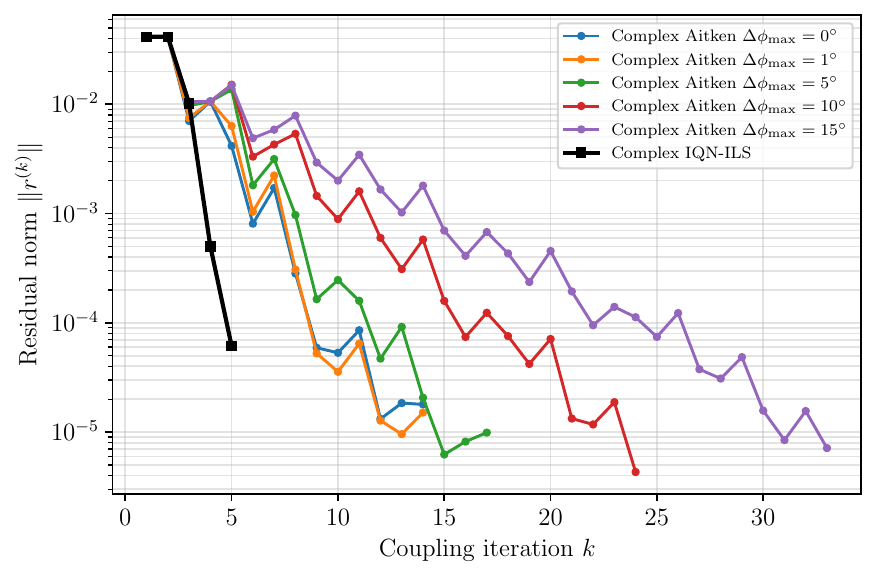}
        \caption{Influence of the maximum admissible phase variation $\Delta\phi_{\max}$ on the residual convergence history, with respect to the real axis}
    \end{subfigure}

    \vspace{0.4cm}

    \begin{subfigure}[t]{0.49\textwidth}
        \centering
        \includegraphics[width=\textwidth]{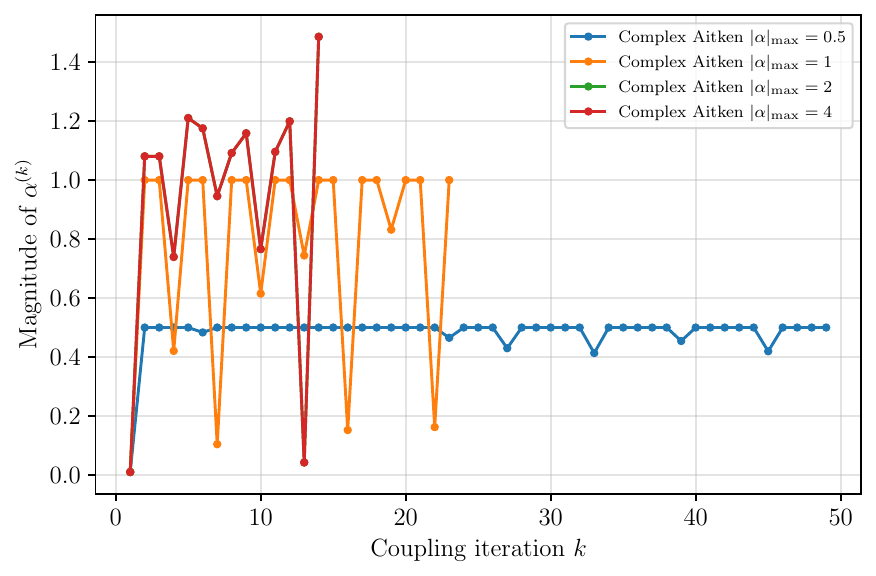}
        \caption{Evolution of the relaxation coefficient magnitude $|\alpha^{(k)}|$ during the coupling iterations.}
    \end{subfigure}
    \hfill
    \begin{subfigure}[t]{0.49\textwidth}
        \centering
        \includegraphics[width=\textwidth]{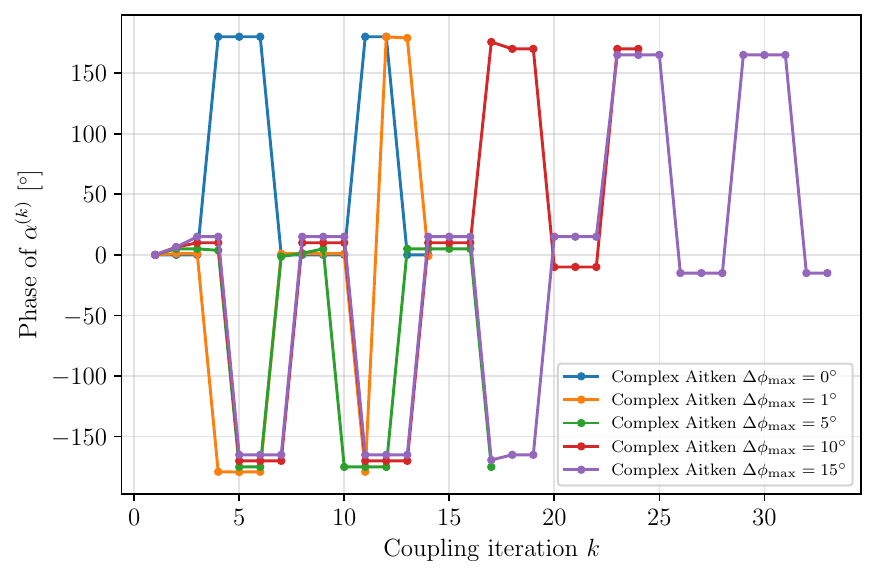}
        \caption{Evolution of the relaxation coefficient phase $\arg(\alpha^{(k)})$ during the coupling iterations.}
    \end{subfigure}

    \caption{Sensitivity analysis of the proposed complex-valued Aitken convergence accelerator for the beam-like vibroacoustic benchmark at the excitation frequency $f = 311\,\mathrm{Hz}$. The left column investigates the influence of the maximum admissible relaxation magnitude $|\alpha|_{\max}$ while fixing the phase variation limit to $\Delta\phi_{\max}=1^\circ$. The right column analyzes the influence of the maximum admissible phase variation $\Delta\phi_{\max}$ while fixing $|\alpha|_{\max}=2$. For all configurations, the initial relaxation parameter is set to $\alpha_0 = 0.01 + 0i$. The residual convergence histories are compared with the complex-valued IQN-ILS accelerator.}
    \label{fig:complex_aitken_sensitivity}

\end{figure}

For the magnitude-sensitivity study, the results show that the admissible bound imposed on the relaxation magnitude directly affects the convergence rate of the partitioned iterations. A very restrictive limit such as $|\alpha|_{\max}=0.5$ leads to a relatively smooth but slow residual reduction, whereas larger admissible values allow stronger corrections and therefore faster residual decay for the considered excitation frequency. The corresponding evolution of $|\alpha^{(k)}|$ further shows that the relaxation coefficient rapidly reaches the prescribed upper bound during the first coupling iterations, indicating that the limiter actively controls the effective relaxation update throughout the iterative process.

The phase-sensitivity study indicates that the convergence behavior of the proposed complex-valued Aitken formulation is strongly influenced by the admissible phase variation of the relaxation parameter. For small phase limits, the relaxation coefficient remains close to the real axis and the residual evolution exhibits a comparatively smooth and stable behavior. Increasing the admissible phase variation allows larger rotations of the complex relaxation factor in the complex plane, thereby modifying the direction of the interface correction applied during the fixed-point iterations.

The observed convergence behavior suggests that, although the interface residuals are inherently complex-valued, improved robustness is obtained when the scalar relaxation parameter remains predominantly aligned with the real axis.

Figure~\ref{fig:complex_accelerator_comparison} further compares the best-performing phase-regularized complex Aitken configuration with the modulus-based Aitken variant and the complex-valued IQN-ILS accelerator. The modulus-based Aitken method computes a real relaxation factor from the component-wise magnitude of the complex residual and applies this single scalar factor to the full complex residual. Although this preserves a unified treatment of the real and imaginary interface quantities, the convergence remains slower than for the phase-regularized complex Aitken method and substantially slower than for IQN-ILS.

\begin{figure}[h!]
    \centering
    \includegraphics[width=0.60\textwidth]
    {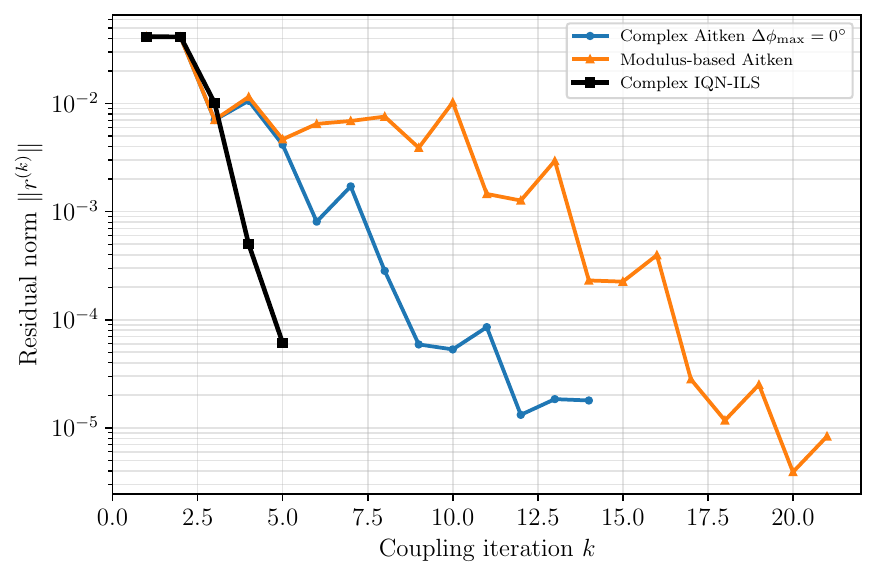}
    \caption{Comparison of the residual convergence histories obtained with the complex-valued IQN-ILS accelerator, the phase-regularized complex Aitken method with $\Delta\phi_{\max}=0^\circ$, and the modulus-based Aitken variant at the excitation frequency $f=311\,\mathrm{Hz}$.}
    \label{fig:complex_accelerator_comparison}
\end{figure}

The complex-valued IQN-ILS result confirms the superior robustness of the quasi-Newton strategy for the considered resonance-dominated case. It reduces the interface residual significantly faster than all tested Aitken configurations and converges within only a few coupling iterations. The results therefore indicate that suitable regularization can improve the robustness of complex Aitken relaxation, but the method remains sensitive to the selected constraints and to the excitation frequency.
}

\subsection{Plane-wave excitation of a spherical shell coupled to interior and exterior acoustic domains}
\label{subsec:example_4}

The final example considers a thin spherical shell separating an enclosed
acoustic fluid from an unbounded exterior acoustic domain. The shell is
excited by a harmonic plane wave propagating through the exterior fluid
and interacts dynamically with the acoustic fields on both sides of its
midsurface. The incident wave excites the shell, whose vibration
generates an outgoing scattered field in the exterior domain and an
acoustic field within the enclosed cavity.

This configuration is used to assess the accuracy and robustness of the
proposed partitioned framework in the presence of strong bidirectional
fluid--structure interaction and independently discretized
fluid--structure interfaces. An analytical solution based on the
spherical-harmonic formulation of Junger and
Feit~\cite{JungerFeit1986} is available for this problem and provides
the exterior acoustic pressure used as the reference solution for the
undamped configuration.

\begin{figure}[!htb]
    \centering
    \includegraphics[width=0.70\textwidth]{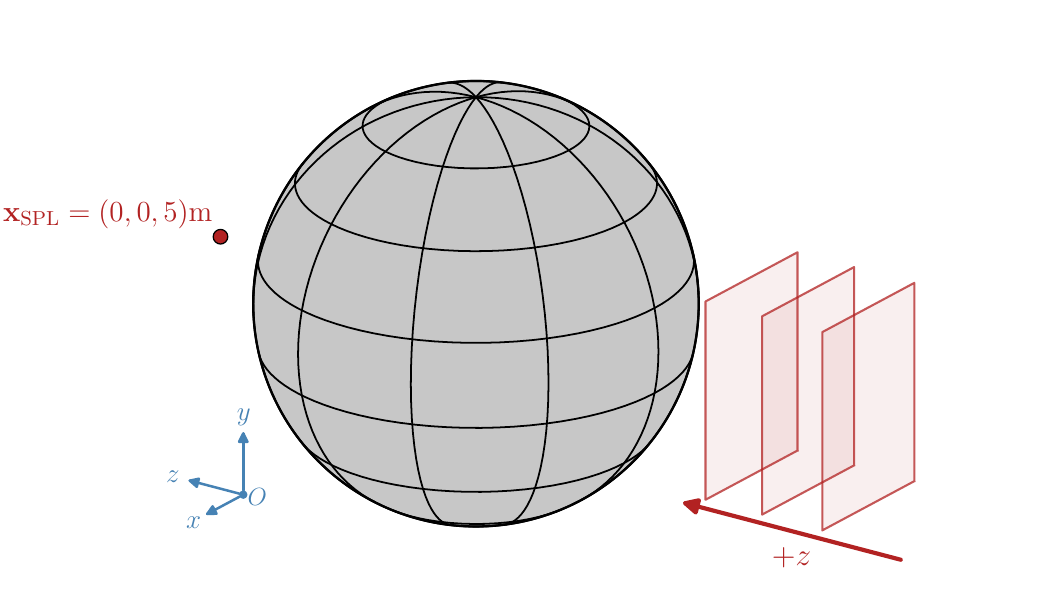}
    \caption{Problem configuration of the spherical shell separating an
    enclosed acoustic fluid from an unbounded exterior fluid. The shell
    is excited by a harmonic plane wave propagating in the positive
    \(z\)-direction, and the exterior acoustic pressure is evaluated at
    \(\mathbf{x}_{\mathrm{obs}}=(0,0,5)\,\mathrm{m}\).}
    \label{fig:example_4}
\end{figure}

The problem configuration is illustrated in
Figure~\ref{fig:example_4}. The shell has midsurface radius \(a\) and
thickness \(t\), and is free in space. A harmonic plane wave of pressure
amplitude \(A\) propagates in the positive \(z\)-direction. The exterior
acoustic pressure is evaluated at the observation point
\(\mathbf{x}_{\mathrm{obs}}=(0,0,5)\,\mathrm{m}\), located on the
positive \(z\)-axis.

The material, geometrical, and acoustic properties employed in this
example are summarized in Table~\ref{tab:sphere_properties}.

\begin{table}[!htb]
\centering
\caption{Material, geometrical, and acoustic properties of the spherical
shell benchmark.}
\label{tab:sphere_properties}
\begin{tabular}{lccr}
\hline
Property & Symbol & Value & Unit \\
\hline
Exterior fluid density
    & \(\rho_{\mathrm{e}}\) & 1000 & kg/m\(^3\) \\
Exterior speed of sound
    & \(c_{\mathrm{e}}\) & 1500 & m/s \\
Interior fluid density
    & \(\rho_{\mathrm{i}}\) & 1000 & kg/m\(^3\) \\
Interior speed of sound
    & \(c_{\mathrm{i}}\) & 1500 & m/s \\
Structural density
    & \(\rho_{\mathrm{s}}\) & 7850 & kg/m\(^3\) \\
Young's modulus
    & \(E\) & 210 & GPa \\
Poisson's ratio
    & \(\nu\) & 0.30 & -- \\
Shell midsurface radius
    & \(a\) & 3.5 & m \\
Shell thickness
    & \(t\) & 0.05 & m \\
Incident pressure amplitude
    & \(A\) & 1 & Pa \\
\hline
\end{tabular}
\end{table}

\begin{figure}[!htb]
    \centering
    \includegraphics[width=0.47\textwidth]{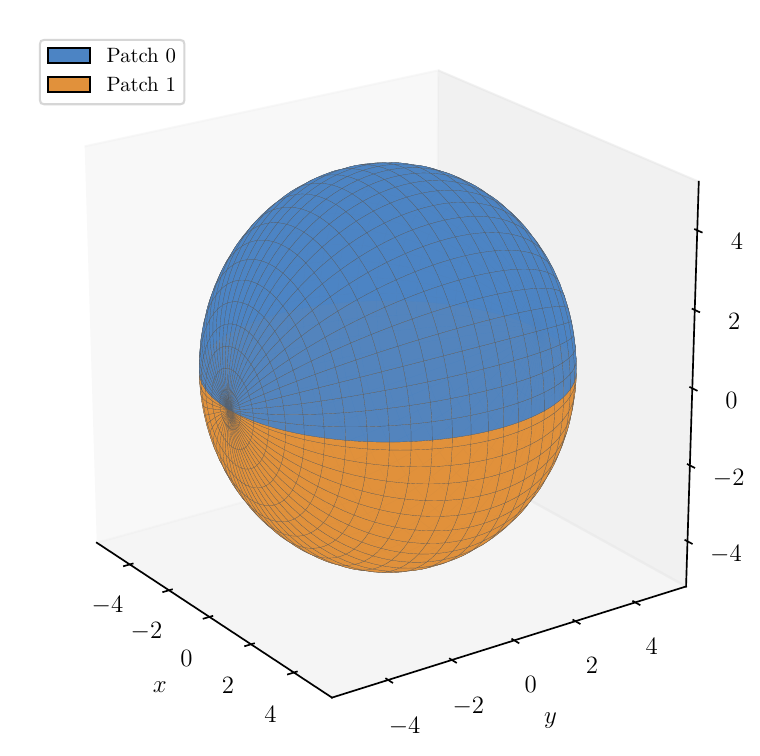}
    \hfill
    \includegraphics[width=0.47\textwidth]{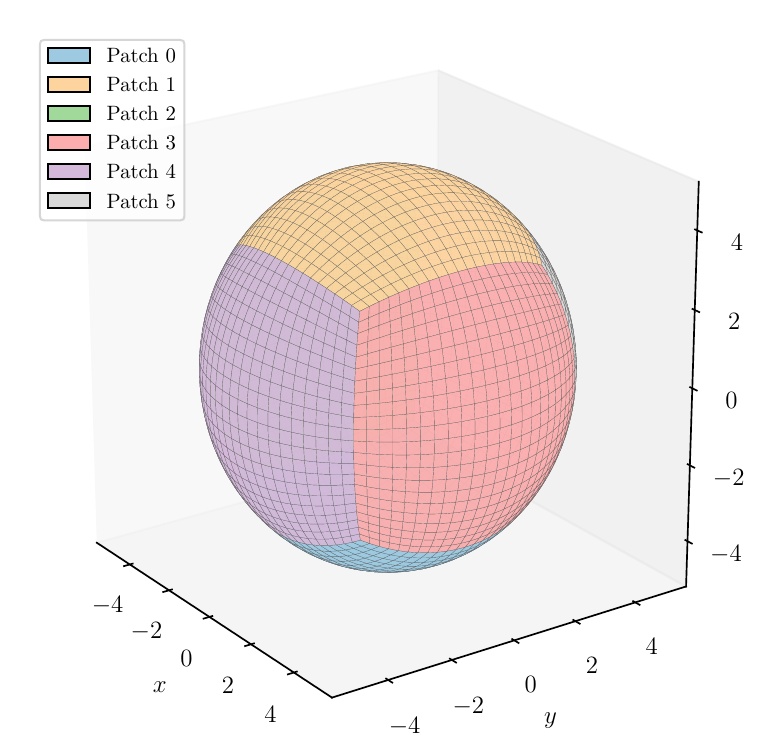}
    \caption{Multipatch discretizations employed for the spherical shell
    benchmark. Left: structural shell represented by two trimmed NURBS
    patches. Right: acoustic boundary represented by six NURBS patches.
    The different patch layouts form a geometrically coincident but
    independently parameterized non-conforming fluid--structure
    interface.}
    \label{fig:sphere_patches}
\end{figure}

For the numerical simulations, independent multipatch CAD
representations are employed for the structural and acoustic domains, as
shown in Figure~\ref{fig:sphere_patches}.

The structural shell is represented by two trimmed quadratic NURBS
patches. Due to repeated knots in the underlying spline
parameterization, the shell midsurface is only \(C^0\)-continuous along
the corresponding knot lines and patch interface. Consequently, the
isogeometric Kirchhoff--Love shell formulation cannot be employed, as it
requires a globally \(C^1\)-continuous displacement field over the shell
midsurface. The structural domain is therefore discretized using the
isogeometric Reissner--Mindlin shell formulation of Benson
et al.~\cite{BENSON2010276}, in which the translational and rotational
degrees of freedom are treated as independent unknowns.

Since both the displacement and rotational fields require
\(C^0\)-continuity, their continuity across the patch interface is
weakly enforced through a penalty formulation. The corresponding
penalty energy is defined as

\begin{equation}
\Pi_{\mathrm{pen}}
=
\frac{1}{2}
\int_{\Gamma_{\mathrm{int}}}
\left(
\alpha_u
\left\|
\llbracket\mathbf{u}\rrbracket
\right\|^2
+
\alpha_\theta
\left\|
\llbracket\boldsymbol{\theta}\rrbracket
\right\|^2
\right)
\,\mathrm{d}\Gamma,
\end{equation}
where
\(\llbracket\mathbf{u}\rrbracket,
\llbracket\boldsymbol{\theta}\rrbracket\in\mathbb{R}^{3}\)
denote the displacement and rotational jumps across the patch interface,
respectively, while \(\alpha_u\) and \(\alpha_\theta\) are the
corresponding displacement and rotational penalty parameters. In the
present simulations, both penalty parameters are set to
\(\alpha_u=\alpha_\theta=10^{14}\).
The resulting contribution is assembled into the structural stiffness
matrix to weakly enforce continuity between adjacent patches. Further
details on the formulation and implementation of the penalty coupling
are given in~\cite{Teschemacher2018}.

The acoustic boundary is represented by six quartic NURBS patches.
Since the acoustic formulation only requires \(C^0\)-continuity of the
pressure field, continuity across adjacent patches is enforced through
the strong patch coupling formulation introduced in
Section~\ref{sec:acoustic_model}. Together with the two-patch structural
representation, this results in a geometrically coincident but
independently parameterized non-conforming fluid--structure interface.

The proposed partitioned framework is evaluated for both undamped and
Rayleigh-damped configurations using the same damping coefficients and
convergence criteria introduced in
Section~\ref{subsec:example_2}. All partitioned simulations employ the
complex IQN--ILS convergence accelerator together with the
nearest-element mapping technique for the transfer of interface
quantities between the structural and acoustic discretizations. The
frequency response is computed over the range
\(20\leq f\leq100~\mathrm{Hz}\) using a frequency increment of
\(\Delta f=0.2~\mathrm{Hz}\).

The structural discretization is obtained through \(k\)-refinement of
the initial quadratic NURBS representation. The basis is first
degree-elevated to cubic order and subsequently refined by knot
insertion to obtain 60 knot spans in each parametric direction of every
patch. The resulting structural model comprises 8978 control points,
each possessing six degrees of freedom corresponding to three
translational and three rotational components.

The acoustic boundary is uniformly refined by knot insertion to obtain
eight knot spans in each parametric direction of every patch. The final
acoustic discretization comprises 728 control points associated with
quartic NURBS basis functions, each carrying a single acoustic degree of
freedom.

\subsubsection{Results and discussion}

\begin{figure}[!htb]
    \centering
    \begin{subfigure}[b]{0.49\textwidth}
        \centering
        \includegraphics[width=\textwidth]
        {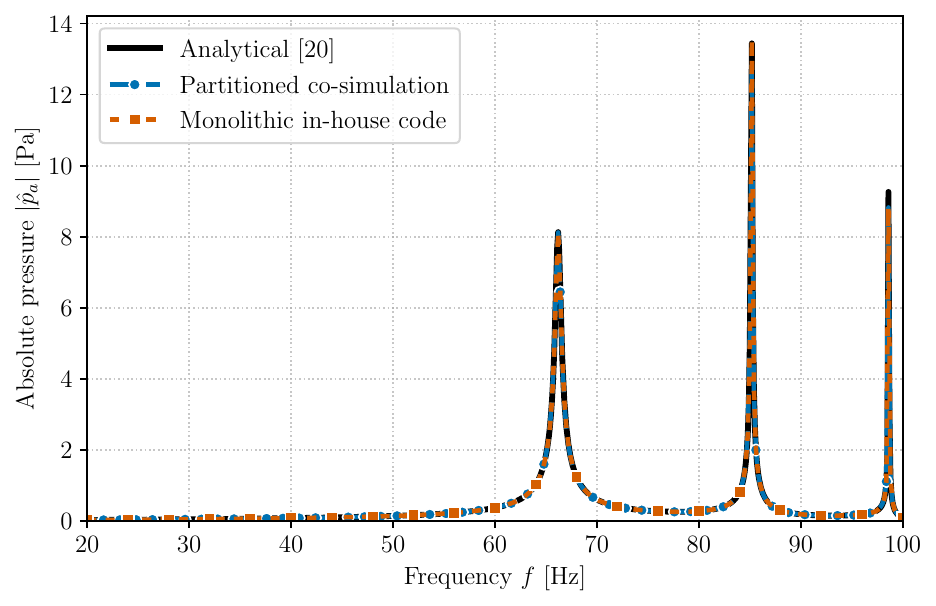}
        \caption{Undamped configuration.}
        \label{fig:sphere_undamped}
    \end{subfigure}
    \hfill
    \begin{subfigure}[b]{0.49\textwidth}
        \centering
        \includegraphics[width=\textwidth]
        {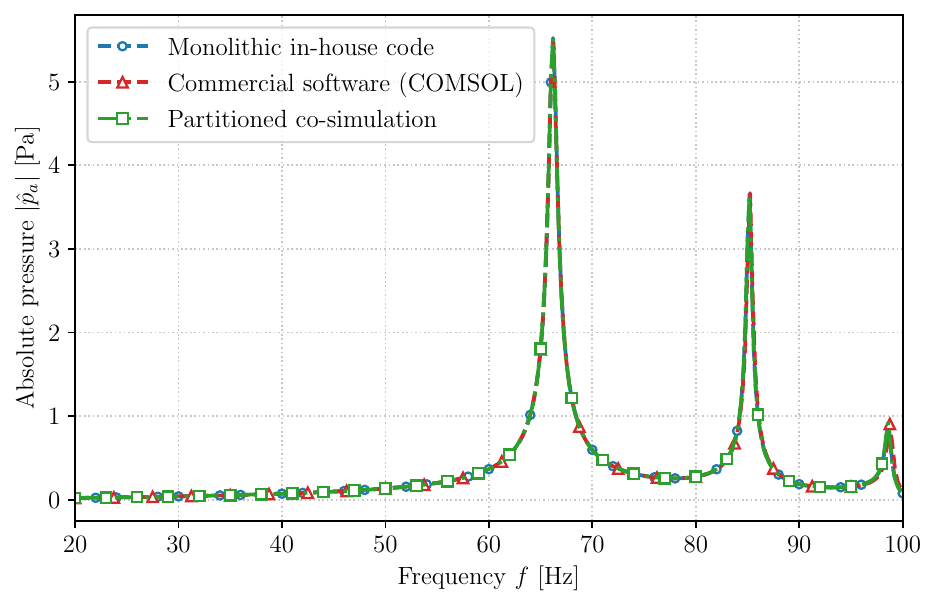}
        \caption{Rayleigh-damped configuration.}
        \label{fig:sphere_damped}
    \end{subfigure}
    \caption{
    Acoustic pressure magnitude at the observation point
    $\mathbf{x}_{\mathrm{obs}}=(0,0,5)\,\mathrm{m}$:
    (a) undamped configuration compared with the analytical solution of
    Junger and Feit~\cite{JungerFeit1986} and the in-house monolithic
    solution, and (b) Rayleigh-damped configuration compared with the
    in-house monolithic and COMSOL Multiphysics solutions.
    }
    \label{fig:sphere_pressure_comparison}
\end{figure}

The resulting acoustic pressure magnitude, $|p|$, at the observation point
\(\mathbf{x}_{\mathrm{obs}}=(0,0,5)\,\mathrm{m}\) is shown in
Figure~\ref{fig:sphere_pressure_comparison} for the undamped and
Rayleigh-damped configurations.

For the undamped configuration,
Figure~\ref{fig:sphere_pressure_comparison}(a) compares the partitioned
solution with both the analytical reference solution of Junger and
Feit~\cite{JungerFeit1986} and the monolithic solution obtained with the
in-house research code. The monolithic model employs a six-patch
representation of the spherical geometry for both the structural and
acoustic domains, with each patch described by a NURBS surface with
$p=q=4$, where $p$ and $q$ denote the polynomial degrees in the two
parametric directions. The structural and acoustic discretizations
comprise 26 and 8 knot spans per patch, respectively. An isogeometric
Kirchhoff--Love shell formulation is used for the structural domain, with
continuity between the structural patches enforced using the multipatch
coupling formulation proposed by Coox et al.~\cite{COOX2017505}. In
contrast, the partitioned model employs a two-patch structural
representation together with an isogeometric Reissner--Mindlin shell
formulation, for which continuity between the structural patches is
weakly enforced using the penalty formulation described above, while
retaining the six-patch representation for the acoustic domain.

Very good agreement is obtained among the analytical, monolithic, and
partitioned solutions over most of the investigated frequency range. The
partitioned model accurately reproduces the overall acoustic pressure
response and the resonance locations, while larger deviations occur
primarily in the vicinity of the resonances. In the undamped case, the
response is particularly sensitive to small perturbations in these
frequency regions. Consequently, differences associated with the
structural formulation and spatial discretization, as well as errors
introduced by the interface mapping and iterative strong coupling
procedure, can be significantly amplified near resonance.

For the Rayleigh-damped configuration,
Figure~\ref{fig:sphere_pressure_comparison}(b) compares the partitioned
solution with the in-house monolithic solution and an independent
reference obtained using COMSOL Multiphysics. In contrast to the proposed
IBRA--IGA-BEM framework, the COMSOL model employs a monolithic FEM--FEM
formulation with quadratic finite elements for both the structural and
acoustic domains, resulting in a total of 109,531 degrees of freedom.
Very good agreement is again obtained over the investigated frequency
range. The partitioned solution closely follows the in-house monolithic
reference, while somewhat larger differences with respect to the COMSOL
solution occur at higher frequencies and in the vicinity of resonance
peaks. These differences should be interpreted in view of the different
numerical formulations and spatial discretizations employed by the
considered models.



\begin{figure}[!htb]
    \centering
    \includegraphics[width=0.60\textwidth]
    {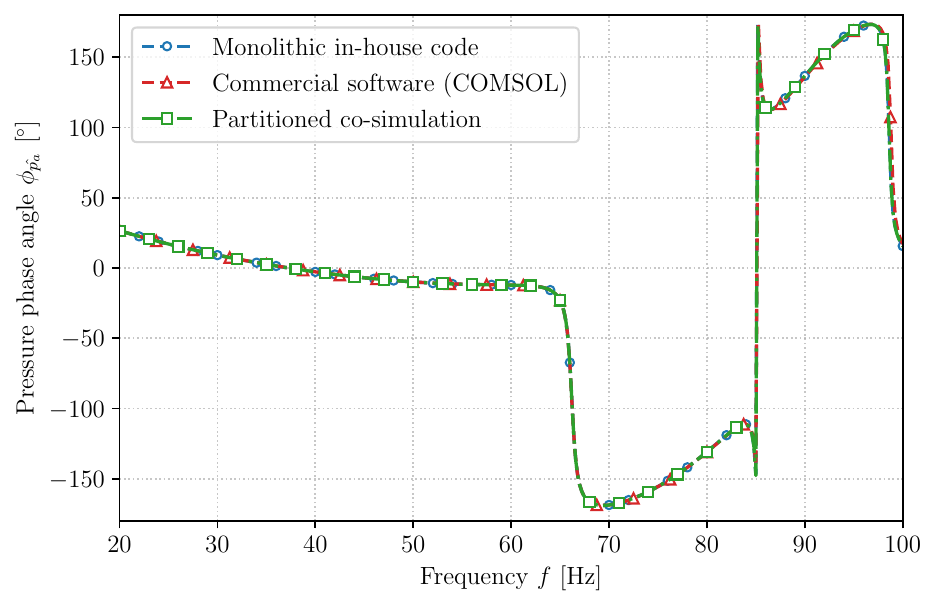}
    \caption{
    Phase angle of the complex acoustic pressure $\hat{p}_a$ at the
    observation point $\mathbf{x}_{\mathrm{obs}}=(0,0,5)\,\mathrm{m}$ for
    the Rayleigh-damped configuration. The partitioned solution is compared
    with the in-house monolithic and COMSOL Multiphysics reference solutions.
    }
    \label{fig:sphere_pressure_phase}
\end{figure}

In addition to the pressure magnitude, the phase of the complex acoustic pressure provides a complementary measure of the agreement between the different solution strategies. Figure~\ref{fig:sphere_pressure_phase} compares the resulting phase angle
obtained with the partitioned co-simulation formulation, the in-house monolithic code,
and COMSOL Multiphysics.

The phase comparison complements the pressure-magnitude results in
Figure~\ref{fig:sphere_pressure_comparison}(b). The partitioned formulation
combined with the proposed complex IQN-ILS accelerator closely reproduces
the phase evolution predicted by the reference solutions, including the
characteristic phase variations associated with the resonant response.
This demonstrates that the proposed complex-valued coupling strategy
reproduces not only the magnitude of the acoustic pressure but also its
phase behavior, thereby accurately capturing the complex-valued acoustic
response.

\begin{figure}[!htb]
    \centering
    \begin{subfigure}[t]{0.48\linewidth}
        \centering
        \includegraphics[width=\linewidth]{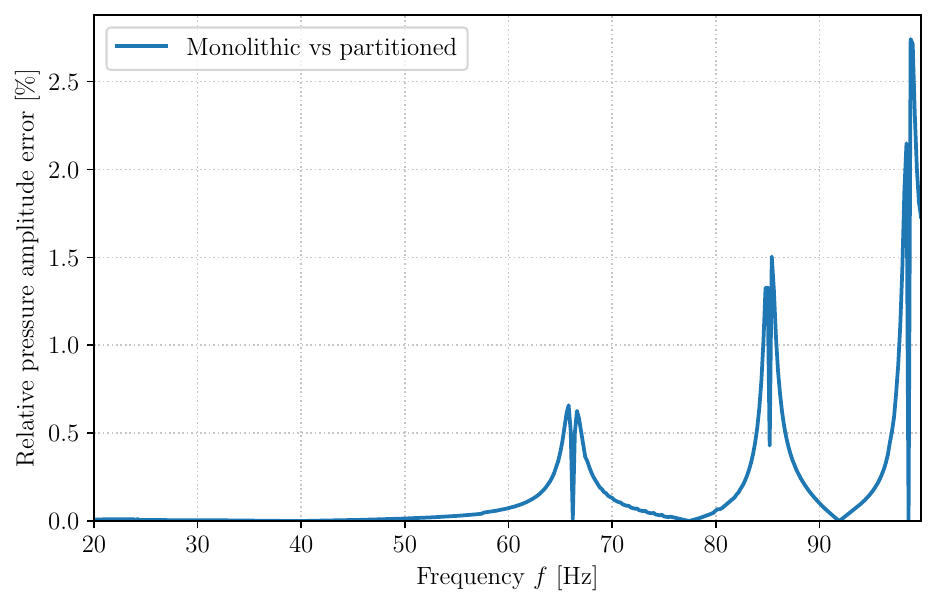}
        \caption{Relative pressure amplitude error for the Rayleigh-damped case.}
    \end{subfigure}
    \hfill
    \begin{subfigure}[t]{0.48\linewidth}
        \centering
        \includegraphics[width=\linewidth]{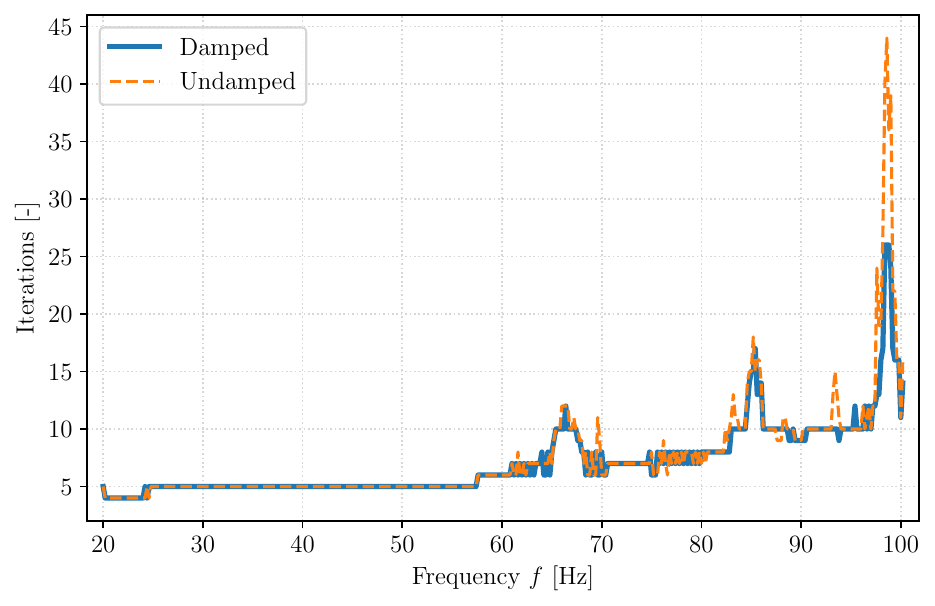}
        \caption{Number of coupling iterations.}
    \end{subfigure}

    \caption{
    Accuracy and convergence behavior of the partitioned solution using
    complex IQN-ILS for the spherical shell subjected to plane-wave excitation:
    (a) relative pressure amplitude error; and (b) number of coupling iterations
    for the undamped and Rayleigh-damped configurations.
    }
    \label{fig:sphere_error_iterations}
\end{figure}

The accuracy of the damped partitioned solution and the convergence
behavior of the strongly coupled procedure are examined in more detail in
Figure~\ref{fig:sphere_error_iterations}. Panel~(a) shows the relative
pressure amplitude error with respect to the in-house monolithic solution. Panel~(b) reports
the number of coupling iterations required for convergence for both the
undamped and Rayleigh-damped configurations.


As shown in Figure~\ref{fig:sphere_error_iterations}(a), the partitioned
and monolithic solutions remain in close agreement in pressure
amplitude over most of the investigated frequency range. The three
peaks in the relative pressure amplitude error correspond to the
resonances, with the largest error occurring at the highest-frequency
resonance. Since the monolithic and partitioned models differ in their
structural representation and shell formulation, and the partitioned
model additionally involves interface mapping and iterative strong
coupling, the observed differences cannot be attributed exclusively
to the coupling procedure. The concentration of error at the
resonances is consistent with the increased sensitivity of the
vibroacoustic response in these regions.

The convergence histories in
Figure~\ref{fig:sphere_error_iterations}(b) show a general increase in the
number of coupling iterations with excitation frequency, together with
pronounced peaks in the vicinity of the structural resonances. For the
undamped structural problem, the dynamic stiffness matrix,
\(\mathbf{K}-\omega^2\mathbf{M}\), becomes singular at the exact
eigenfrequencies and increasingly ill-conditioned in their vicinity.
Small perturbations in the transferred interface quantities can therefore
produce comparatively large changes in the structural response, making
the strongly coupled partitioned iterations more difficult to converge.
The simultaneous increase in the pressure differences and coupling
iterations toward the upper end of the frequency range further reflects
the increased numerical sensitivity of the coupled problem in this
region.

Rayleigh damping regularizes the structural response in the vicinity of
resonance and consequently reduces the pronounced peaks in the number of
coupling iterations. Away from these frequency regions, the undamped and
damped configurations exhibit comparable convergence behavior. Overall,
the results demonstrate that the proposed partitioned framework remains
robust over the investigated frequency range, while highlighting the
increased numerical sensitivity of the strongly coupled problem near
structural resonances and the beneficial effect of structural damping on
the convergence behavior.
\ignore{
\subsection{Example 4: Hemispherical shell subjected to an incident plane wave}
\label{subsec:example_4}

Finally, a simply supported hemispherical shell subjected to the combined action of an incident acoustic plane wave and a surface load is considered. In contrast to the previous plate configurations, this example introduces a curved structural interface and a distributed acoustic excitation in a fully three-dimensional vibroacoustic setting. The hemispherical geometry leads to increased complexity in both the acoustic wave propagation and the interface coupling procedure.

The surrounding acoustic medium is modeled as air with density $\rho_a = 1.225~\mathrm{kg/m^3}$ and speed of sound $c = 340~\mathrm{m/s}$. The shell is excited by an incident acoustic plane wave propagating in the positive $x$-direction. In addition, a harmonic structural pressure load acting in the positive $z$-direction is applied to the shell surface, as illustrated in Figure~\ref{fig:example_4_hemisphere}. The resulting problem therefore involves a fully bidirectional interaction between the structural vibrations and the scattered acoustic field.

\begin{figure}[h!]
\centering
\tdplotsetmaincoords{65}{120}
\begin{tikzpicture}[tdplot_main_coords, scale=3.5, line cap=round, line join=round]

\definecolor{HemiBlue}{RGB}{0,92,170}
\definecolor{SupportGray}{RGB}{90,90,90}
\definecolor{LoadRed}{RGB}{180,60,60}
\definecolor{LoadYellow}{RGB}{255,165,0}

\tikzset{
  axisArrow/.style={
    draw=LoadRed,
    -{Latex[length=2.2mm,width=1.6mm]},
    line width=0.9pt
  }
}

\pgfmathsetmacro{\R}{1.0}
\pgfmathsetmacro{\Nlat}{28}
\pgfmathsetmacro{\Nmer}{7}

\draw[axisArrow] (0,0,0) -- (1.5,0,0) node[below=1mm, text=LoadRed] {$x$};
\draw[axisArrow] (0,0,0) -- (0,1.5,0) node[anchor=south, text=LoadRed] {$y$};
\draw[axisArrow] (0,0,0) -- (0,0,1.5) node[anchor=south, text=LoadRed] {$z$};

\draw[SupportGray, dashed, line width=0.9pt]
  plot[domain=120:310, samples=120, variable=\t]
  ({\R*cos(\t)}, {\R*sin(\t)}, {0});

\draw[SupportGray, line width=1.25pt]
  plot[domain=-50:120, samples=120, variable=\t]
  ({\R*cos(\t)}, {\R*sin(\t)}, {0});

\foreach \k in {1,...,\Nlat}{
  \pgfmathsetmacro{\phiA}{(\k-1)*90/\Nlat}
  \pgfmathsetmacro{\phiB}{\k*90/\Nlat}

  \pgfmathsetmacro{\zA}{\R*sin(\phiA)}
  \pgfmathsetmacro{\rA}{\R*cos(\phiA)}
  \pgfmathsetmacro{\zB}{\R*sin(\phiB)}
  \pgfmathsetmacro{\rB}{\R*cos(\phiB)}

  \pgfmathsetmacro{\op}{0.015 + 0.11*(\k/\Nlat)}

  \path[fill=HemiBlue, fill opacity=\op, draw=none]
    plot[domain=0:360, samples=120, variable=\t]
      ({\rB*cos(\t)}, {\rB*sin(\t)}, {\zB})
    --
    plot[domain=360:0, samples=120, variable=\t]
      ({\rA*cos(\t)}, {\rA*sin(\t)}, {\zA})
    -- cycle;
}

\foreach \m in {0,...,\Nmer}{
  \pgfmathsetmacro{\az}{\m*180/\Nmer}
  \draw[HemiBlue!85, line width=0.75pt]
    plot[domain=0:90, samples=120, variable=\v]
      ({\R*cos(\v)*cos(\az)}, {\R*cos(\v)*sin(\az)}, {\R*sin(\v)});
}

\foreach \phi in {15,30,45,60,75}{
  \pgfmathsetmacro{\z}{\R*sin(\phi)}
  \pgfmathsetmacro{\rr}{\R*cos(\phi)}
  \draw[HemiBlue!55, line width=0.55pt]
    plot[domain=0:360, samples=140, variable=\t]
      ({\rr*cos(\t)}, {\rr*sin(\t)}, {\z});
}

\draw[SupportGray, dashed, line width=0.7pt] (0,0,0) -- (\R,0,0);
\node[LoadRed, font=\small, anchor=south] at (0.25,-0.11,0) {$R=0.25\;m$};

\pgfmathsetmacro{\Nsupp}{36}
\pgfmathparse{int(\Nsupp-1)}\edef\NsuppLast{\pgfmathresult}

\foreach \k in {0,...,\NsuppLast}{
  \pgfmathsetmacro{\ang}{360*\k/\Nsupp}
  \coordinate (SS\k) at ({\R*cos(\ang)},{\R*sin(\ang)},0);
}

\begin{scope}[tdplot_screen_coords, draw=SupportGray, scale=0.09, transform shape]
  \foreach \k in {0,...,\NsuppLast}{
    \support{2}{SS\k}[0]
  }
\end{scope}

\definecolor{WaveGreen}{RGB}{60,150,110}

\pgfmathsetmacro{\xLeft}{-2.35}
\pgfmathsetmacro{\Nfront}{3}
\pgfmathsetmacro{\dx}{0.22}
\pgfmathsetmacro{\ySpan}{0.40}
\pgfmathsetmacro{\zSpan}{0.32}
\pgfmathsetmacro{\zOffset}{0.35}

\foreach \i in {0,...,\Nfront}{
  \pgfmathsetmacro{\xx}{\xLeft + \i*\dx}
  \draw[WaveGreen, line width=0.85pt, opacity=0.9]
    (\xx, -\ySpan, \zOffset) --
    (\xx,  \ySpan, \zOffset) --
    (\xx,  \ySpan, {\zOffset+\zSpan}) --
    (\xx, -\ySpan, {\zOffset+\zSpan}) -- cycle;
}

\draw[WaveGreen, -Latex, line width=1.1pt]
  (\xLeft-0.20, 0, {\zOffset+0.16})
  -- (\xLeft+0.35, 0, {\zOffset+0.16});

\node[font=\small, text=WaveGreen, anchor=south]
  at (\xLeft+0.05, 0.0, {\zOffset+\zSpan+0.12})
  {plane wave};

\pgfmathsetmacro{\xp}{-1.5}
\pgfmathsetmacro{\yp}{0.0}
\pgfmathsetmacro{\zp}{0.25}

\fill[LoadRed] (\xp,\yp,\zp) circle[radius=0.8pt];

\node[font=\small, text=LoadRed, anchor=west]
  at (\xp,\yp,\zp)
  {SPL Probe @ $[-0.94\;m,\,0.0\;m,\,0.34\;m]$};


\pgfmathsetmacro{\ArrowLen}{0.22}

\foreach \phi in {18,32,46,60,74}{
  \foreach \theta in {0,45,...,315}{

    \pgfmathsetmacro{\x}{\R*cos(\phi)*cos(\theta)}
    \pgfmathsetmacro{\y}{\R*cos(\phi)*sin(\theta)}
    \pgfmathsetmacro{\z}{\R*sin(\phi)}

    \draw[LoadYellow, -Latex, line width=0.6pt]
      (\x,\y,{\z+\ArrowLen})
      --
      (\x,\y,\z);
  }
}

\draw[LoadYellow, -Latex, line width=1.1pt]
  (0,0,{\R+0.35}) -- (0,0,{\R+0.08});

\node[
  font=\small,
  text=LoadYellow,
  anchor=south
] at (0,-0.35,{\R+0.18}) {$f_z = 50\;Pa$};

\end{tikzpicture}
\caption{Simply supported hemispherical shell subjected to a harmonic surface pressure load and an incident acoustic plane wave excitation.}
\label{fig:example_4_hemisphere}
\end{figure}

The hemispherical shell is modeled as an undamped, homogeneous isotropic elastic structure with thickness $t = 0.005~\mathrm{m}$, Young’s modulus $E = 170.0 \times 10^{9}~\mathrm{Pa}$, Poisson ratio $\nu = 0.33$, and density $\rho_s = 4500~\mathrm{kg/m^3}$. The structural domain is discretized using quadratic isogeometric Kirchhoff--Love shell elements with approximately 20 knot spans in each parametric direction, while the acoustic interface is discretized using quadratic isogeometric boundary elements with only 2 knot spans per parametric direction. The vibroacoustic response is evaluated over the frequency range $0 \leq f \leq 3000\,\mathrm{Hz}$.

Interface quantities are exchanged using the nearest-neighbor mapping strategy described in Section~\ref{sec:nearest_neighbor_mapping}. In this example, the nearest-neighbor mapper is preferred because the hemispherical shell discretization is only $C^0$-continuous along several knot lines, which can deteriorate the robustness of mapping strategies based on geometric projections. In particular, the associated nonlinear Newton-Raphson projection procedure requires derivative information of the underlying geometry, which may become less robust in the vicinity of continuity reductions.

Figure~\ref{fig:air_hemisphere_complex_iqnils} shows the results obtained for the hemispherical shell using the complex IQN-ILS accelerator. The strongly coupled partitioned solution reproduces the monolithic SPL response with very good agreement over the complete investigated frequency range. Moreover, the number of coupling iterations remains bounded by four for all frequencies, demonstrating the robustness and efficiency of the complex IQN-ILS formulation for this fully three-dimensional curved vibroacoustic configuration.

Figure~\ref{fig:air_hemisphere_complex_aitken} shows the corresponding results obtained with two Aitken-based acceleration strategies. The first strategy applies the relaxation using the absolute values of the real and imaginary residual components, while the second one corresponds to a complex Aitken formulation in which the phase of the relaxation parameter with respect to the real axis is limited to $1^\circ$. Both Aitken variants reproduce the monolithic SPL response well for most of the investigated frequency range. However, at two frequencies the coupling iterations fail to converge. The corresponding SPL values are therefore not physically meaningful and appear as spurious deviations in the frequency response.

The comparison shows that, although suitably constrained Aitken-type accelerators can provide accurate results when convergence is achieved, their robustness remains limited for this challenging hemispherical benchmark. In contrast, the complex IQN-ILS accelerator converges reliably over the entire frequency sweep and preserves excellent agreement with the monolithic reference solution while requiring at most five coupling iterations.

\begin{figure}[h!]
    \centering
    
    \begin{subfigure}[t]{0.49\textwidth}
        \centering
        \includegraphics[width=\textwidth]{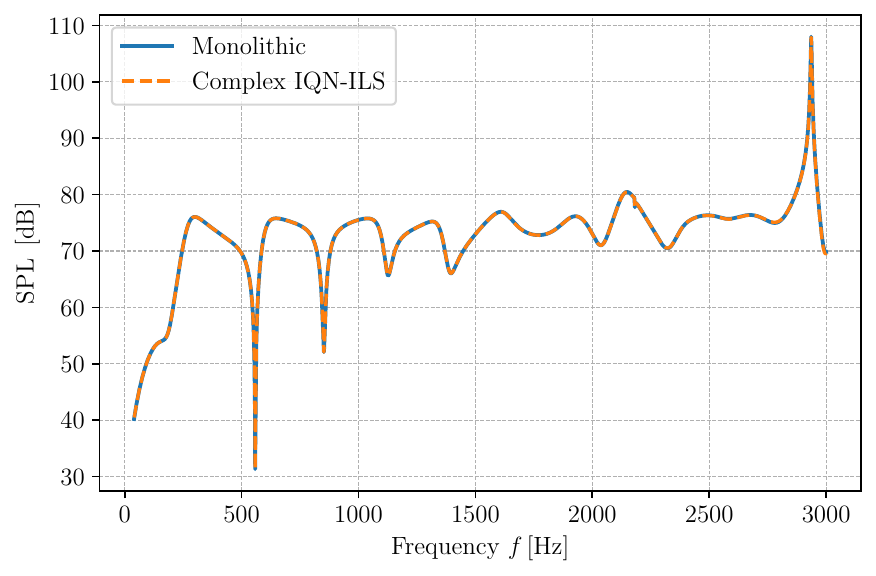}
        \caption{Comparison of the monolithic and strongly coupled partitioned SPL responses obtained using the complex IQN-ILS accelerator.}
        \label{fig:air_hemisphere_complex_iqnils_spl}
    \end{subfigure}
    \hfill
    \begin{subfigure}[t]{0.49\textwidth}
        \centering
        \includegraphics[width=\textwidth]{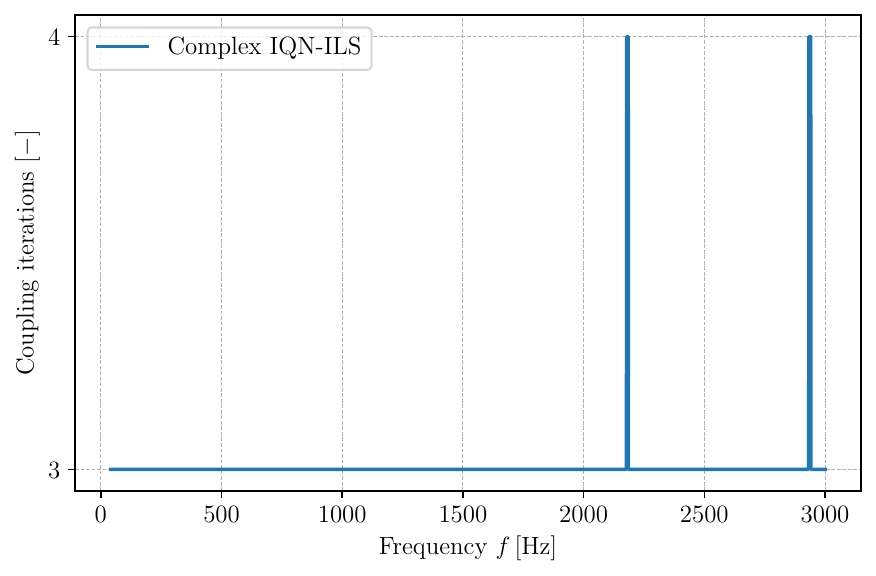}
        \caption{Corresponding number of coupling iterations required for convergence at each excitation frequency.}
        \label{fig:air_hemisphere_complex_iqnils_iterations}
    \end{subfigure}
    
    \caption{Hemispherical shell benchmark using the complex IQN-ILS accelerator.}
    \label{fig:air_hemisphere_complex_iqnils}
\end{figure}

\begin{figure}[h!]
    \centering
    
    \begin{subfigure}[t]{0.49\textwidth}
        \centering
        \includegraphics[width=\textwidth]{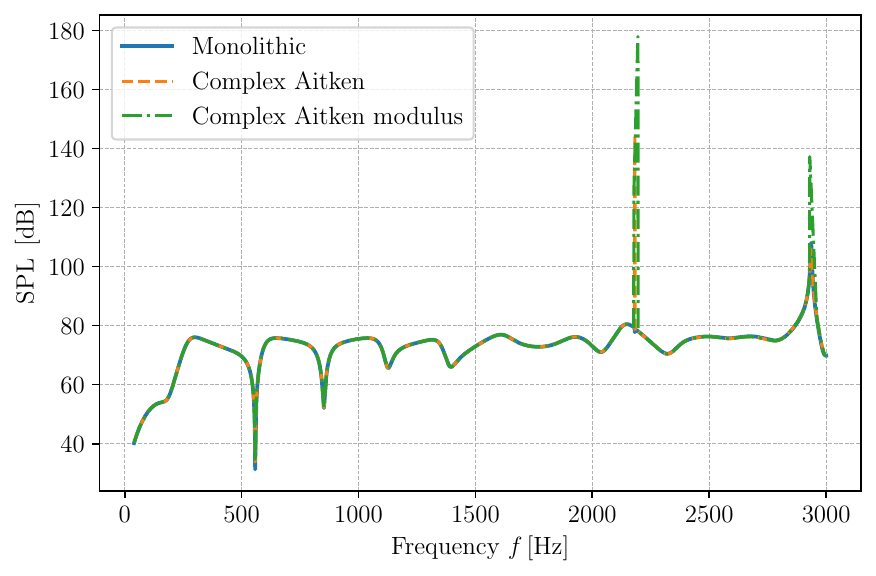}
        \caption{Comparison of the monolithic and strongly coupled partitioned SPL responses obtained using the complex Aitken and complex modulus Aitken accelerators.}
        \label{fig:air_hemisphere_complex_aitken_spl}
    \end{subfigure}
    \hfill
    \begin{subfigure}[t]{0.49\textwidth}
        \centering
        \includegraphics[width=\textwidth]{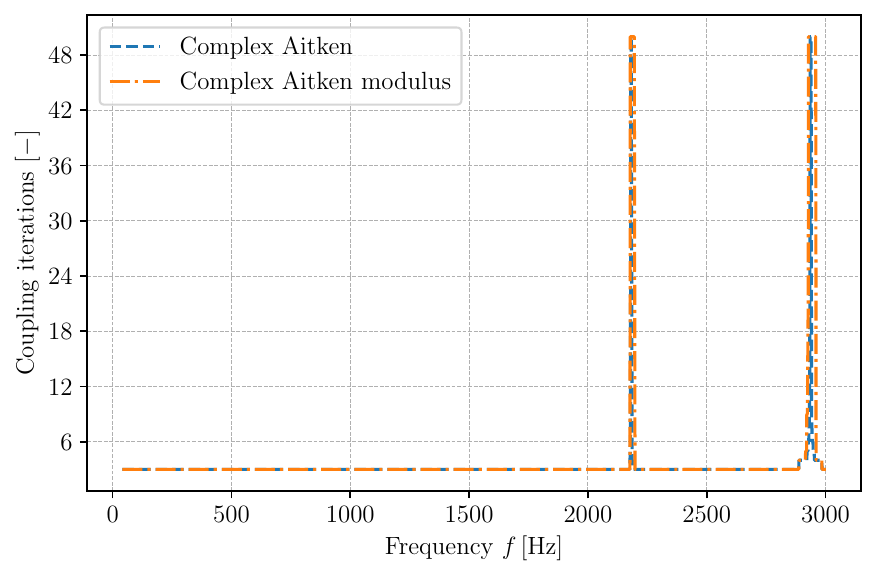}
        \caption{Corresponding number of coupling iterations required for convergence at each excitation frequency.}
        \label{fig:air_hemisphere_complex_aitken_iterations}
    \end{subfigure}
    
    \caption{Hemispherical shell benchmark using the complex Aitken accelerator and the complex modulus variant.}
    \label{fig:air_hemisphere_complex_aitken}
\end{figure}

Finally, the hemispherical shell benchmark is reconsidered using water as the surrounding acoustic medium while keeping the structural configuration unchanged. Compared to the air-loaded case, the significantly higher density and acoustic impedance of water ($c = 1500\;\mathrm{m/s}$, $\rho = 1000\;\mathrm{kg/m^3}$) lead to substantially stronger fluid-structure interaction effects, providing a considerably more challenging benchmark for the proposed partitioned coupling framework.

Figure~\ref{fig:water_hemisphere_spl} compares the SPL responses obtained with the monolithic reference solution and the different strongly coupled partitioned formulations for the water-loaded hemispherical shell. The complex IQN-ILS accelerator remains in very good agreement with the monolithic solution over the complete investigated frequency range, despite the substantially stronger vibroacoustic interaction induced by the surrounding fluid. In contrast, both Aitken-based formulations, namely the complex Aitken and the complex modulus variants, reproduce the monolithic response reasonably well only at lower frequencies. Starting from approximately $f \approx 1200,\mathrm{Hz}$, both approaches fail to converge, leading to non-physical deviations in the corresponding SPL response.

The corresponding number of coupling iterations for all investigated convergence accelerators is shown in Figure~\ref{fig:water_hemisphere_iterations}. Compared to the air-loaded configuration, the number of iterations required by the complex IQN-ILS formulation increases progressively with frequency, reflecting the stronger fluid--structure interaction and the increased sensitivity of the coupled problem. Nevertheless, the complex IQN-ILS accelerator converges robustly for all investigated frequency steps. In contrast, both Aitken-based formulations exhibit rapidly increasing iteration counts and eventually lose convergence beyond approximately $1200,\mathrm{Hz}$, consistent with the non-physical SPL values observed in Figure~\ref{fig:water_hemisphere_spl}. Moreover, for the Aitken-based formulations the simulation eventually crashes around this frequency range, further highlighting the lower robustness of these acceleration strategies for strongly coupled water-loaded vibroacoustic problems.

\begin{figure}[h!]
    \centering
    
    \begin{subfigure}[t]{0.49\textwidth}
        \centering
        \includegraphics[width=\textwidth]{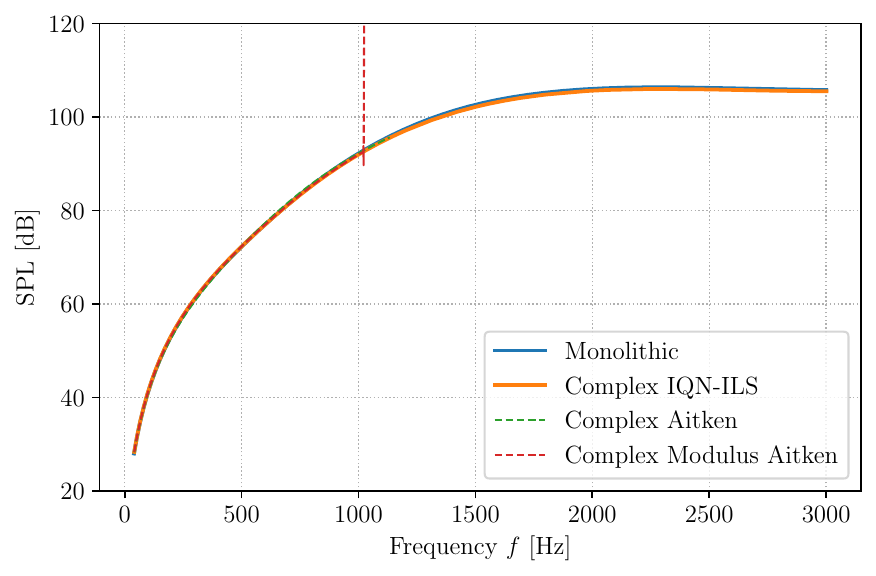}
        \caption{Comparison of the monolithic and strongly coupled partitioned SPL responses obtained using the complex IQN-ILS, complex Aitken, and complex modulus Aitken accelerators.}
        \label{fig:water_hemisphere_spl}
    \end{subfigure}
    \hfill
    \begin{subfigure}[t]{0.49\textwidth}
        \centering
        \includegraphics[width=\textwidth]{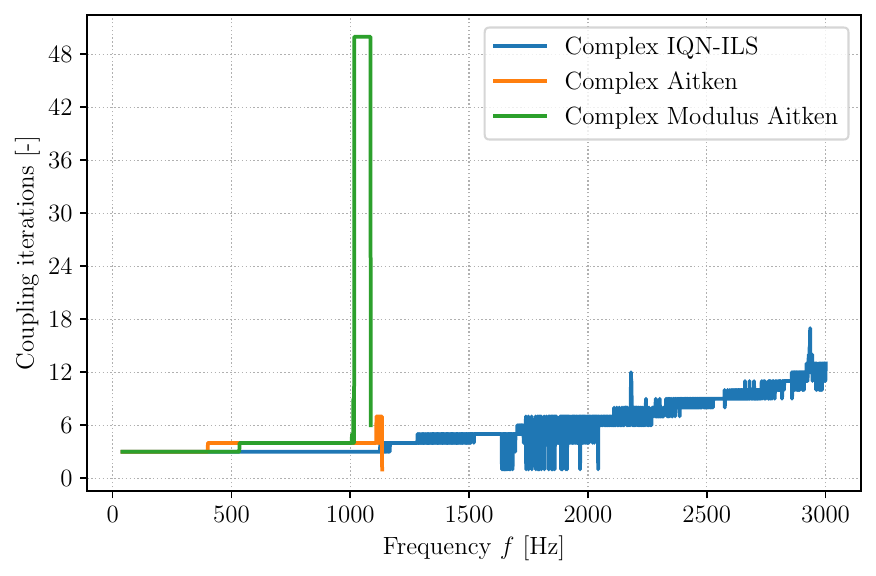}
        \caption{Corresponding number of coupling iterations required for convergence at each excitation frequency. While the complex IQN-ILS formulation converges robustly over the complete investigated frequency range, both Aitken-based approaches fail to converge beyond approximately $1200\,\mathrm{Hz}$.}
        \label{fig:water_hemisphere_iterations}
    \end{subfigure}
    
    \caption{Water-loaded hemispherical shell benchmark.}
    \label{fig:water_hemisphere}
\end{figure}
}

\section{Conclusions}
\label{sec:conclusions}

This work presented a partitioned co-simulation strategy for frequency-domain vibroacoustic analysis combining isogeometric B-Rep
structural models (IBRA) with an external IGA-BEM acoustic solver for
unbounded domains. The proposed framework enables a fully modular and
non-intrusive coupling procedure, while supporting
fully CAD-integrated workflows based on spline-native structural and
acoustic interface representations.

The numerical examples demonstrated very good agreement between the
partitioned and monolithic vibroacoustic solutions for both one-way and
two-way coupled problems. In particular, the proposed strategy accurately captured the acoustic
pressure response and the resonance behavior of the investigated structures
over the considered frequency range.

The mapping studies highlighted the importance of the interface transfer
operator for non-conforming spline-based interfaces. Among the
investigated approaches, the nearest-element mapper provided the most
accurate and reliable transfer behavior for strongly non-conforming
discretizations and highly oscillatory interface fields.

For strongly coupled vibroacoustic problems, the coupling iterations
became highly sensitive near structural resonance frequencies. Since
most existing co-simulation infrastructures exchange interface data as
real-valued vectors, frequency-domain complex quantities are commonly
transferred by separating their real and imaginary components. This
naturally motivates applying standard convergence accelerators
independently to both fields. However, the presented numerical results
showed that this separated-field treatment leads to a deterioration of
the convergence behavior and significantly reduced robustness,
particularly for strongly coupled configurations and near resonance
frequencies.

To address this limitation, complex-valued extensions of the Aitken and
IQN-ILS accelerators were introduced by reconstructing the complex
interface residuals in the co-simulation tool and performing the convergence acceleration directly
in the complex domain. The results demonstrated that consistently
accounting for the coupled amplitude-phase relation of the interface
fields significantly improves the stability and convergence behavior of
the partitioned solution procedure. Among the investigated strategies,
the complex IQN-ILS formulation provided the best overall performance in
terms of accuracy, robustness, and iteration count.

Overall, the proposed partitioned IGA-BEM framework provides a flexible
and accurate alternative to monolithic vibroacoustic formulations while
preserving solver modularity and supporting fully CAD-integrated
analysis of coupled structural-acoustic systems with non-conforming
spline-based interfaces.

\appendix

\section{Implementation of the remote-controlled acoustic participant}
\label{app:remote_controlled_participant}

The remote-controlled execution strategy adopted for the acoustic
participant is implemented through a Python interface layer that
registers a set of callback functions in \texttt{CoSimIO}. This layer
acts as a wrapper between the co-simulation infrastructure and the
MATLAB-based acoustic BEM solver, handling data exchange, solver
execution, and communication with the structural participant.
Listing~\ref{lst:acoustic_participant} summarizes the essential
structure of the acoustic participant used in the present work.

\begin{lstlisting}[
    style=remotecontrol,
    caption={Remote-controlled acoustic BEM participant.},
    label={lst:acoustic_participant}]
# acoustic solver initializes ...

CoSimIO.Connect(...)  # establish connection

# define functions to be registered
def ImportData(identifier):
    if identifier == "disp_real":
        CoSimIO.ImportData(Re(u_Gamma))

    if identifier == "disp_imag":
        CoSimIO.ImportData(Im(u_Gamma))


def SolveSolutionStep():
    # reconstruct complex interface displacement
    u_Gamma = Re(u_Gamma) + i * Im(u_Gamma)

    # solve acoustic BEM problem
    f_Gamma = BEM(u_Gamma, omega)

    # split complex acoustic load
    Re(f_Gamma), Im(f_Gamma)


def ExportData(identifier):
    if identifier == "load_real":
        CoSimIO.ExportData(Re(f_Gamma))

    if identifier == "load_imaginary":
        CoSimIO.ExportData(Im(f_Gamma))


def ExportMesh():
    # construct acoustic interface mesh from integration points
    X_Gamma = ComputeIntegrationPoints()
    CoSimIO.ExportMesh(X_Gamma)


def FinalizeSolutionStep():
    # acoustic postprocessing
    SPL = Postprocess(f_Gamma, omega)


# after defining the functions they are registered in CoSimIO
CoSimIO.Register(ImportData)
CoSimIO.Register(SolveSolutionStep)
CoSimIO.Register(ExportData)
CoSimIO.Register(ExportMesh)
CoSimIO.Register(FinalizeSolutionStep)

# after all functions are registered, execution control is handed over
CoSimIO.Run(...)

CoSimIO.Disconnect(...)  # stop connection
\end{lstlisting}

\section*{Declarations}
\paragraph{\textbf{Conflict of interest}} 
The authors have no competing interests to declare that are relevant to the content of this article.

\section*{Data availability}
\noindent Data will be made available on request.

\section*{Acknowledgments}
\noindent The authors gratefully acknowledge the \emph{Design for IGA-type discretization workflows (GECKO)} project. The Design for IGA-type discretization workflows has received funding from the European Union’s Horizon Europe research and Innovation programme under grant agreement No.~101073106, Call: HORIZON-MSCA-2021-DN-01. Views and opinions expressed are however those of the authors and do not necessarily reflect those of the European Union. The European Union cannot be held responsible for them.


 \bibliographystyle{elsarticle-num} 
 \bibliography{camarotti_paper-refs}

@article{Kiendl2009IGA_KL,
  author  = {Kiendl, J. and Bletzinger, K.-U. and Linhard, J. and W{\"u}chner, R.},
  title   = {Isogeometric shell analysis with Kirchhoff-Love elements},
  journal = {Computer Methods in Applied Mechanics and Engineering},
  volume  = {198},
  number  = {49--52},
  pages   = {3902--3914},
  year    = {2009},
  doi     = {10.1016/j.cma.2009.08.013}
}

@article{irons1969,
author = {Irons, Bruce M. and Tuck, Robert C.},
title = {A version of the Aitken accelerator for computer iteration},
journal = {International Journal for Numerical Methods in Engineering},
volume = {1},
number = {3},
pages = {275-277},
doi = {https://doi.org/10.1002/nme.1620010306},
url = {https://onlinelibrary.wiley.com/doi/abs/10.1002/nme.1620010306},
eprint = {https://onlinelibrary.wiley.com/doi/pdf/10.1002/nme.1620010306},
year = {1969}
}

@article{DEGROOTE2009793,
title = {Performance of a new partitioned procedure versus a monolithic procedure in fluid–structure interaction},
journal = {Computers \& Structures},
volume = {87},
number = {11},
pages = {793-801},
year = {2009},
note = {Fifth MIT Conference on Computational Fluid and Solid Mechanics},
issn = {0045-7949},
doi = {https://doi.org/10.1016/j.compstruc.2008.11.013},
url = {https://www.sciencedirect.com/science/article/pii/S0045794908002605},
author = {Joris Degroote and Klaus-Jürgen Bathe and Jan Vierendeels}
}

@article{DEGROOTE2010446,
title = {Performance of partitioned procedures in fluid–structure interaction},
journal = {Computers \& Structures},
volume = {88},
number = {7},
pages = {446-457},
year = {2010},
issn = {0045-7949},
doi = {https://doi.org/10.1016/j.compstruc.2009.12.006},
url = {https://www.sciencedirect.com/science/article/pii/S0045794909003022},
author = {Joris Degroote and Robby Haelterman and Sebastiaan Annerel and Peter Bruggeman and Jan Vierendeels}
}

@article{Kuttler2008,
  author = {K{\"u}ttler, U. and Wall, W. A.},
  title = {Fixed-point fluid--structure interaction solvers with dynamic relaxation},
  journal = {Computational Mechanics},
  year = {2008},
  volume = {43},
  number = {1},
  pages = {61--72}
}

@article{brebbia1978boundary,
  title={The boundary element method for engineers},
  author={Brebbia, Carlos Alberto},
  journal={Pentech Press},
  year={1978}
}

@book{banerjee1981boundary,
  title={Boundary element methods in engineering science},
  author={Banerjee, P.K. and Banerjee, P.K. and Butterfield, R.},
  isbn={9780070841208},
  lccn={80040983},
  year={1981},
  publisher={McGraw-Hill Book Company (UK)}
}

@book{fish2007first,
  title={A first course in finite elements},
  author={Fish, Jacob and Belytschko, Ted},
  volume={1},
  year={2007},
  publisher={Wiley New York}
}

@article{HUGHES20054135,
	title = {Isogeometric analysis: CAD, finite elements, NURBS, exact geometry and mesh refinement},
	journal = {Computer Methods in Applied Mechanics and Engineering},
	volume = {194},
	number = {39},
	pages = {4135-4195},
	year = {2005},
	issn = {0045-7825},
	doi = {https://doi.org/10.1016/j.cma.2004.10.008},
	author = {T.J.R. Hughes and J.A. Cottrell and Y. Bazilevs}
}

@article{SIMPSON2014265,
	title = {Acoustic isogeometric boundary element analysis},
	journal = {Computer Methods in Applied Mechanics and Engineering},
	volume = {269},
	pages = {265-290},
	year = {2014},
	issn = {0045-7825},
	doi = {https://doi.org/10.1016/j.cma.2013.10.026},
	author = {R.N. Simpson and M.A. Scott and M. Taus and D.C. Thomas and H. Lian}
}

@article{COOX2017186,
	title = {An isogeometric indirect boundary element method for solving acoustic problems in open-boundary domains},
	journal = {Computer Methods in Applied Mechanics and Engineering},
	volume = {316},
	pages = {186-208},
	year = {2017},
	note = {Special Issue on Isogeometric Analysis: Progress and Challenges},
	issn = {0045-7825},
	doi = {https://doi.org/10.1016/j.cma.2016.05.039},
	author = {Laurens Coox and Onur Atak and Dirk Vandepitte and Wim Desmet}
}

@article{marburg2002six,
	title={Six boundary elements per wavelength: Is that enough?},
	author={Marburg, Steffen},
	journal={Journal of Computational Acoustics},
	volume={10},
	number={01},
	pages={25--51},
	year={2002}, 
	doi = {10.1142/S0218396X02001401},
	publisher={World Scientific}
}

@article{diwan2019pollution,
	title={Pollution studies for high order isogeometric analysis and finite element for acoustic problems},
	author={Diwan, Ganesh C and Mohamed, M Shadi},
	journal={Computer Methods in Applied Mechanics and Engineering},
	volume={350},
	pages={701--718},
	year={2019},
	doi = {https://doi.org/10.1016/j.cma.2019.03.031},
	publisher={Elsevier}
}

@article{burnett1994three,
  title={A three-dimensional acoustic infinite element based on a prolate spheroidal multipole expansion},
  author={Burnett, David S},
  journal={The Journal of the Acoustical Society of America},
  volume={96},
  number={5},
  pages={2798--2816},
  year={1994},
  publisher={Acoustical Society of America}
}

@article{astley2000numerical,
  title={Numerical studies of conjugated infinite elements for acoustical radiation},
  author={Astley, RJ and Hamilton, JA},
  journal={Journal of Computational Acoustics},
  volume={8},
  number={01},
  pages={1--24},
  year={2000},
  publisher={World Scientific}
}

@article{berenger1994perfectly,
  title={A perfectly matched layer for the absorption of electromagnetic waves},
  author={Berenger, Jean-Pierre},
  journal={Journal of computational physics},
  volume={114},
  number={2},
  pages={185--200},
  year={1994},
  publisher={Elsevier}
}

@article{SICKLINGER2015134,
title = {Fully coupled co-simulation of a wind turbine emergency brake maneuver},
journal = {Journal of Wind Engineering and Industrial Aerodynamics},
volume = {144},
pages = {134-145},
year = {2015},
note = {Selected papers from the 6th International Symposium on Computational Wind Engineering CWE 2014},
issn = {0167-6105},
doi = {https://doi.org/10.1016/j.jweia.2015.03.021},
url = {https://www.sciencedirect.com/science/article/pii/S0167610515000811},
author = {Stefan Sicklinger and Christopher Lerch and Roland Wüchner and Kai-Uwe Bletzinger}
}

@article{liu2018isogeometric,
  title={Isogeometric FEM-BEM coupled structural-acoustic analysis of shells using subdivision surfaces},
  author={Liu, Zhaowei and Majeed, Musabbir and Cirak, Fehmi and N. Simpson, Robert},
  journal={International Journal for Numerical Methods in Engineering},
  volume={113},
  number={9},
  pages={1507--1530},
  year={2018},
  publisher={Wiley Online Library}
}

@article{wu2021isogeometric,
  title={Isogeometric symmetric {FE-BE} coupling method for acoustic-structural interaction},
  author={Wu, YH and Dong, CY and Yang, HS and Sun, FL},
  journal={Applied Mathematics and Computation},
  volume={393},
  pages={125758},
  year={2021},
  publisher={Elsevier}
}

@article{wu20203d,
  title={A 3D isogeometric {FE-IBE} coupling method for acoustic-structural interaction problems with complex coupling models},
  author={Wu, YH and Dong, CY and Yang, HS},
  journal={Ocean Engineering},
  volume={218},
  pages={108183},
  year={2020},
  publisher={Elsevier}
}

@article{sicklinger2014,
author = {Sicklinger, S. and Belsky, V. and Engelmann, B. and Elmqvist, H. and Olsson, H. and Wüchner, R. and Bletzinger, K.-U.},
title = {Interface Jacobian-based Co-Simulation},
journal = {International Journal for Numerical Methods in Engineering},
volume = {98},
number = {6},
pages = {418-444},
doi = {https://doi.org/10.1002/nme.4637},
url = {https://onlinelibrary.wiley.com/doi/abs/10.1002/nme.4637},
eprint = {https://onlinelibrary.wiley.com/doi/pdf/10.1002/nme.4637},
year = {2014}
}

@phdthesis{bucher2024,
	author = {Bucher, Philipp Lukas},
	title = {CoSimulation and Mapping for large scale engineering applications},
	year = {2024},
	school = {Technische Universität München},
	pages = {200},
	language = {en},
	note = {},
	url = {https://mediatum.ub.tum.de/1714023},
	doi = {},
}

@Article{dadvand2010,
  author  = {P. Dadvand and R. Rossi and E. O{\~n}ate},
  title   = {An Object-oriented Environment for Developing Finite Element Codes for Multi-disciplinary Applications},
  journal = {Archives of Computational Methods in Engineering},
  year    = {2010},
  volume  = {17},
  number  = {3},
  pages   = {253--297},
  doi = {https://doi.org/10.1007/s11831-010-9045-2}
}

@article{dadvand2013,
  author  = {P. Dadvand and R. Rossi and M. Gil and X. Martorell and J. Cotela and E. Juanpere and S.R. Idelsohn and E. O{\~n}ate},
  title   = {Migration of a generic multi-physics framework to HPC environments},
  journal = {Computers \& Fluids},
  volume  = {80},
  pages   = {301--309},
  year    = {2013},
  doi = {https://doi.org/10.1016/j.compfluid.2012.02.004}
}

@article{Breitenberger2015,
  author    = {Michael Breitenberger and Andreas Apostolatos and Philipp Bucher and Roland Wüchner and Kai-Uwe Bletzinger},
  title     = {{\href{https://doi.org/10.1016/j.cma.2014.09.033}{Analysis in computer aided design: Nonlinear isogeometric B-Rep analysis of shell structures}}},
  journal   = {Computational Methods in Applied Mechanics and Engineering},
  volume    = {284},
  pages     = {401--457},
  year      = {2015}
}

@article{Teschemacher2018,
  author    = {Tobias Teschemacher and Anna M. Bauer and Thomas Oberbichler and Michael Breitenberger and Riccardo Rossi and Roland Wüchner and Kai-Uwe Bletzinger},
  title     = {{\href{https://doi.org/10.1186/s40323-018-0109-4}{Realization of CAD-integrated shell simulation based on isogeometric B-Rep analysis}}},
  journal   = {Advanced Modeling and Simulation in Engineering Sciences},
  volume    = {5},
  pages     = {19},
  year      = {2018}
}

@article{Teschemacher2022,
  author    = {Tobias Teschemacher and Anna M. Bauer and Ricky Aristio and Manuel Meßmer and Roland Wüchner and Kai-Uwe Bletzinger},
  title     = {{\href{https://doi.org/10.1007/s00366-022-01732-4}{Concepts of data collection for the CAD-integrated isogeometric analysis}}},
  journal   = {Engineering with Computers},
  volume    = {38},
  number    = {6},
  pages     = {5675--5693},
  year      = {2022}
}

@article{Apostolatos2021,
  author    = {Apostolatos, Athanasios and Emiroğlu, Alperen and Shayegan, Sadegh and others},
  title     = {An isogeometric b-rep mortar-based mapping method for non-matching grids in fluid-structure interaction},
  journal   = {Advanced Modeling and Simulation in Engineering Sciences},
  volume    = {8},
  number    = {9},
  year      = {2021},
  doi       = {10.1186/s40323-021-00190-9},
  url       = {https://doi.org/10.1186/s40323-021-00190-9}
}

@article{bentley1975kdtree,
  author  = {Bentley, Jon Louis},
  title   = {Multidimensional Binary Search Trees Used for Associative Searching},
  journal = {Communications of the ACM},
  volume  = {18},
  number  = {9},
  pages   = {509--517},
  year    = {1975},
  doi     = {10.1145/361002.361007}
}

@article{meagher1982octree,
  author  = {Meagher, Donald},
  title   = {Geometric Modeling Using Octree Encoding},
  journal = {Computer Graphics and Image Processing},
  volume  = {19},
  number  = {2},
  pages   = {129--147},
  year    = {1982},
  doi     = {10.1016/0146-664X(82)90104-6}
}

@book{JungerFeit1986,
  author    = {Junger, M. C. and Feit, D.},
  title     = {Sound, Structures, and Their Interaction},
  publisher = {MIT Press},
  address   = {Cambridge, MA},
  year      = {1986},
  isbn      = {9780262100371}
}

@article{Everstine1990,
  author  = {Everstine, G. C. and Henderson, F. M.},
  title   = {Coupled finite element/boundary element approach for fluid--structure interaction},
  journal = {The Journal of the Acoustical Society of America},
  volume  = {87},
  number  = {5},
  pages   = {1938--1947},
  year    = {1990},
  doi     = {10.1121/1.399186}
}

@article{RodriguezTembleque2015,
title = {Partitioned solution strategies for coupled BEM–FEM acoustic fluid–structure interaction problems},
journal = {Computers \& Structures},
volume = {152},
pages = {45-58},
year = {2015},
issn = {0045-7949},
doi = {https://doi.org/10.1016/j.compstruc.2015.02.018},
url = {https://www.sciencedirect.com/science/article/pii/S0045794915000565},
author = {Luis Rodríguez-Tembleque and José A. González and Antonio Cerrato}
}

@article{Bunting2021,
    author = {Bunting, Gregory and Miller, Scott T.},
    title = {Partitioned Coupling for Structural Acoustics},
    journal = {Journal of Vibration and Acoustics},
    volume = {142},
    number = {1},
    pages = {011012},
    year = {2019},
    month = {11},
    issn = {1048-9002},
    doi = {10.1115/1.4045215},
    url = {https://doi.org/10.1115/1.4045215},
    eprint = {https://asmedigitalcollection.asme.org/vibrationacoustics/article-pdf/142/1/011012/6448035/vib_142_1_011012.pdf},
}

@inproceedings{Kersschot2020,
  author    = {Kersschot, Jurgen and Denayer, Herv{\'e} and De Roeck, Wim and Desmet, Wim},
  title     = {Simulation of Strong Vibro-Acoustic Coupling Effects in Ducts Using a Partitioned Approach in the Time Domain},
  booktitle = {Proceedings of the International Conference on Noise and Vibration Engineering (ISMA 2020)},
  year      = {2020},
  address   = {Leuven, Belgium},
  pages      = {285--294}
}

@article{preCICEv2,
  author = {Chourdakis, G and Davis, K and Rodenberg, B and Schulte, M and Simonis, F and Uekermann, B and Abrams, G and Bungartz, HJ and Cheung Yau, L and Desai, I and Eder, K and Hertrich, R and Lindner, F and Rusch, A and Sashko, D and Schneider, D and Totounferoush, A and Volland, D and Vollmer, P and Koseomur, OZ},
  title = {{preCICE} v2: A sustainable and user-friendly coupling library [version 2; peer review: 2 approved]
  },
  journal = {Open Research Europe},
  volume = {2},
  year = {2022},
  number = {51},
  doi = {10.12688/openreseurope.14445.2},
  url = {https://doi.org/10.12688/openreseurope.14445.2}
}

@book{zienkiewicz2005finite,
  author    = {Zienkiewicz, O. C. and Taylor, R. L. and Zhu, J. Z.},
  title     = {The Finite Element Method: Its Basis and Fundamentals},
  edition   = {7},
  publisher = {Butterworth-Heinemann},
  address   = {Oxford},
  year      = {2013},
  isbn      = {9781856176338}
}

@book{bathe2006finite,
  author    = {Bathe, Klaus-J{\"u}rgen},
  title     = {Finite Element Procedures},
  publisher = {Klaus-J{\"u}rgen Bathe},
  address   = {Watertown, MA},
  year      = {2006},
  isbn      = {9780979004900}
}

@article{Delaisse2023QuasiNewton,
  author  = {A. Delaiss{\'e} and J. Degroote},
  title   = {Quasi-Newton Methods for Partitioned Simulation of Fluid-Structure Interaction},
  journal = {Archives of Computational Methods in Engineering},
  volume  = {30},
  pages   = {2515--2558},
  year    = {2023}
}

@article{BENSON2010276,
title = {Isogeometric shell analysis: The Reissner–Mindlin shell},
journal = {Computer Methods in Applied Mechanics and Engineering},
volume = {199},
number = {5},
pages = {276-289},
year = {2010},
note = {Computational Geometry and Analysis},
issn = {0045-7825},
doi = {https://doi.org/10.1016/j.cma.2009.05.011},
url = {https://www.sciencedirect.com/science/article/pii/S0045782509001820},
author = {D.J. Benson and Y. Bazilevs and M.C. Hsu and T.J.R. Hughes},
}

@article{COOX2017235,
title = {A robust patch coupling method for NURBS-based isogeometric analysis of non-conforming multipatch surfaces},
journal = {Computer Methods in Applied Mechanics and Engineering},
volume = {316},
pages = {235-260},
year = {2017},
note = {Special Issue on Isogeometric Analysis: Progress and Challenges},
issn = {0045-7825},
doi = {https://doi.org/10.1016/j.cma.2016.06.022},
url = {https://www.sciencedirect.com/science/article/pii/S0045782516306120},
author = {Laurens Coox and Francesco Greco and Onur Atak and Dirk Vandepitte and Wim Desmet}
}

@article{COOX2017505,
title = {A flexible approach for coupling NURBS patches in rotationless isogeometric analysis of Kirchhoff–Love shells},
journal = {Computer Methods in Applied Mechanics and Engineering},
volume = {325},
pages = {505-531},
year = {2017},
issn = {0045-7825},
doi = {https://doi.org/10.1016/j.cma.2017.07.022},
url = {https://www.sciencedirect.com/science/article/pii/S0045782517305674},
author = {Laurens Coox and Florian Maurin and Francesco Greco and Elke Deckers and Dirk Vandepitte and Wim Desmet}
}

@article{MatthiesSteindorf2003,
    author  = {Matthies, Hermann G. and Steindorf, Jan},
    title   = {Partitioned strong coupling algorithms for fluid--structure interaction},
    journal = {Computers \& Structures},
    volume  = {81},
    pages   = {805--812},
    year    = {2003},
    doi     = {10.1016/S0045-7949(02)00409-1}
}

@article{Mehl2016,
    author  = {Mehl, Miriam and Uekermann, Benjamin and Bijl, Hester
               and Blom, Fred and Bungartz, Hans-Joachim
               and van Zuijlen, Alexander H.},
    title   = {Parallel coupling numerics for partitioned
               fluid--structure interaction simulations},
    journal = {Computers \& Mathematics with Applications},
    volume  = {71},
    number  = {4},
    pages   = {869--891},
    year    = {2016},
    doi     = {10.1016/j.camwa.2015.12.025}
}





\end{document}